\documentclass[12pt]{article}
\usepackage{amsfonts}
\usepackage{amsmath, amsthm}
\usepackage{epsfig}
\usepackage{algorithm}

\usepackage{soul}
\usepackage[normalem]{ulem}
\usepackage{amsfonts}
\usepackage{subcaption}
\usepackage{epsfig}
\usepackage{booktabs}
\usepackage{lscape}
\usepackage{multirow}
\usepackage{longtable}
\usepackage{rotating}
\usepackage{array}
\usepackage{multicol}
\usepackage{graphicx}
\usepackage{bigstrut}
\usepackage{setspace}
\usepackage{epstopdf}
\usepackage{mathrsfs}
\usepackage{amsfonts}
\usepackage{epsfig}
\usepackage{booktabs}
\usepackage{threeparttable}
\usepackage{epstopdf}
\usepackage{lscape}
\usepackage{multirow}

\usepackage{algorithmic}
\usepackage{amsmath,verbatim,color,amssymb,epsfig}
\usepackage{bm}
\usepackage{amsfonts}
\usepackage{epsfig}
\usepackage{multirow}
\usepackage{graphicx}
\usepackage{array}
\usepackage[table]{xcolor}
\definecolor{Gray}{gray}{0.80}
\usepackage{lscape}
\usepackage{cite}
\usepackage{natbib}
\usepackage[normalem]{ulem}
\usepackage[colorlinks=true,urlcolor=blue,citecolor=black,linkcolor=blue,bookmarks=true]{hyperref}
\usepackage{caption}
\begin{document}
\def\eqx"#1"{{\label{#1}}}
\def\eqn"#1"{{\ref{#1}}}

\makeatletter % make @ act like a letter
\@addtoreset{equation}{section}
\makeatother  % make @ act like a non-letter

\def\yincomment#1{\vskip 2mm\boxit{\vskip 2mm{\color{red}\bf#1} {\color{blue}\bf --Yin\vskip 2mm}}\vskip 2mm}
\def\squarebox#1{\hbox to #1{\hfill\vbox to #1{\vfill}}}
\def\boxit#1{\vbox{\hrule\hbox{\vrule\kern6pt
          \vbox{\kern6pt#1\kern6pt}\kern6pt\vrule}\hrule}}

\newcommand{\blue}[1]{\textcolor{blue}{{#1}}}
\def\theequation{\thesection.\arabic{equation}}
\newcommand{\ds}{\displaystyle}

\newcommand{\bJ}{\mbox{\bf J}}
\newcommand{\bF}{\mbox{\bf F}}
\newcommand{\bM}{\mbox{\bf M}}
\newcommand{\bR}{\mbox{\bf R}}
\newcommand{\bZ}{\mbox{\bf Z}}
\newcommand{\bX}{\mbox{\bf X}}
\newcommand{\bx}{\mbox{\bf x}}
\newcommand{\bQ}{\mbox{\bf Q}}
\newcommand{\bH}{\mbox{\bf H}}
\newcommand{\bh}{\mbox{\bf h}}
\newcommand{\bz}{\mbox{\bf z}}
\newcommand{\ba}{\mbox{\bf a}}
\newcommand{\be}{\mbox{\bf e}}
\newcommand{\bG}{\mbox{\bf G}}
\newcommand{\bB}{\mbox{\bf B}}
\newcommand{\bb}{\mbox{\bf b}}
\newcommand{\bA}{\mbox{\bf A}}
\newcommand{\bC}{\mbox{\bf C}}
\newcommand{\bI}{\mbox{\bf I}}
\newcommand{\bD}{\mbox{\bf D}}
\newcommand{\bU}{\mbox{\bf U}}
\newcommand{\bc}{\mbox{\bf c}}
\newcommand{\bd}{\mbox{\bf d}}
\newcommand{\bs}{\mbox{\bf s}}
\newcommand{\bS}{\mbox{\bf S}}
\newcommand{\bV}{\mbox{\bf V}}
\newcommand{\bv}{\mbox{\bf v}}
\newcommand{\bW}{\mbox{\bf W}}
\newcommand{\bw}{\mbox{\bf w}}
\newcommand{\bg}{\mbox{\bf g}}
\newcommand{\bu}{\mbox{\bf u}}
\def\bb{{\bf b}}

\newcommand{\bcU}{\boldsymbol{\cal U}}
\newcommand{\bbeta}{\boldsymbol{\beta}}
\newcommand{\bdelta}{\boldsymbol{\delta}}
\newcommand{\bDelta}{\boldsymbol{\Delta}}
\newcommand{\boldeta}{\boldsymbol{\eta}}
\newcommand{\bxi}{\boldsymbol{\xi}}
\newcommand{\bGamma}{\boldsymbol{\Gamma}}
\newcommand{\bSigma}{\boldsymbol{\Sigma}}
\newcommand{\balpha}{\boldsymbol{\alpha}}
\newcommand{\bOmega}{\boldsymbol{\Omega}}
\newcommand{\btheta}{\boldsymbol{\theta}}
\newcommand{\bmu}{\boldsymbol{\mu}}
\newcommand{\bnu}{\boldsymbol{\nu}}
\newcommand{\bgamma}{\boldsymbol{\gamma}}

\newtheorem{thm}{Theorem}[section]
\newtheorem{lem}{Lemma}[section]
\newtheorem{rem}{Remark}[section]
\newtheorem{cor}{Corollary}[section]
\newcolumntype{L}[1]{>{\raggedright\let\newline\\\arraybackslash\hspace{0pt}}m{#1}}
\newcolumntype{C}[1]{>{\centering\let\newline\\\arraybackslash\hspace{0pt}}m{#1}}
\newcolumntype{R}[1]{>{\raggedleft\let\newline\\\arraybackslash\hspace{0pt}}m{#1}}

\newcommand{\tabincell}[2]{\begin{tabular}{@{}#1@{}}#2\end{tabular}}

\baselineskip=24pt
\begin{center}
{\Large \bf
$P$-value: A Bless or A Curse for Evidence-Based Studies?
}
%Is it necessary to control the type I error
%rate in Bayesian adaptive sequential designs?}
\end{center}

\vspace{2mm}
\begin{center}
{\bf Haolun Shi and Guosheng Yin$^{*}$ }
\end{center}

\begin{center}

Department of Statistics and Actuarial Science\\
The University of Hong Kong\\
Pokfulam Road, Hong Kong\\

\vspace{2mm}

{\em *email}: gyin@hku.hk\\
\end{center}
\noindent{Abstract.}
As a convention, $p$-value is often computed in frequentist hypothesis testing
and compared with the nominal significance level of 0.05 to determine
whether or not to reject the null hypothesis.
The smaller the $p$-value, the more significant the statistical test.
We consider both one-sided and two-sided hypotheses in the
composite hypothesis setting. For one-sided hypothesis tests,
we establish the equivalence of $p$-value and the Bayesian posterior
probability of the null hypothesis, which
renders $p$-value an explicit interpretation of how strong
the data support the null. For two-sided hypothesis tests of
a point null, we recast the problem as a combination of two one-sided hypotheses alone the opposite directions and put forward the notion of a two-sided posterior probability,
which also has an equivalent relationship with the (two-sided) $p$-value.
Extensive simulation studies are conducted to demonstrate the
Bayesian posterior probability interpretation for the $p$-value.
Contrary to common criticisms of the use of $p$-value in evidence-based
studies, we justify its utility and reclaim its importance from the Bayesian
perspective, and recommend the continual use of $p$-value in hypothesis testing.
After all, $p$-value is not all that bad.

\vspace{0.5cm}

\newpage
\section{Introduction}

Hypothesis testing is ubiquitous in modern statistical applications, which
permeates many different fields such as biology, medicine,
phycology, economics, and engineering etc.
As a critical component of the hypothesis testing procedure (Lehmann and Romano, 2005),
$p$-value is defined as the probability of observing the
random data as or more extreme than the observed
given the null hypothesis being true.
In general, the statistical significance level
or the type I error rate is set at 5\%, so that a $p$-value below
5\% is considered significant leading to rejection of the null hypothesis,
and that above 5\% insignificant resulting in failure to reject the null.

Although $p$-value is the most commonly used summary measure for evidence
or strength in the data regarding the null hypothesis, it
has been the center of controversies and debates for decades.
To clarify ambiguities surrounding $p$-value,
the American Statistical Association (2016) gave statements on $p$-value
and, in particular, the second point states that ``$P$-values do not measure
the probability that the studied
hypothesis is true, or the probability that the data
were produced by random chance alone.'' It is often argued
that $p$-value only gives information on how incompatible
the data are with the null hypothesis, but it does not provide
any information on how likely the data would occur under
the alternative hypothesis.

Extensive investigations have been conducted on the inadequacy of the $p$-value.
Rosenthal and Rubin (1983) studied how $p$-value can be adjusted to
allow for greater power when an order of importance exists on the hypothesis tests.
Royall (1986) investigated the effect of sample size on $p$-value. Schervish (1996)
described computation of the $p$-value for one-sided point null hypotheses, and also
discussed the intermediate interval hypothesis.
Hung et al. (1997) studied the behavior of
$p$-value under the alternative hypothesis,
which depends on both the true value of
the tested parameter and sample size.
Rubin (1998) proposed an alternative randomization-based
$p$-value for double-blind trials with non-compliance.
Sackrowitz and Samuel-Cahn (1999) promoted more
widespread use of the expected $p$-value in practice.
Donahue (1999) suggested that the distribution of the $p$-value
under the alternative hypothesis provide more information for rejection of implausible
alternative hypotheses. As there is a widespread notion that medical research
is interpreted mainly based on $p$-value,
Ioannidis (2005) claimed that most of the published findings are false.
Hubbard and Lindsay (2008) showed that $p$-value
tends to exaggerate the evidence against the null hypothesis.
Simmons et al. (2011) demonstrated that $p$-value is subject to
manipulation to achieve the threshold of 0.05 and cautioned against its use.
Nuzzo (2014) gave an editorial on why $p$-value alone
cannot serve as adequate statistical evidence for inference.

Criticisms on $p$-value and null hypothesis significance
testing have become even more contentious in recent years.
If the key words ``misuse of $p$-value'' or ``ban $p$-value'' are
used in Google search, millions of queries can be found
to attack and bash $p$-value.
More seriously, several journals, e.g.,
{\em Basic and Applied Social Psychology} and {\em Political Analysis},
have made claims to ban the use of $p$-value in their publications
(Trafimow and Marks, 2015; Gill, 2018).
The controversy over $p$-value has recently been reignited,
which is more centered around the proposals to adjust,
abandon or provide alternatives to $p$-value.
Fidler et al. (2004) and Ranstam (2012) recommended use of
the confidence interval as an alternative to $p$-value,
and Cumming (2014) called for abandoning
$p$-value in favor of reporting the confidence interval.
Colquhoun (2014) investigated the issue of misinterpretation of
$p$-value as a culprit for the high false discovery rate.
Concato and Hartigan (2016) suggested that $p$-value should not be the
primary focus of attention or the sole basis for evaluation of scientific results.
McShane et al. (2017) recommended that the role of $p$-value
as a threshold for screening scientific findings should be demoted, and that
$p$-value should not take priority over other statistical measures.
In the aspect of reproducibility concerns of scientific research,
Johnson (2013) traced one major cause of nonreproducibility as
the routine use of the null hypothesis testing procedure.
Leek et al. (2017) proposed abandonment of $p$-value thresholding and
transparent reporting of false positive risk as remedies to the replicability
issue in science. Benjamin et al. (2018) recommended shifting the
significance threshold from 0.05 to 0.005, while Trafimow et al.
(2018) argued that such a shift is futile and unacceptable.

Bayesian approaches are often advocated as a solution to the
crisis resulting from abusing the $p$-value.
Goodman (1999) strongly supported use of the Bayes factor in contrast to
$p$-value as a measure of evidence for medical evidence-based
research. Rubin (1984) proposed the predictive $p$-value as the tail-area
probability of the posterior predictive distribution, and Meng (1994)
further studied its properties.
In the applications to psychology, Wagenmakers (2007)
revealed the issues with $p$-value and recommended
use of the Bayesian information criterion instead.
In an effort to support the wider use of Bayesian statistics,
Lee (2010) demonstrated that Bayesian approaches
provide a superior alternative to the frequentist methods using $p$-values.
Alongside its ban on $p$-value, the journal of
{\em Basic and Applied Social Psychology} gave endorsement of
Bayesian approaches (Trafimow and Marks, 2015).
Briggs (2017) proposed that $p$-value should be proscribed and be substituted with the Bayesian posterior probability, while Savalei and Dunn (2015) expressed skepticism on the utility of abandoning $p$-value and resorting to alternative hypothesis testing paradigms, such as the Bayesian approach, in solving the reproducibility issue.

On the other hand, extensive research has been conducted in an attempt to reconcile
or account for the differences between frequentist and Bayesian
hypothesis testing approaches (Berger, 2003; and Bayarri and Berger, 2004).
For hypothesis testing, Berger and Sellke (1987),
Berger and Delampady (1987), and Casella and Berger (1987)
investigated the relationships between $p$-value and
the Bayesian measure of evidence against
the null hypothesis. In particular, they provided an in-depth study of one-sided hypothesis
testing and point null cases, and also discussed the posterior probability of the null hypothesis
with respect to various prior distributions including the mixture prior distribution with a
point mass at the null and the other more broad distribution over the alternative (Lindley, 1957). Sellke, Bayarri, and Berger (2001) proposed to calibrate $p$-value
for testing precise null hypotheses.
%Based on the uniformly most powerful Bayesian test (Johnson, 2013b),
%the size of a frequentist hypothesis test can
%be equated with the evidence threshold in a Bayesian test, and
%$p$-value with the Bayes factor.

Although $p$-value is often regarded as an inadequate and insufficient representation
of statistical evidence, it did not stall the scientific advancement
in the past years.
Jager and Leek (2014) surveyed high-profile medical journals
and estimated the rate of false discoveries in the medical
literature using reported $p$-values as the data, which
led to a conclusion that the medical literature remains a reliable
record of scientific progress.
Murtaugh (2014) defended the use of $p$-value based on the ground
that it is closely linked to the confidence interval and to the difference in
Akaike's information criterion.
Despite the fact that Bayesian alternatives are often recommended
as superior solutions to the various notorious drawbacks of $p$-value,
in many common cases, $p$-value in fact has a simple and clear Bayesian interpretation.
We present the relationship between the frequentist $p$-value and
Bayesian posterior probability in several commonly encountered settings
in clinical trials,
and show that in both one-sided and two-sided hypothesis tests,
asymptotic equivalence, sometimes exact equivalence, can be established.
Although in terms of definition, $p$-value is not the probability
that the null hypothesis is true, contrary to the conventional notion,
it does have a close correspondence to the Bayesian posterior probability of the null hypothesis being true. Based on the theoretical results of Dudley and Haughton (2002), we present several
cases where $p$-value and the posterior probability of the null
are equivalent for one-sided tests. Further, we extend such equivalence
results to two-sided hypothesis testing problems,
where most of the controversies and discrepancies lie.
In particular, we introduce
the notion of two-sided posterior probability which matches the $p$-value
from a two-sided hypothesis test.
After all, we conclude that $p$-value is not all that bad.

The rest of the paper is organized as follows. In Section 2,
we present a motivating example that shows the similarity in
operating characteristics of a frequentist hypothesis test and a
Bayesian counterpart using the posterior probability. In Section 3,
we show that $p$-value and the posterior probability have an
equivalence relationship for the case of binary outcomes. In Section 4, we present such equivalence properties for univariate normal data with known and unknown variances respectively,
and in Section 5, we develop similar results for hypothesis tests involving
multivariate data.
%normal data.
Finally, Section 6 concludes with some remarks.

\section{Motivating Example}
The use of binary endpoint is common in clinical trial design. Frequentist design typically utilizes an exact binomial test or $Z$-test based on normal approximation, and Bayesian design often bases the decision on the posterior probabilities.
As a motivating example, we consider a two-arm clinical trial comparing the response rate of an experimental drug $p_E$ versus that of the standard drug $p_S$.
We are interested in testing a one-sided hypothesis,
\begin{equation}\label{binaryh}
H_0\mbox{:} \ p_E \le p_S \quad  {\rm versus} \quad H_1\mbox{:} \ p_E > p_S.
\end{equation}
When there is sufficient evidence to support $H_1$, we would reject $H_0$ and claim
that the experimental treatment is superior.

Under the frequentist approach, we
construct a $Z$-test statistic,
\begin{equation}\label{binaryztest}
 Z = \frac{\hat p_E-\hat p_S}{[{\{ {{\hat p_E}(1-{\hat p_E})} + {{\hat p_S}(1-{\hat p_S})}  \}/n}]^{1/2}},
 \end{equation}
 where $n$ is the sample size per arm, $\hat p_E=y_E/n$ and $\hat p_S=y_S/n$ are the sample proportions, $y_E$ and $y_S$ are the numbers of responders in the respective arms.
 We reject the null hypothesis if $Z > z_{\alpha}$, where $z_{\alpha}$ is the $100(1-\alpha)$th
percentile of the standard normal distribution.

Under the Bayesian framework,
we assume beta prior distributions for $p_E$ and $p_S$, i.e.,
$p_E\sim{\rm Beta}(a_E,b_E)$ and $p_S \sim {\rm Beta}(a_S,b_S)$.
The binomial likelihood function for group $g$
can be formulated as
$${P}({y_g}|{p_g}) = {\displaystyle {n \choose y_g}}
{p_g^{y_g}}{(1 - p_g)^{n - y_g}},
\quad g=E, S.$$
The posterior distribution of $p_g$
is given by
\begin{eqnarray}
{p_g}|{y_g}&\sim& {\rm Beta}(a_g+y_g,b_g+n-y_g),\nonumber
\end{eqnarray}
for which the density function is denoted by $f({p_g}|{y_g})$.
Let $\eta$ be a prespecified cutoff probability boundary. We declare
treatment superiority if the posterior probability of $p_E$ greater than $p_S$ exceeds threshold $\eta$.
Based on the posterior probability, we can construct a Bayesian decision rule
so that the experimental treatment is declared as superior if
\begin{equation}\label{binarybtest}
\Pr(H_1|y_E,y_S) = \Pr(p_E>p_S|y_E,y_S)> \eta,
\end{equation}
where
\[
\Pr({p_E} > {p_S}|{y_E},{y_S})
=\int_0^1{\int_{{p_S}}^1{f({p_E}|{y_E})f({p_S}|y_S)}}d{p_E}d{p_S}.
\]
Otherwise, we fail to declare treatment superiority, i.e., fail to reject the null hypothesis.

To maintain the frequentist type I error rate at $\alpha$, we need to set $\eta = 1-\alpha$. The exact probabilities of committing type I and type II errors under the frequentist design are respectively given by
$$
\alpha={\sum\limits_{{y_E}=0}^{{n}}{\sum\limits_{{y_S}=0}^{{n}}}}{P}({y_E}|{p_E=p_S})
{P}({y_S}|{p_S})I(Z>z_\alpha),$$
and
$$\beta={\sum\limits_{{y_E}=0}^{{n}}{\sum\limits_{{y_S}=0}^{{n}}}}{P}({y_E}|{p_E=p_S+\delta})
{{P}}({y_S}|{p_S})I(Z<z_\alpha),
$$
where $\delta$ is the desired treatment difference and $I(\cdot)$ is the indicator function.
The exact error rates under the Bayesian test can
be derived similarly by replacing
$Z>z_\alpha$ with $\Pr(p_E>p_S|y_E,y_S) > 1-\alpha$ inside the indicator function.

\begin{figure}[htb]
\begin{center}
\includegraphics[height=8cm,width=8cm]{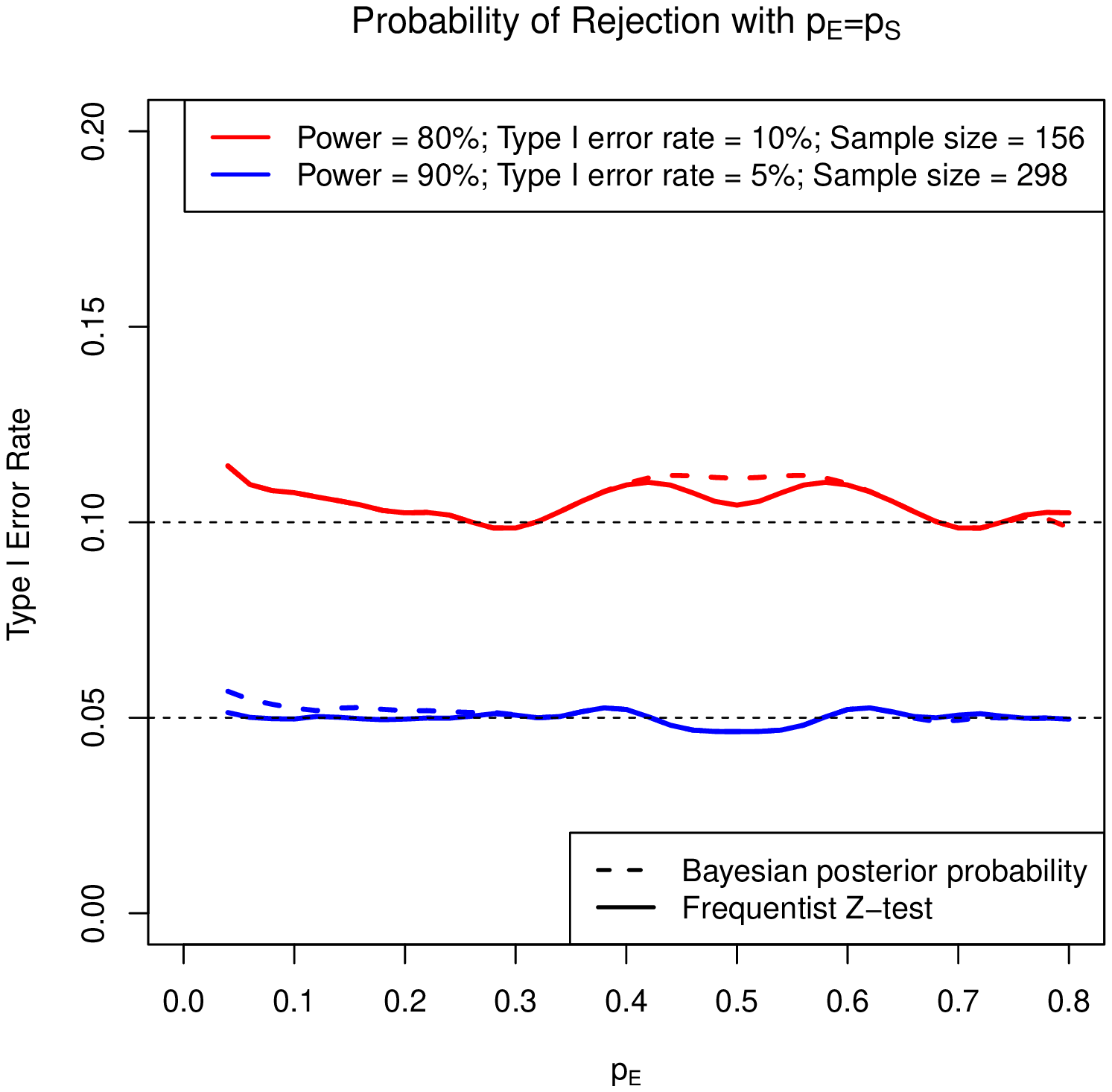}
\includegraphics[height=8cm,width=8cm]{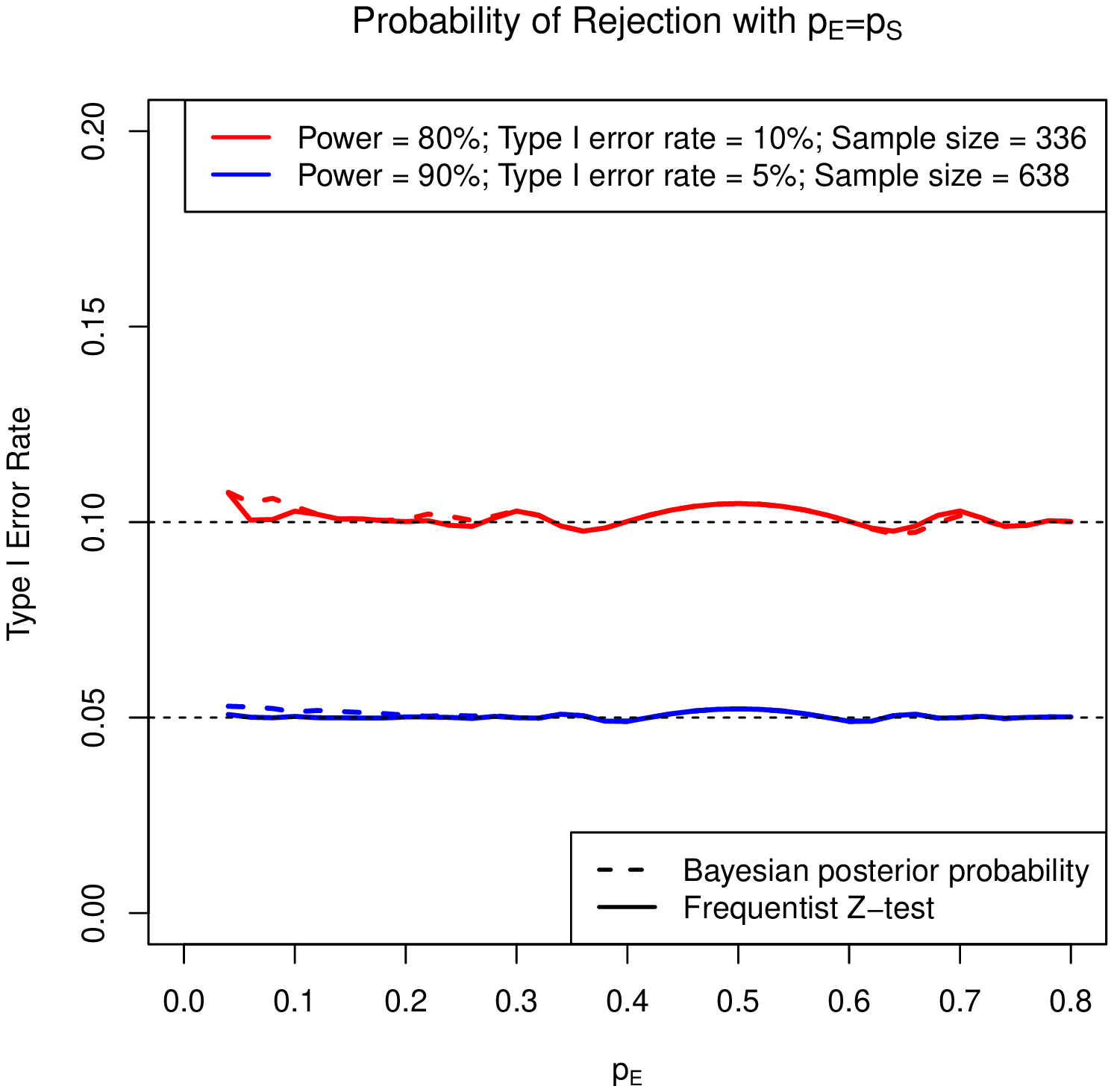}\\
\includegraphics[height=8cm,width=8cm]{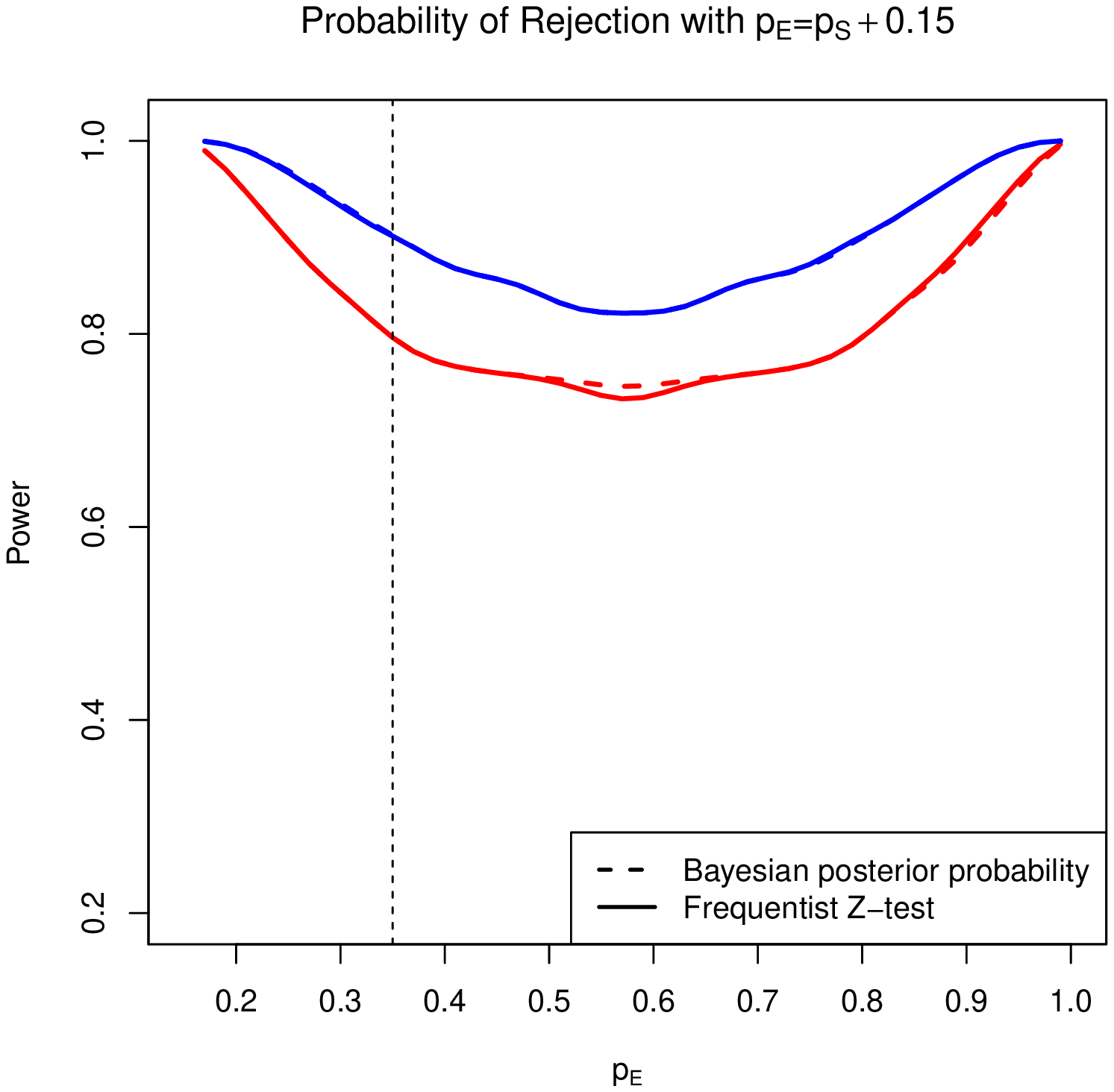}
\includegraphics[height=8cm,width=8cm]{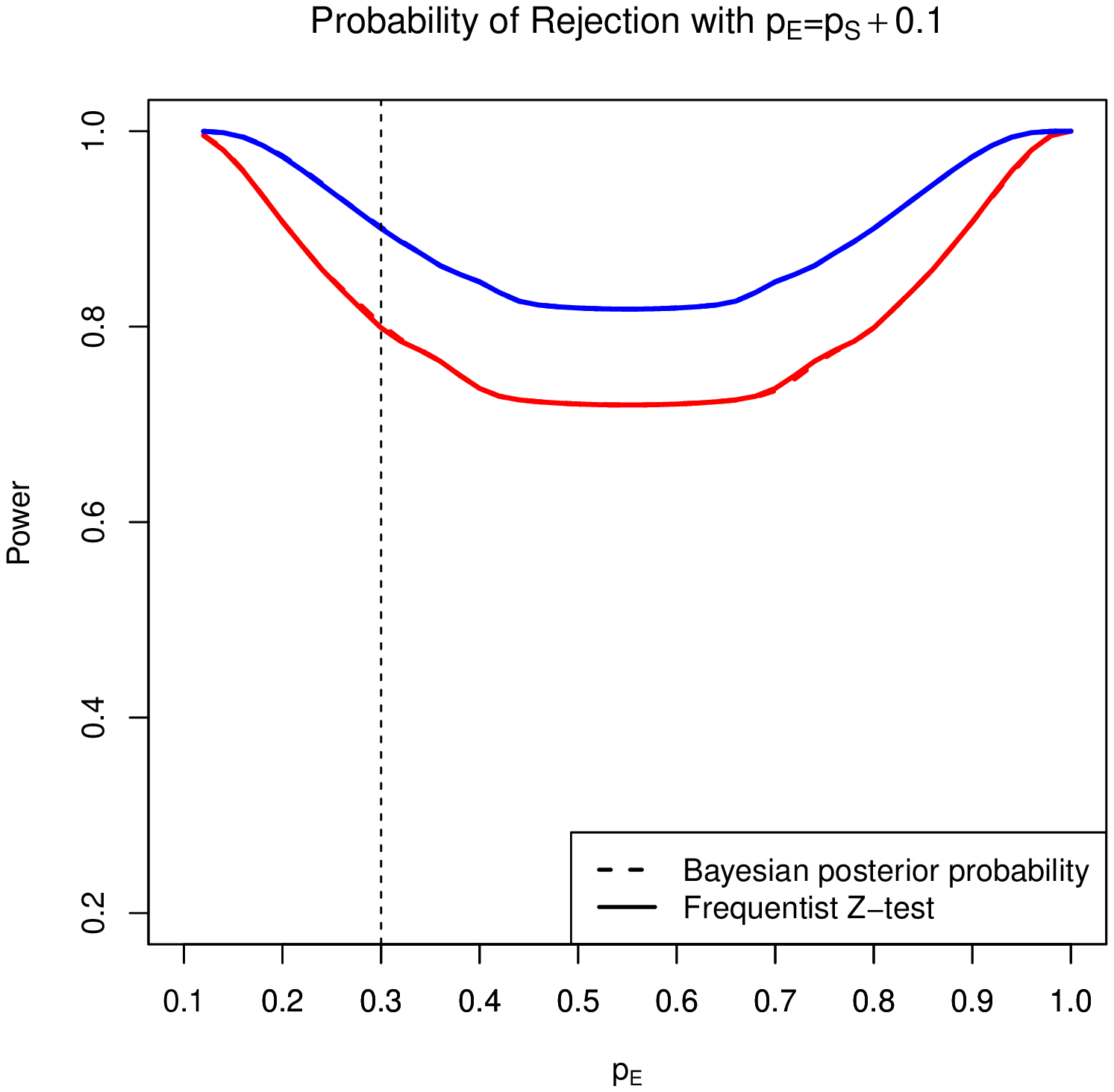}
\end{center}
\caption{Comparison of the type I error rate and power under the frequentist $Z$-test and Bayesian test based on the posterior probability for detecting treatment difference $\delta = 0.15$ (left) and $\delta=0.1$ (right).
}
\label{comparet1}
\end{figure}

As a numerical study, we consider a two-arm randomized trial with
  a type I error rate of 10\% and 5\% and target power of 80\% and 90\% when $(p_S,p_E)=(0.2,0.3)$ and $(p_S,p_E)=(0.2,0.35)$, respectively.
Under equal randomization, to achieve the desired power, the required sample size per arm is
$$
n=\frac{(z_{\alpha}+z_{\beta})^2}{\delta^2}\{p_E(1- p_E)+p_S(1- p_S)\},
$$
where we take $\delta=0.1$ and 0.15.
Under the Bayesian design, we assume non-informative prior
distributions, $p_S \sim {\rm Beta}(0.2,0.8)$ and $p_E\sim{\rm Beta}(0.2,0.8)$. For comparison, we compute the type I error rate and power for both the
Bayesian test with $\eta=1-\alpha$ and the frequentist
$Z$-test with a critical value $z_{1-\alpha}$. As shown in
Figure \ref{comparet1}, both designs
produce similar operating characteristics: the type I error rate can
be maintained at the nominal level, and the power attains the target level of 80\% or 90\% at
the specified values of $(p_S,p_E)$. It is worth noting that
because the endpoints are binary and the trial outcomes are discrete,
exact calibration of the empirical type I error rate to the nominal level is not possible, particularly when the sample size is small.
When we adopt a larger sample size by setting the type I error rate to be 5\% and the target power to be 90\%, the empirical type I error rate is closer to the nominal level as shown in the blue lines.

\section{Hypothesis Test for Binary Data}
\subsection{Two-Sample Hypothesis Test}
We first study the relationship between $p$-value and the posterior probability in a two-arm randomized clinical trial with dichotomous outcomes.
We consider the one-sided hypothesis test in (\ref{binaryh}), and under the frequentist $Z$-test for two proportions given by (\ref{binaryztest}),
the $p$-value is
 $$
 p\mbox{-value}_1 = 1-\Phi(Z),
 $$
where $\Phi(\cdot)$ denotes the cumulative distribution function (CDF) of the standard normal distribution. At the significance level of $\alpha$, we reject the null hypothesis if $p$-value is smaller than $\alpha$.

In the Bayesian paradigm, we base our decision on the posterior probability, as given in (\ref{binarybtest}). We reject the null hypothesis if the posterior probability of $p_E\le p_S$ is smaller than $\alpha$,
$${\rm PoP}_1= \Pr(p_E \le p_S|y_E,y_S)<\alpha.$$

\begin{figure}[htb]
\begin{center}
\includegraphics[height=5cm,width=5cm]{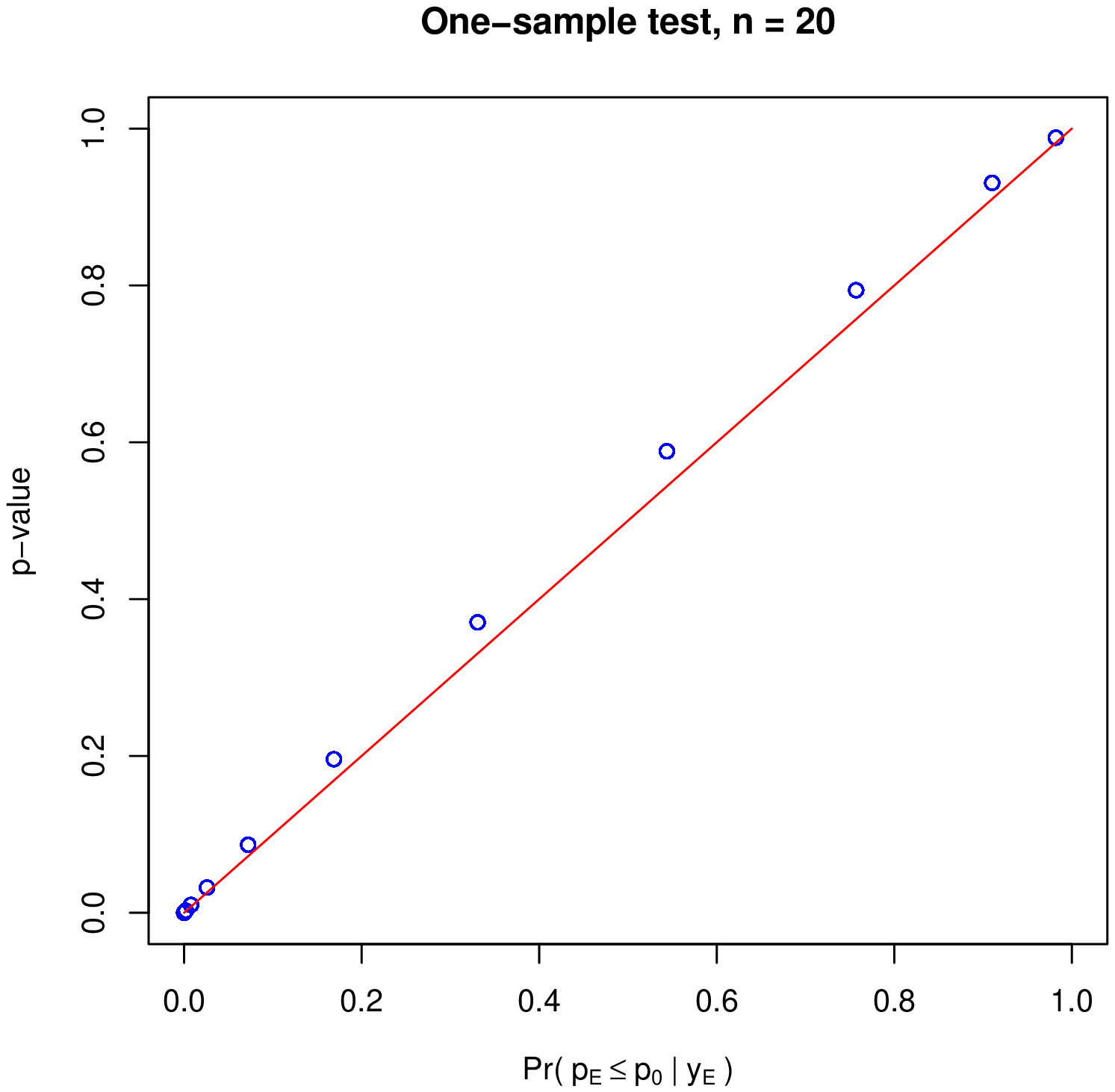}
\includegraphics[height=5cm,width=5cm]{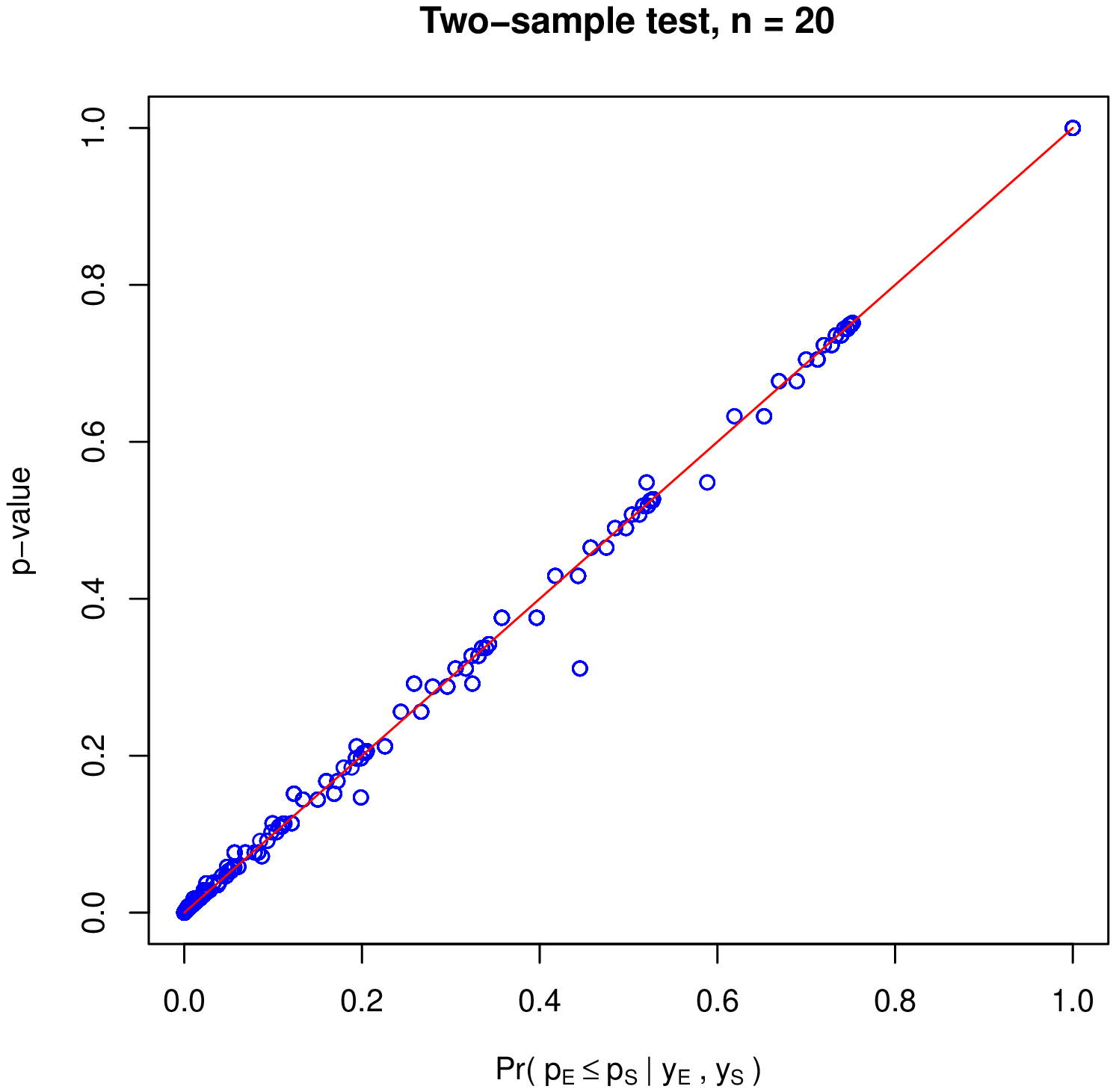}\\
\includegraphics[height=5cm,width=5cm]{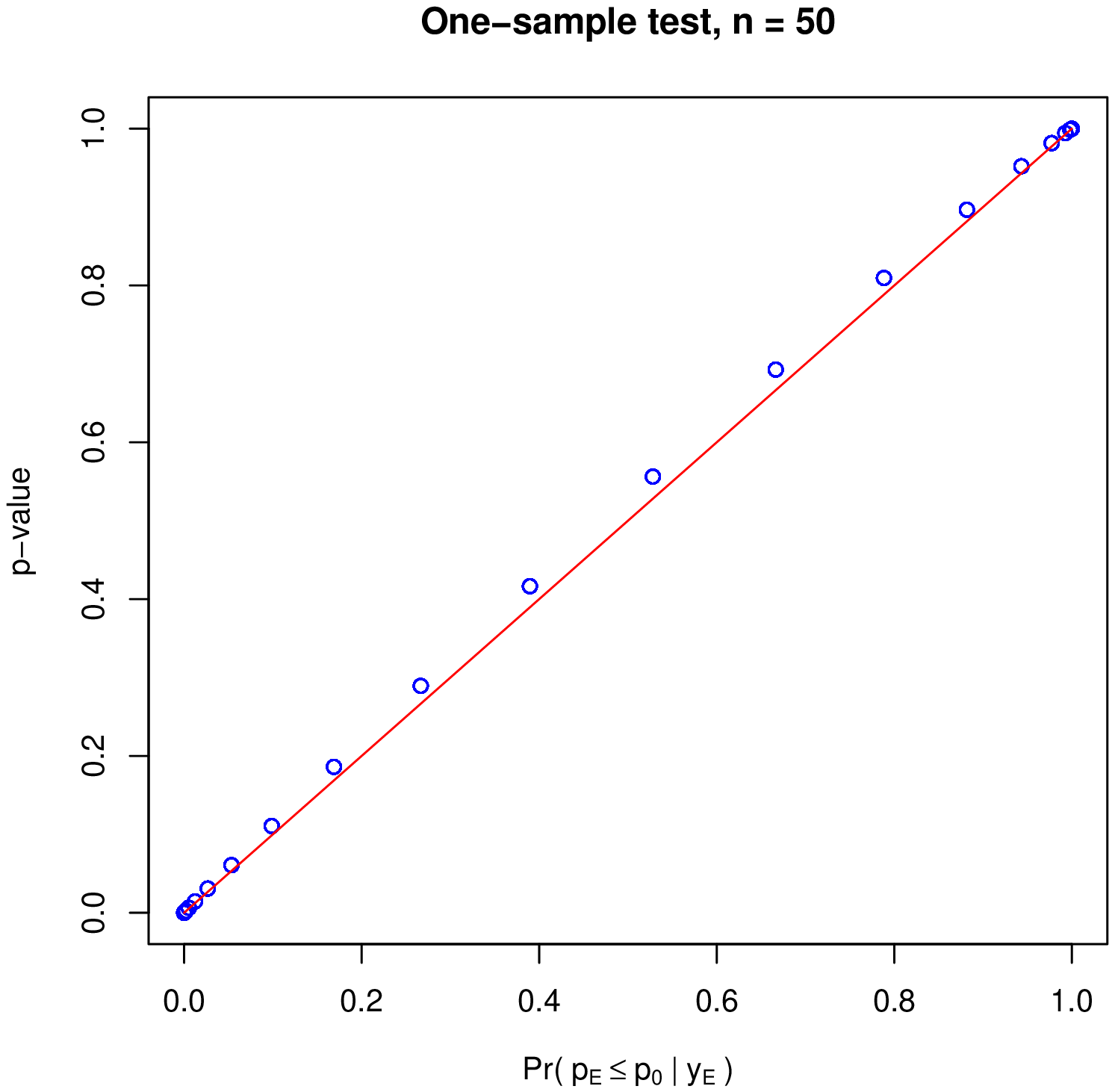}
\includegraphics[height=5cm,width=5cm]{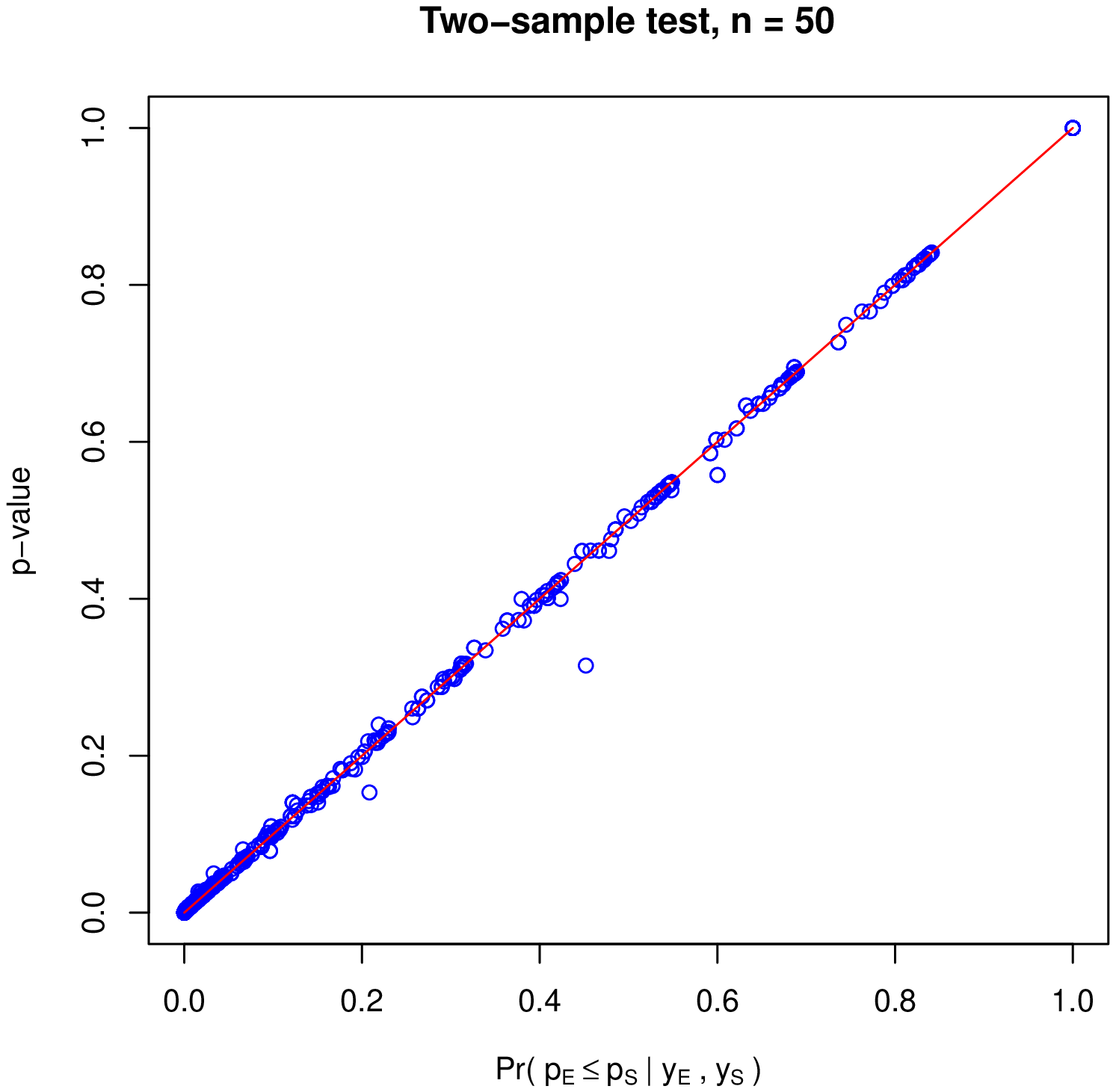}\\
\includegraphics[height=5cm,width=5cm]{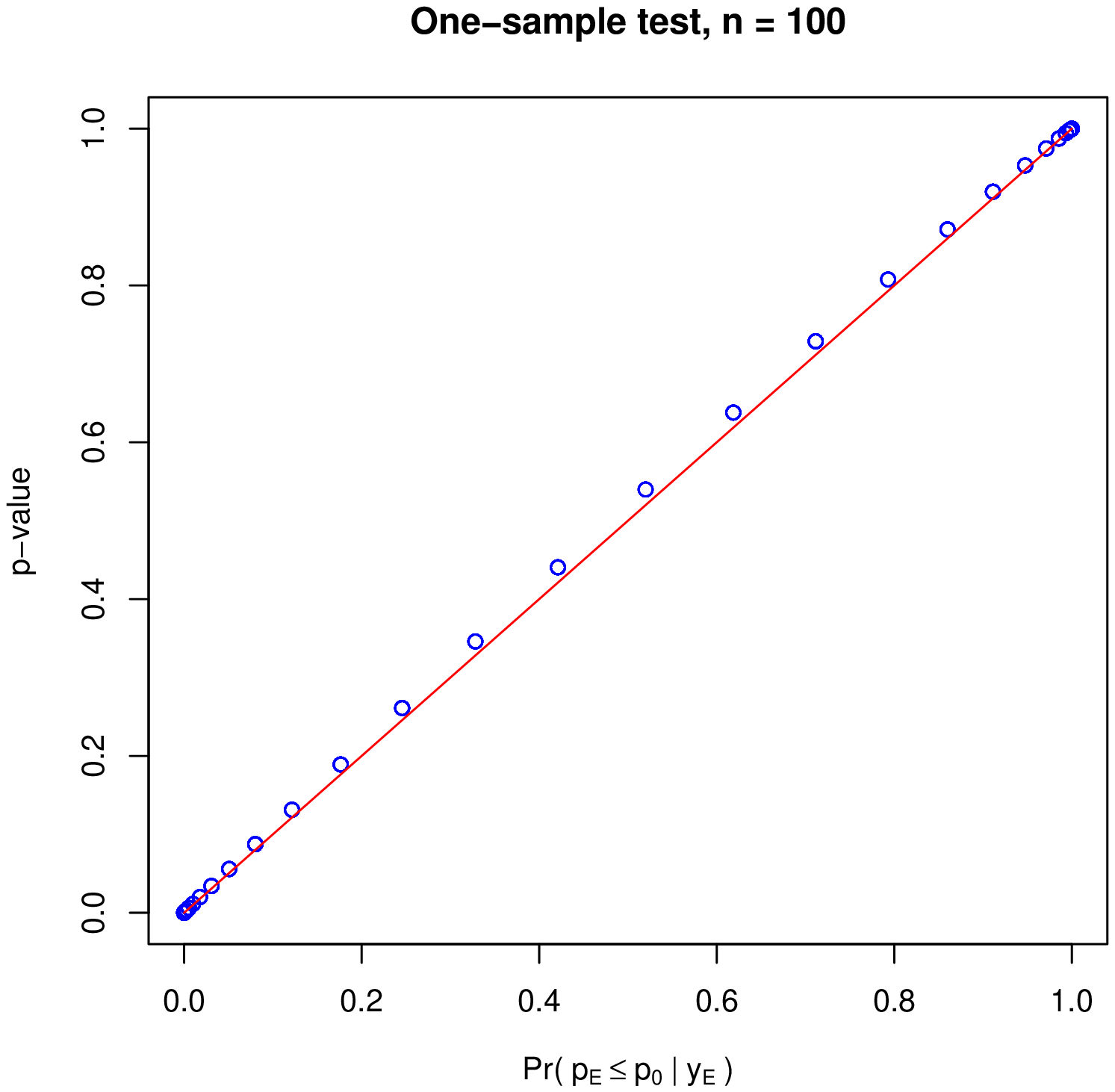}
\includegraphics[height=5cm,width=5cm]{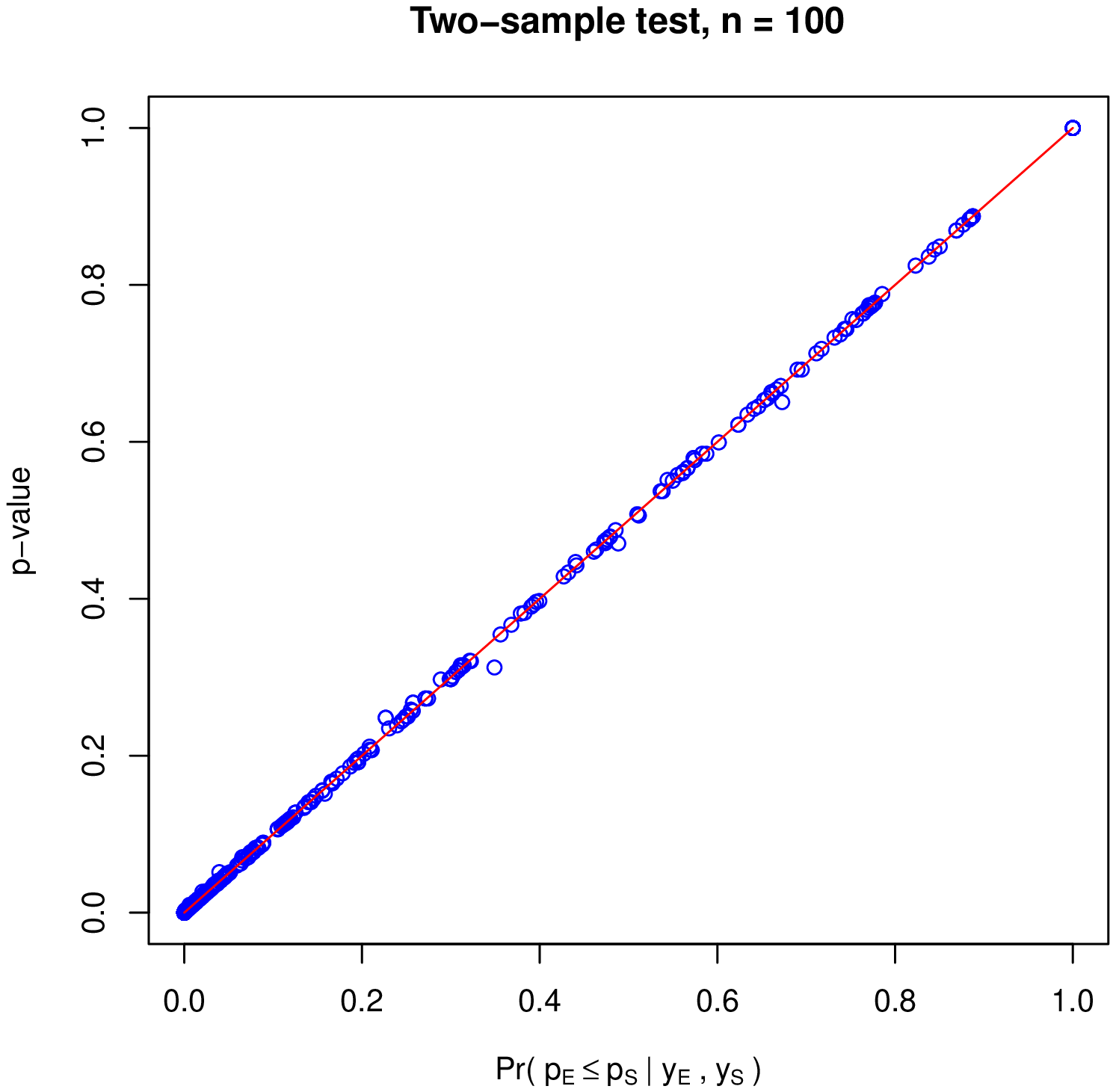}\\
\includegraphics[height=5cm,width=5cm]{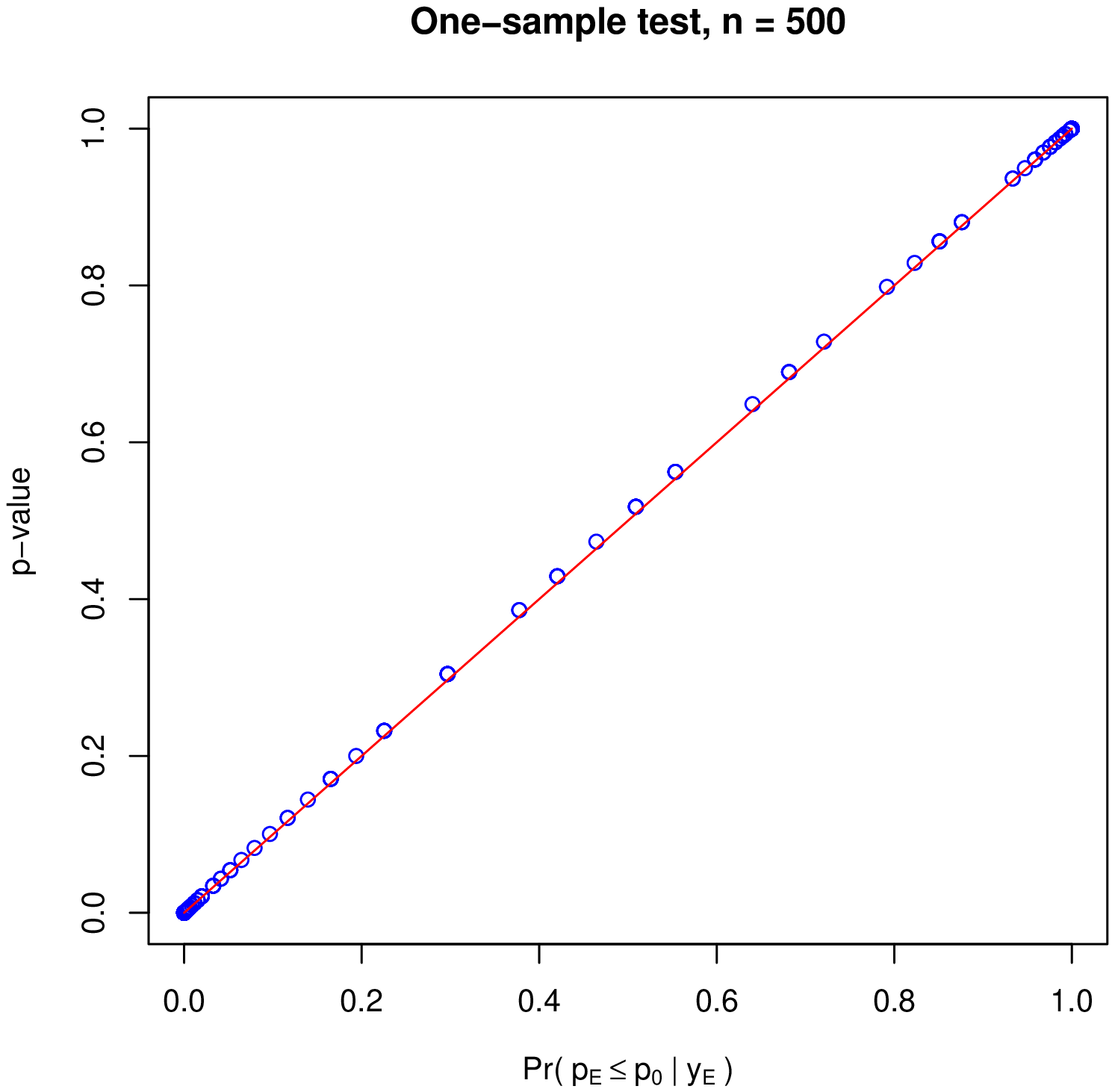}
\includegraphics[height=5cm,width=5cm]{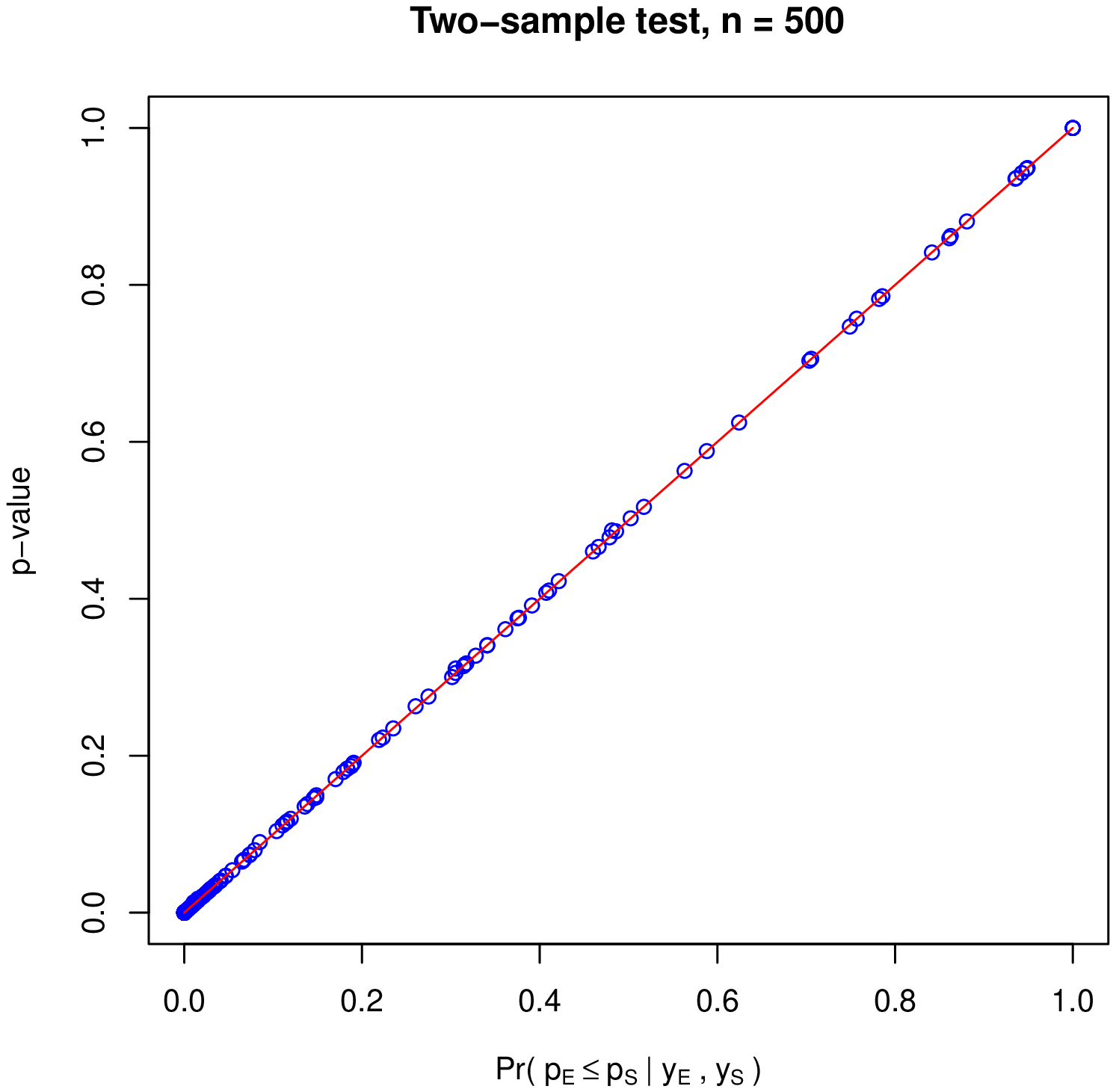}\\

\end{center}
\caption{The relationship between $p$-value and the posterior probability over 1000 replications under one-sided one-sample and two-sample hypothesis tests with binary outcomes under sample sizes of 20, 50, 100 and 500 per arm, respectively.
}
\label{os}
\end{figure}

As a numerical study, we set $n=20$, 50, 100 and 500, and randomly draw integers between 0 and $n$ to be the values for $y_E$ and $y_S$, and for each replication we compute the posterior probability of the null hypothesis $\Pr(H_0|y_E,y_S)$ and the $p$-value. As shown in Figure \ref{os}, all the paired values lie very close to the straight line of $y=x$, indicating the equivalence between the $p$-value and posterior probability of the null.

\begin{figure}[htb]
\begin{center}
\includegraphics[height=8cm,width=8cm]{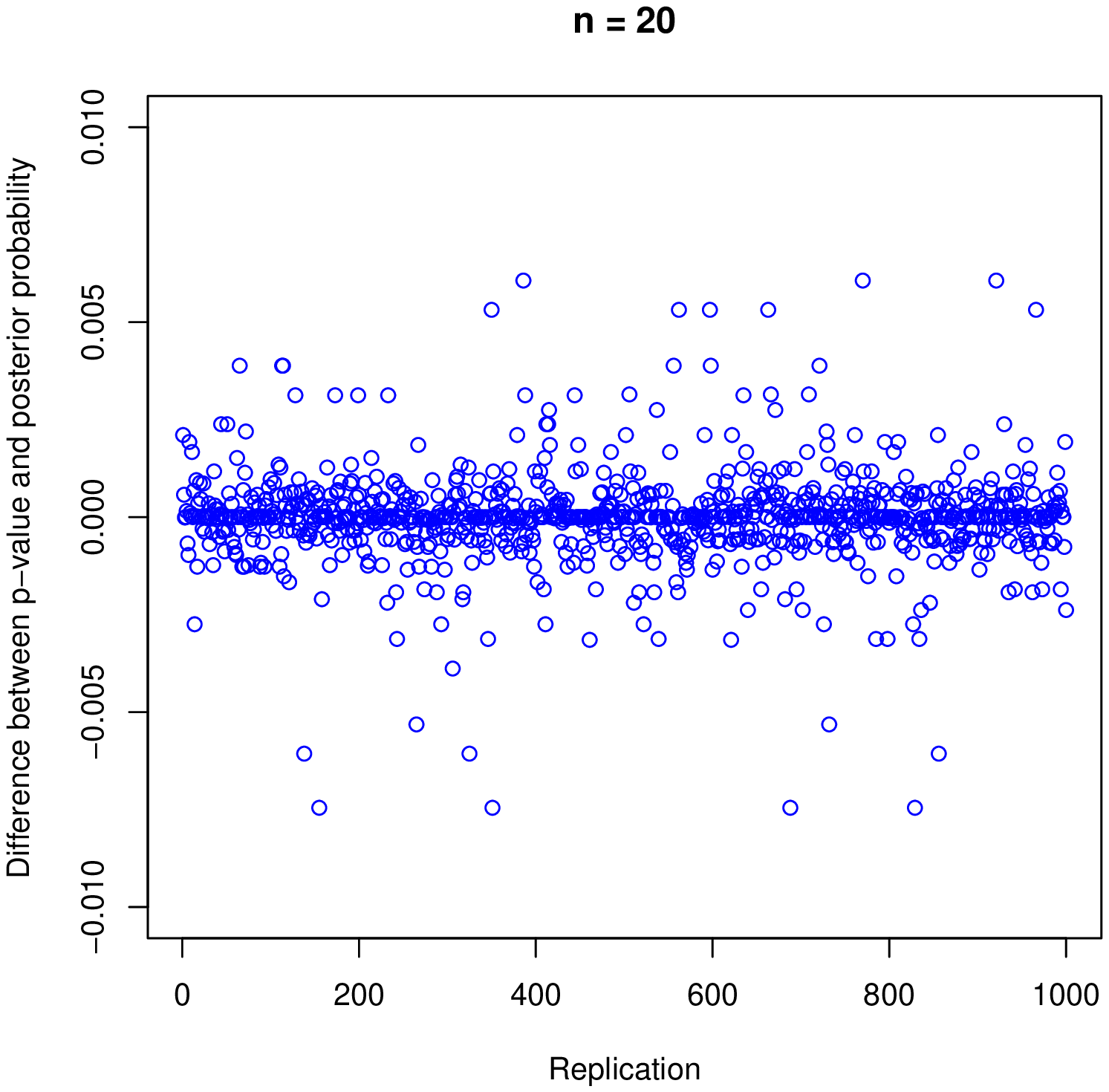}
\includegraphics[height=8cm,width=8cm]{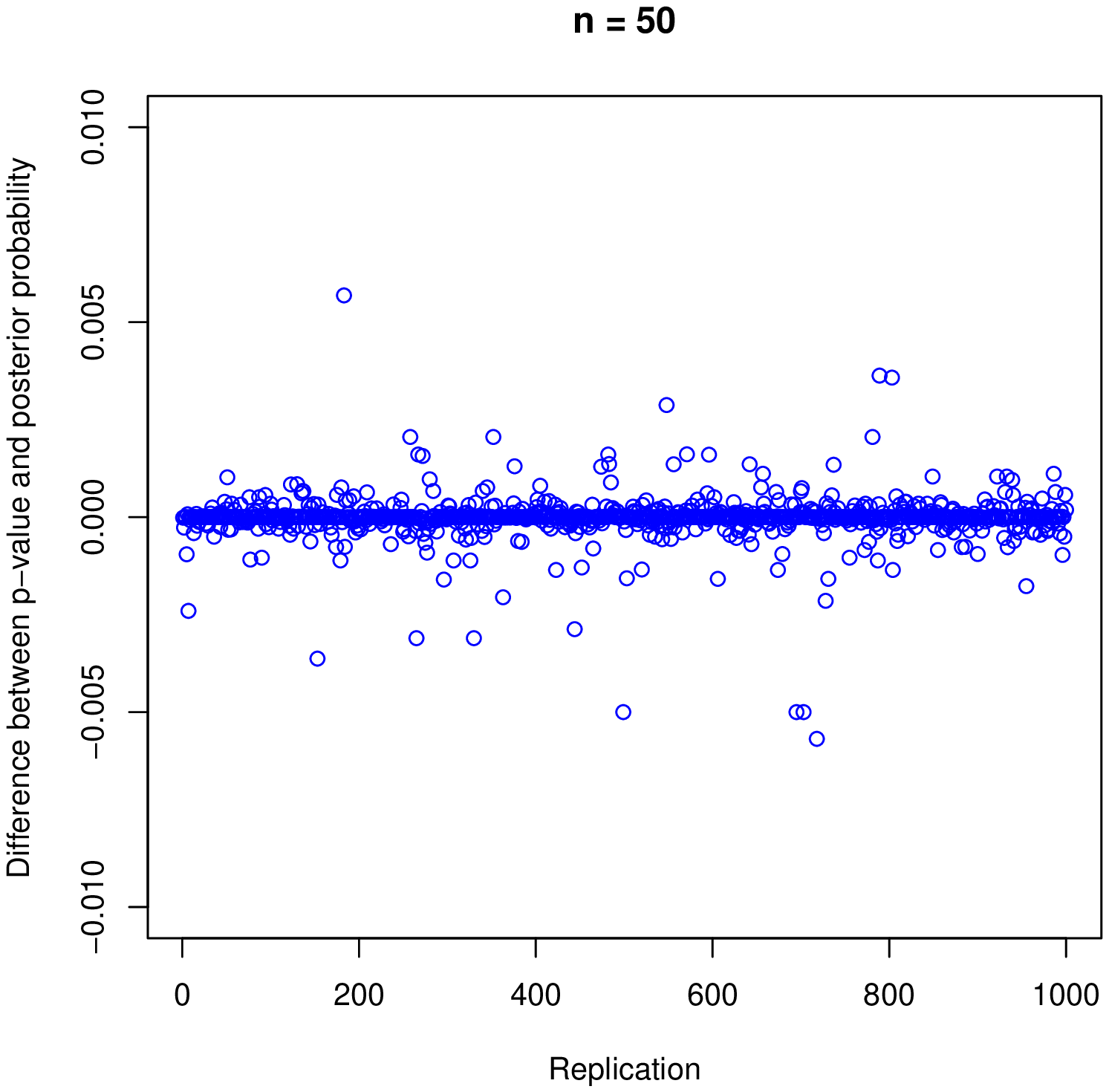}\\
\includegraphics[height=8cm,width=8cm]{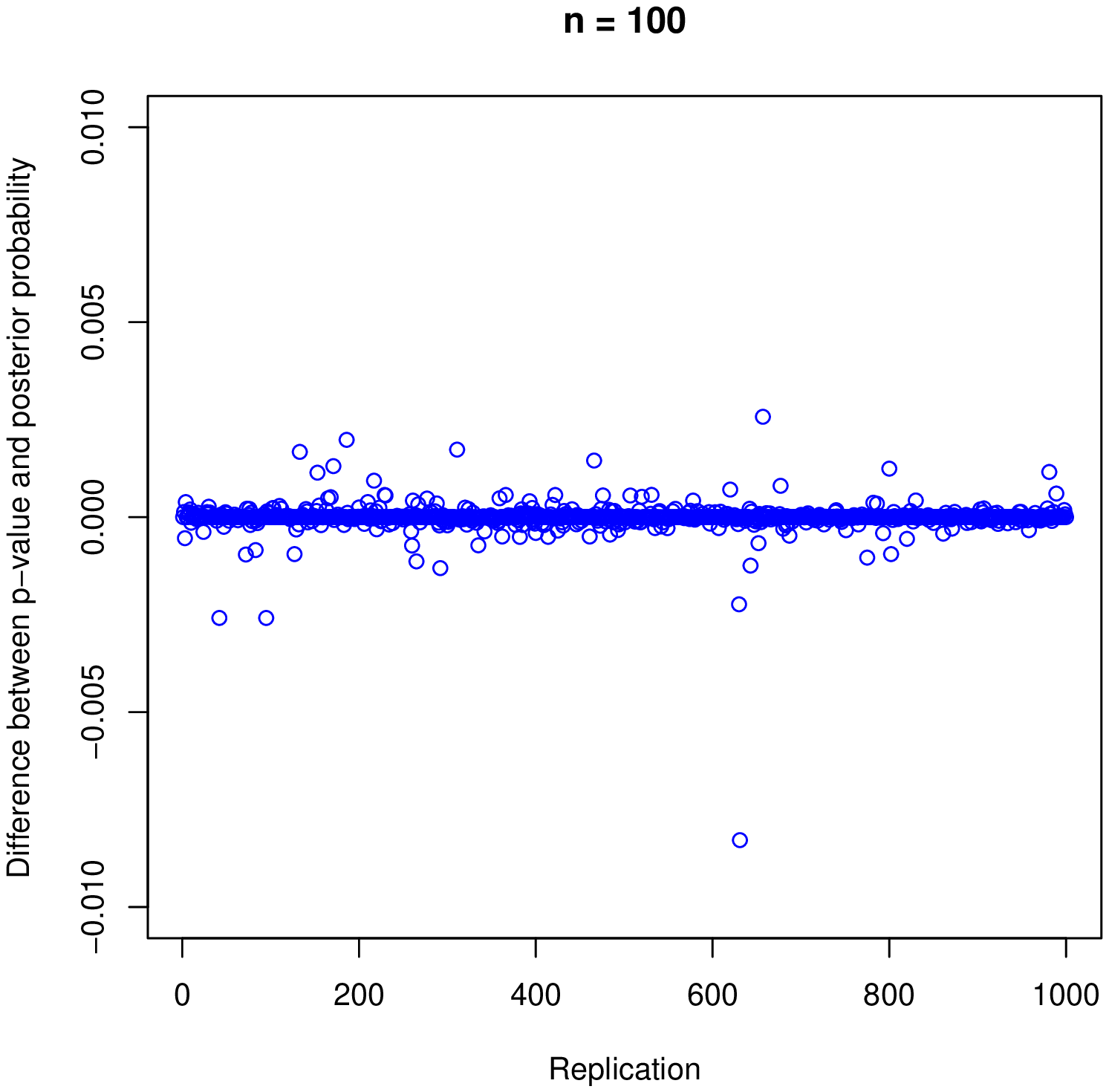}
\includegraphics[height=8cm,width=8cm]{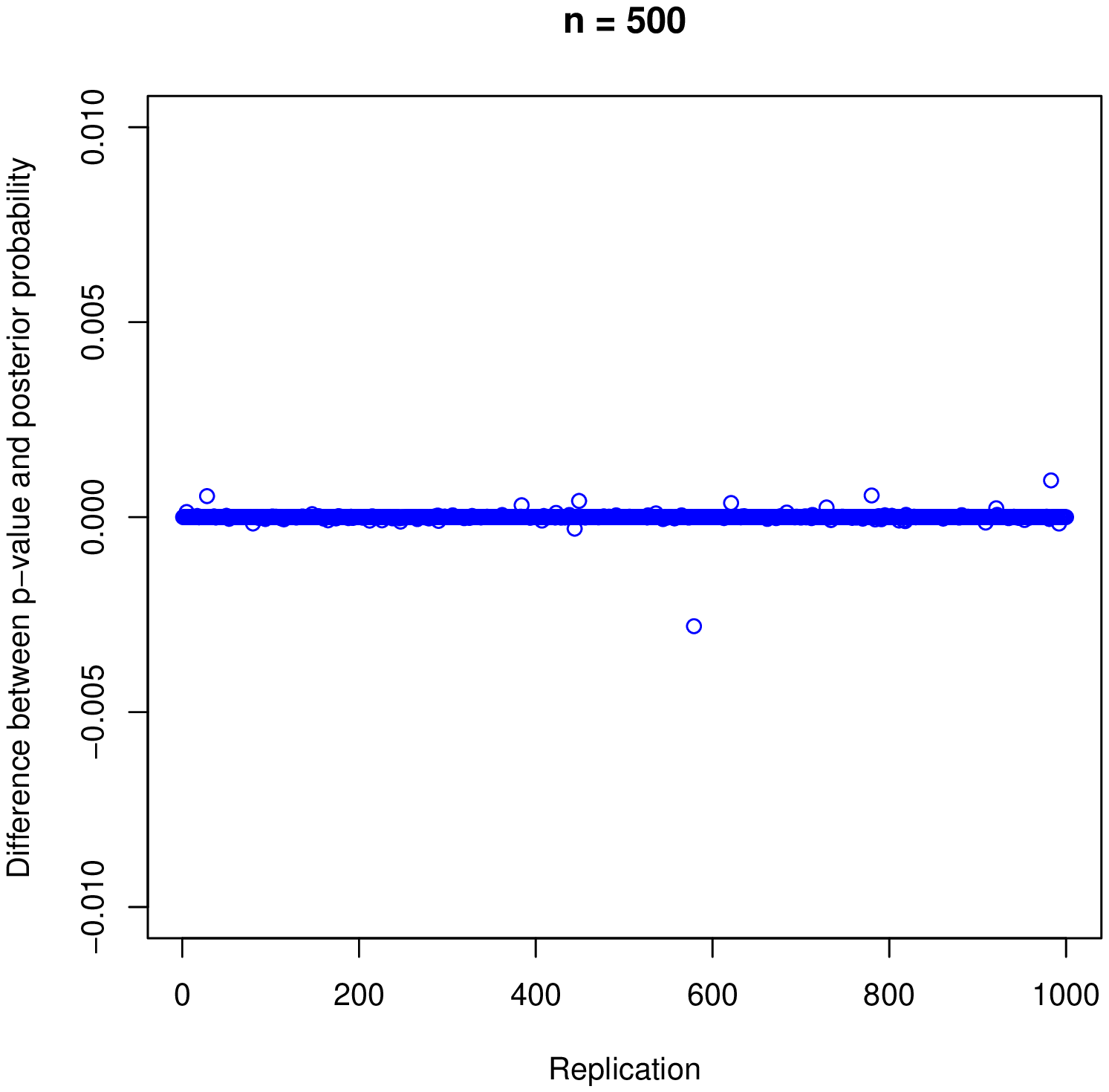}
\end{center}
\caption{The differences between $p$-values and posterior probabilities over 1000 replications in one-sided two-sample hypothesis tests with binary outcomes under sample sizes of 20, 50, 100 and 500, respectively.
}
\label{three}
\end{figure}

Figure \ref{three} shows the differences between $p$-values and posterior probabilities $\Pr({p_E} \le {p_S}|{y_E},{y_S})$ under sample sizes of 20, 50, 100 and 500, respectively. As sample size increases, the differences diminish toward 0, corroborating the asymptotic equivalence between $p$-value and the posterior probability.

For two-sided hypothesis tests,  we are interested in
examining whether there is any difference in the treatment effect between the experimental drug and the standard drug,
\begin{equation*}\label{Hypothesis}
H_0\mbox{:} \ p_E = p_S \quad  {\rm versus} \quad H_1\mbox{:} \ p_E \neq p_S.
\end{equation*}

 The $p$-value under the two-sided hypothesis test is
 \begin{eqnarray*}
 p \mbox{-value}_2 &=& 2-2\Phi(|Z|) =  2[1-{\rm max}\{\Phi(Z),\Phi(-Z)  \}].
 \end{eqnarray*}

 It is worth emphasizing that under the frequentist paradigm, the two-sided test can be viewed as a combination of two one-sided tests along the opposite directions. Therefore, to construct an equivalent counterpart under the Bayesian paradigm, we may regard the problem as two opposite one-sided Bayesian test and compute the posterior probabilities of the two opposite hypotheses; this approach to Bayesian hypothesis testing is different from the one commonly adopted in the literature, where a prior probability mass is imposed on the point null, e.g., see Berger and Sellke (1987), Berger and Delampady (1987), and Berger (2003).

\begin{figure}[htb]
\begin{center}
\includegraphics[height=6.5cm,width=6.5cm]{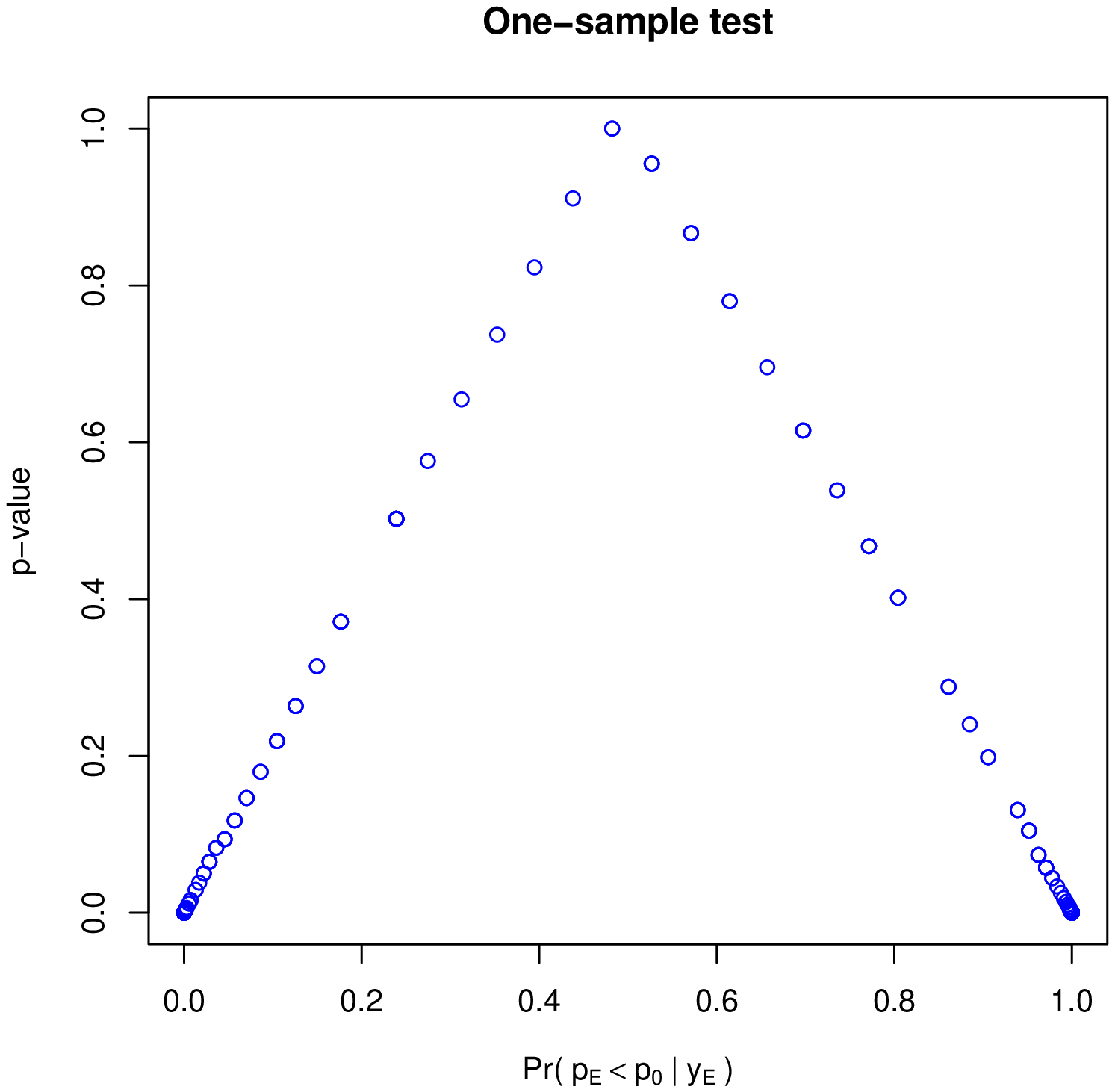}
\includegraphics[height=6.5cm,width=6.5cm]{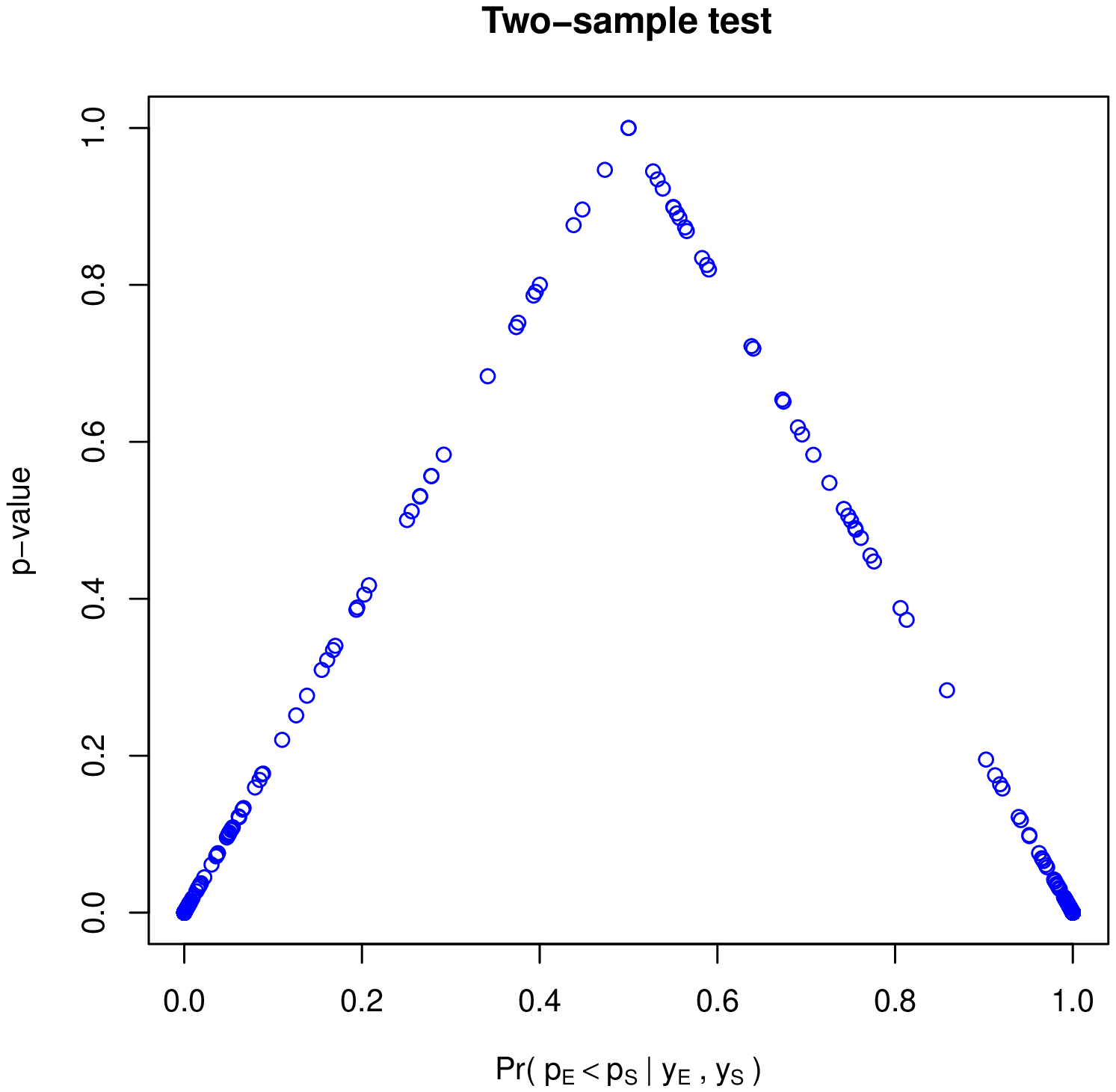}
\\
\includegraphics[height=6.5cm,width=6.5cm]{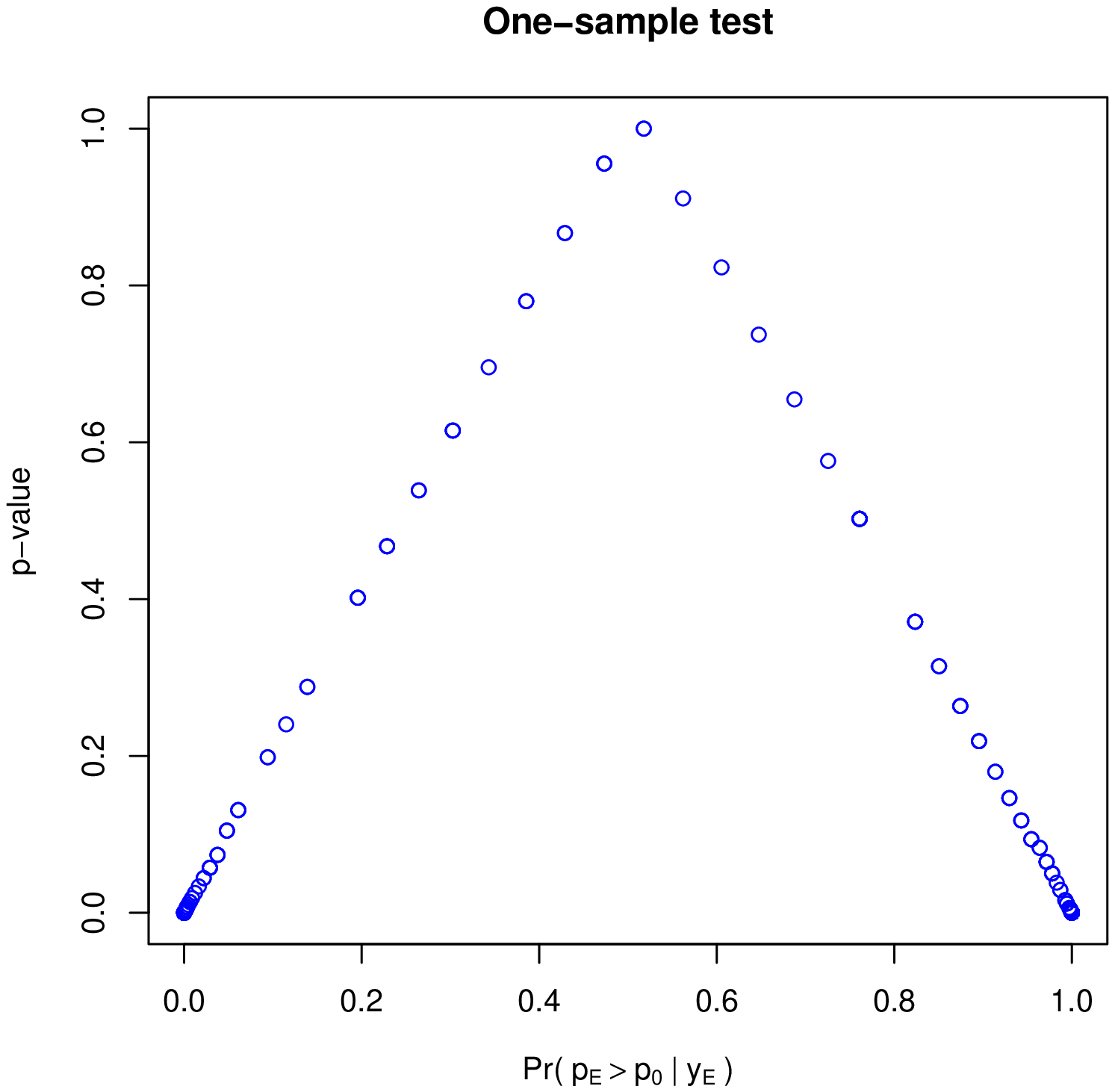}
\includegraphics[height=6.5cm,width=6.5cm]{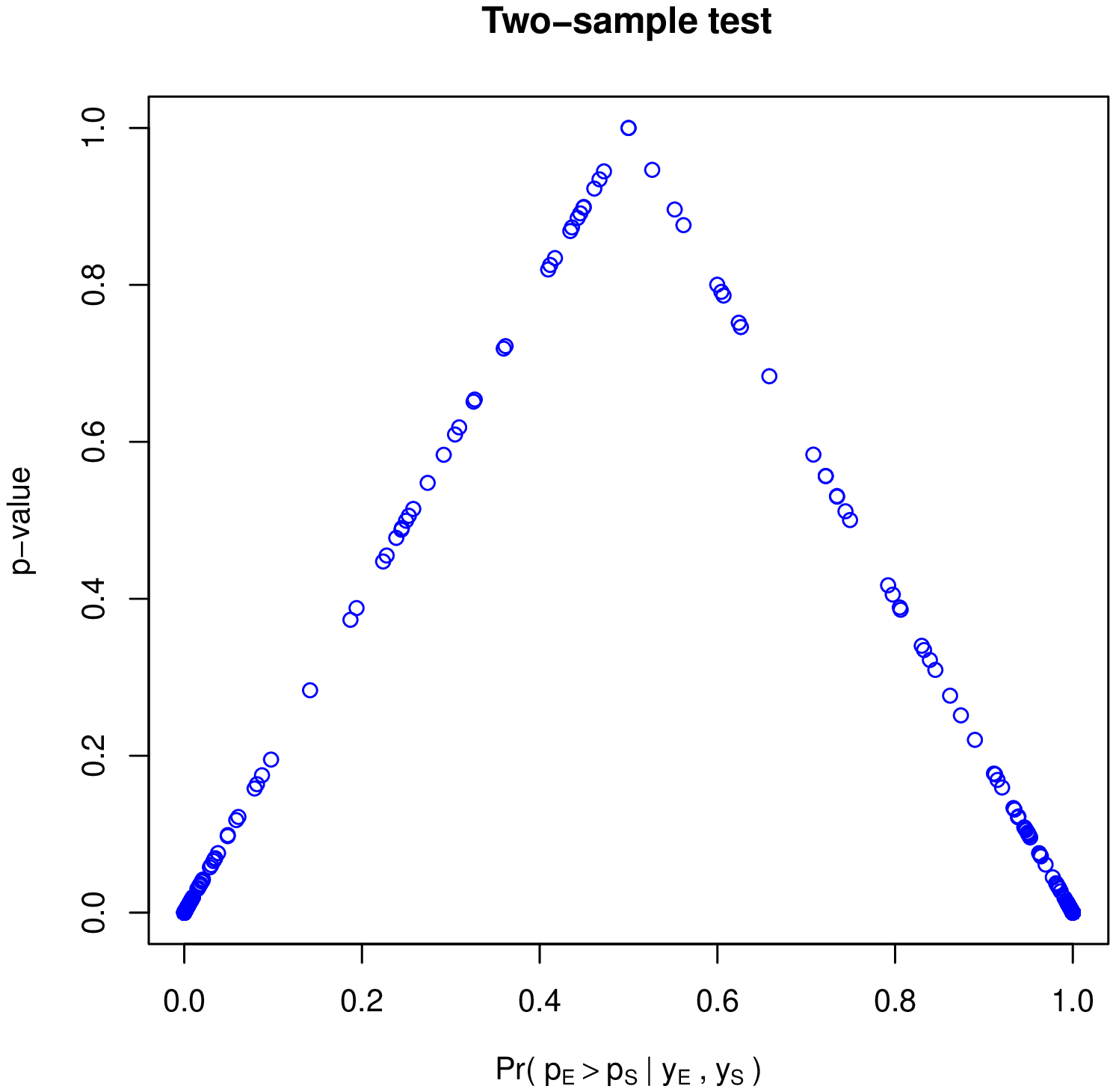}
\\
\includegraphics[height=6.5cm,width=6.5cm]{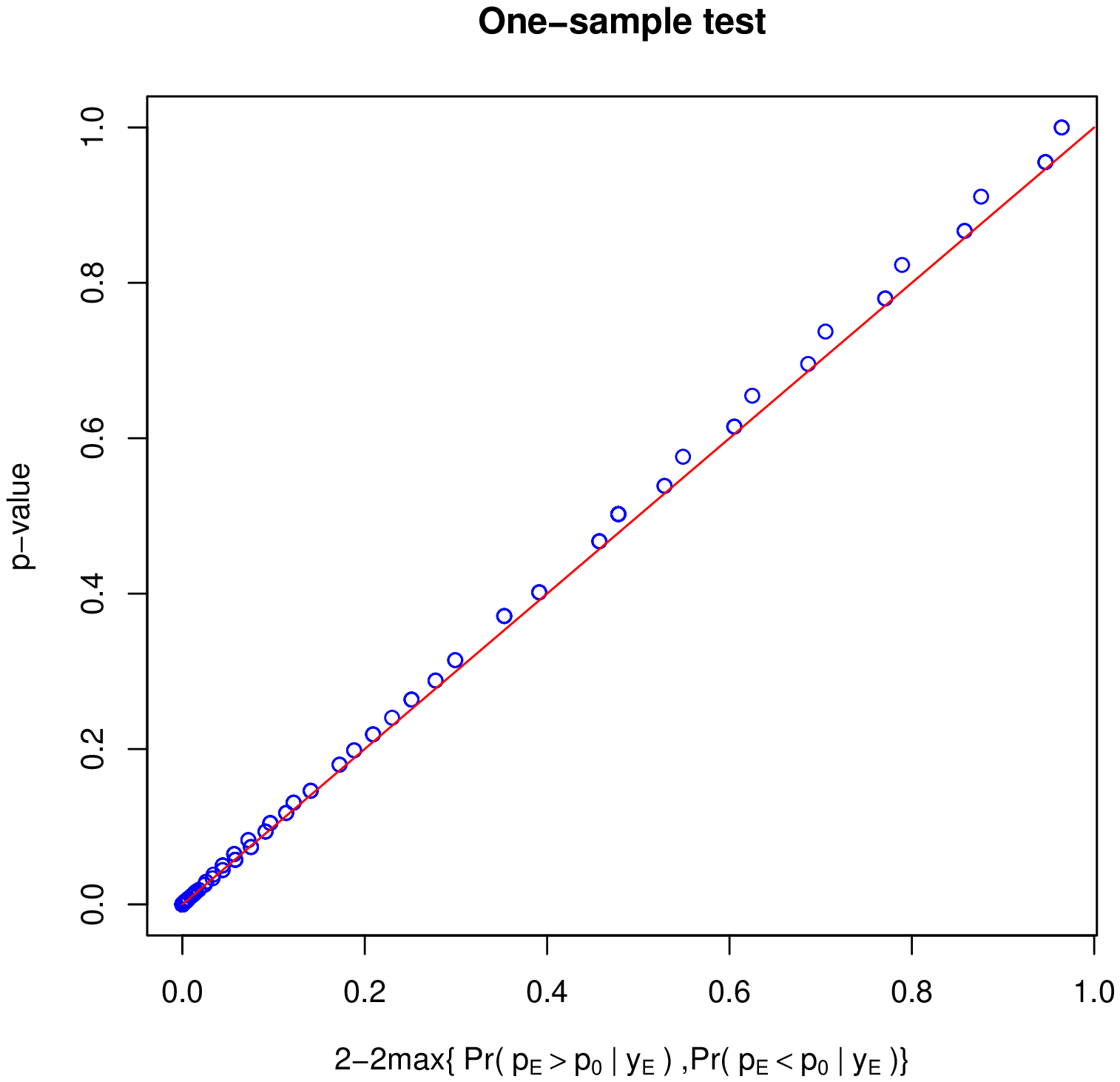}
\includegraphics[height=6.5cm,width=6.5cm]{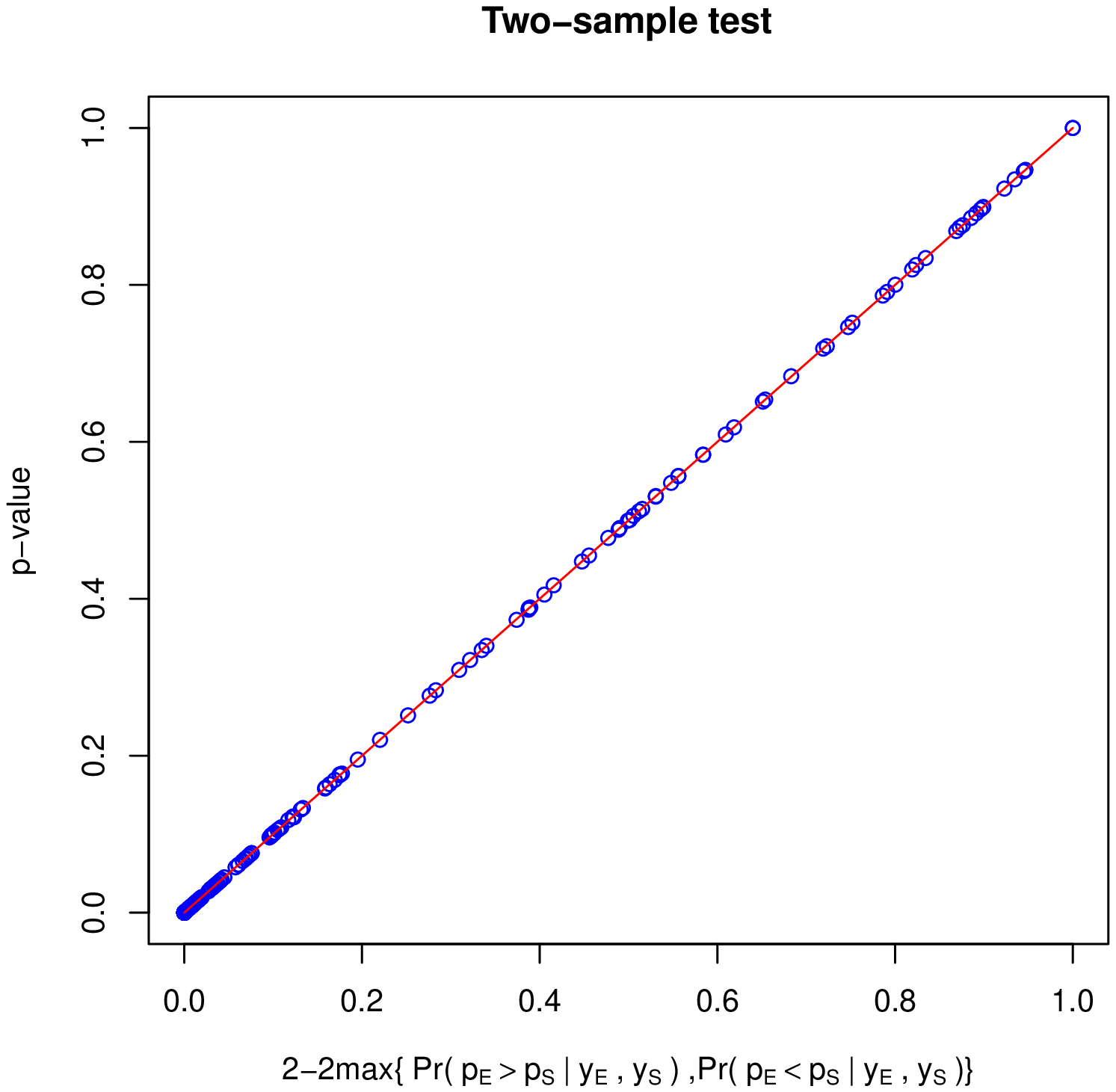}
\end{center}
\caption{The relationship between $p$-value and the posterior probability over 1000 replications under two-sided one-sample and two-sample hypothesis tests with binary outcomes under sample size of 500 per arm.}
\label{ts}
\end{figure}

If we define the two-sided posterior probability (${\rm PoP}_2$) as
$${\rm PoP_2} = 2[1-{\rm max}\{\Pr({p_E} > {p_S}|{y_E},{y_S}), \Pr({p_E} < {p_S}|{y_E},{y_S})\}],$$
then its relationship with $p$-value is similar to that of one-sided hypothesis testing as shown in Figure \ref{ts}.

The equivalence of the $p$-value and the posterior probability in the case of binary outcomes can be established by applying the Bayesian central limit theorem. Under large sample size, the posterior distribution of $p_E$ and $p_S$ can be approximated as
$$
p_g|y_g \sim {\rm N}({\hat p_g},{\hat p_g}(1-{\hat p_g}) / n), \quad g=E,S.
$$
As $y_E$ and $y_S$ are independent, the posterior distribution of $p_E-p_S$ can be derived as
$$p_E-p_S | y_E,y_S \sim {\rm N}({\hat p_E} - {\hat p_S} , \{ {{\hat p_E}(1-{\hat p_E})} + {{\hat p_S}(1-{\hat p_S})}  \}/n ).$$
Therefore, the posterior probability of $p_E \le p_S$ is
$$
{\rm PoP}_1= \Pr(p_E \le p_S|y_E,y_S) \approx  \Phi\bigg(-\frac{{\hat p_E} - {\hat p_S}}{[\{ {{\hat p_E}(1-{\hat p_E})} + {{\hat p_S}(1-{\hat p_S})}  \}/n]^{1/2}}\bigg) = \Phi(-Z),
$$
which is equivalent to $p \mbox{-value}_1 =1-\Phi(Z)= \Phi(-Z)$.
The equivalence relationship for a two-sided test can be derived along similar lines.

More generally, Dudley and Haughton (2002) proved that under mild regularity conditions, the posterior probability of a half space converges to the standard normal CDF transformation of the likelihood ratio test statistic. In a one-sided hypothesis test, the posterior probability of the half space is $\Pr(H_1|D) = 1-{\rm PoP}_1$, whereas the standard normal CDF transformation of the likelihood ratio test statistic equals to one minus $p \mbox{-value}_1$, an therefore ${\rm PoP}_1$ and $p \mbox{-value}_1$ are asymptotically equivalent.

\subsection{One-Sample Hypothesis Test}
In a single-arm clinical trial with dichotomous outcomes, we are interested in
examining whether the response rate of the experimental drug $p_E$ exceeds a prespecified threshold $p_0$, by formulating a one-sided hypothesis test,
\begin{equation*}\label{Hypothesis}
H_0\mbox{:} \ p_E \le p_0 \quad  {\rm versus} \quad H_1\mbox{:} \ p_E > p_0.
\end{equation*}

In the frequentist paradigm, the $p$-value can be computed based on the exact binomial test. In the Bayesian paradigm, we assume a beta prior distribution for $p_E$, e.g., $p_E\sim{\rm Beta}(a_E,b_E)$.
The posterior distribution of $p_E$
is given by $
{p_E}|{y_E}\sim {\rm Beta}(a_E+y_E,b_E+n-y_E)$,
for which the density function
is denoted by $f({p_E}|{y_E})$.
Based on the posterior probability, we can construct a Bayesian decision rule
so that the experimental treatment is declared as promising if
$$\Pr(H_1|y_E) = \Pr(p_E>p_0|y_E)> \eta,$$
where
\[
\Pr({p_E} > {p_0}|{y_E})
=\int_{p_0}^1 f(p_E|y_E)d{p_E}.
\]
Otherwise, we fail to declare treatment efficacy.
As a result, the one-sided posterior probability is defined as
$${\rm PoP}_1=\Pr(H_0|y_E) = \Pr(p_E \le p_0|y_E).$$

For two-sided hypothesis tests, we are interested in
examining whether the response rate of the experimental drug is different from $p_0$,
\begin{equation*}\label{Hypothesis}
H_0\mbox{:} \ p_E = p_0 \quad  {\rm versus} \quad H_1\mbox{:} \ p_E \neq p_0.
\end{equation*}
The $p$-value can be computed based on the exact binomial test.
If we define the two-sided posterior probability,
$${\rm PoP_2} = 2[1-{\rm max}\{\Pr({p_E} > {p_0}|{y_E}), \Pr({p_E} < {p_0}|{y_E})\}],$$
 then its relationship with $p$-value is similar to that of one-sided hypothesis testing as shown in Figure \ref{ts}.

In a numerical study, we set $n=20$, 50, 100 and 500, $p_0=0.2$, and randomly draw integers between 0 and $n$ to be the values of $y_E$. We assume a noninformative prior for $p_E$, i.e., $a_E = 1$ and $b_E = 1$. Figure \ref{os} shows the relationship between the posterior probability of the null hypothesis $\Pr(H_0|y_E)$ and the $p$-value, which clearly indicates that all the points lie very close to the straight line of $y=x$.

\section{Hypothesis Test for Normal Data}
\subsection{Hypothesis Test with Known Variance}

In a two-arm randomized clinical trial with normal endpoints, we are interested in comparing the means of the outcomes between the experimental and standard arms.
Let $n$ denote the sample size for each arm, and let $D = \{(y_{E1},y_{S1}),\ldots,(y_{En},y_{Sn})  \}$ denote the paired data under the experimental and standard treatments. Assume $y_{Ei} \sim {\rm N}(\mu_E, \sigma^2)$ and $y_{Si} \sim {\rm N}(\mu_S, \sigma^2)$ with  unknown means $\mu_E$ and $\mu_S$ but a known variance $\sigma^2 = 1$.
Let $\bar y_E = \sum_{i=1}^{n} y_{Ei}/n$ and $\bar y_S = \sum_{i=1}^{n} y_{Si}/n$ denote the sample means, and let $\theta = \mu_E-\mu_S$ and $\hat \theta = \bar y_E-\bar y_S$ denote the true and the observed treatment difference, respectively.

Considering the one-sided hypothesis test,
\begin{equation*}\label{Hypothesis}
H_0\mbox{:} \ \theta \le 0 \quad  {\rm versus} \quad H_1\mbox{:} \ \theta > 0,
\end{equation*}
the frequentist $Z$-test statistic is formulated as
${\hat \theta}/{\sqrt{2/n}}$,
which follows the standard normal distribution under the null hypothesis. Therefore, the $p$-value under the one-sided hypothesis test is given by
\begin{align*}
  p \mbox{-value}_1 &= \Pr(Z\ge  \hat \theta \sqrt{n/2}|H_0)
   = 1 - \Phi( \hat \theta \sqrt{n/2}),
\end{align*}
where $Z$ denotes the standard normal random variable.

In the Bayesian paradigm, if we assume an improper flat prior distribution, $p(\theta) \propto 1$, the posterior distribution of $\theta$ is
$$\theta|D \sim {\rm N}(\hat \theta, 2/n).$$
Therefore, the posterior probability of $\theta$ smaller or equal to 0 is
$$
{\rm PoP}_1 = \Pr(\theta \le 0 | D) = 1- \Phi( \hat \theta  \sqrt{n/2}).
$$
Under such an improper prior distribution of $\theta$, we can establish an exact equivalence relationship between $p$-value and $\Pr(\theta \le 0 | D)$.

Under the two-sided hypothesis test, $H_0:\theta=0$ versus $H_1:\theta \neq 0$, the $p$-value is given by
\begin{align*}
  p \mbox{-value}_2 &= 2[1- {\rm max}\{\Pr(Z\ge z|H_0),\Pr(Z \le z|H_0)\} ]\\
   &= 2 - 2 {\rm max} \{\Phi( \hat \theta \sqrt{n/2}),\Phi(- \hat \theta \sqrt{n/2})\}.
\end{align*}
Correspondingly, the two-sided posterior probability is defined as
\begin{align*}
{\rm PoP}_2 &= 2[1 - {\rm max}\{\Pr(\theta < 0 | D), \Pr(\theta > 0 | D)\}]\\
 &= 2 - 2 {\rm max} \{\Phi( \hat \theta \sqrt{n/2}),\Phi(- \hat \theta \sqrt{n/2})\},
\end{align*}
which is exactly the same as the (two-sided) $p$-value.

\subsection{Hypothesis Test with Unknown Variance}
In a more general setting, we consider the case where $\mu_E$, $\mu_S$ and $\sigma$ are all unknown parameters. We define $x_i = y_{Ei}-y_{Si}$, which follows the normal distribution ${\rm N}(\theta, 2\sigma^2)$. For notational simplicity, let $\nu = 2\sigma^2$ and we are interested in modeling the joint posterior distribution of $\theta$ and $\nu$.

In the frequentist paradigm, Student's $t$-test statistic is
$$
T = \frac{\hat \theta }{\sqrt{\sum_{i=1}^{n}(x_i-\hat \theta)^2 / \{(n-1)n\}}}.
$$
Therefore, the $p$-value under the one-sided hypothesis test is
\begin{align*}
  p \mbox{-value}_1  &= 1 - F_{t_{n-1}}(T),
\end{align*}
where $F_{t_{n-1}}(\cdot)$ denotes the CDF of  Student's $t$ distribution with $n-1$ degrees of freedom.

In the Bayesian paradigm, if we assume Jeffreys' prior for $\theta$ and $\nu$,
$
p(\theta,\nu) \propto \nu^{-{3}/{2}}
$,
the corresponding posterior distribution is
$$
p(\theta,\nu|D) \propto \nu^{-(n+3)/2}\exp\bigg\{ -\frac{\sum_{i=1}^{n}(x_i-\hat \theta)^2 + n(\hat \theta - \theta)^2  }{2\nu}  \bigg\},
$$
which matches the normal-inverse-chi-square distribution,
$$(\theta,\nu)|D \sim {\rm N}\mbox{--Inv} \ {\chi^{2}} \bigg(\hat \theta, n, n, \sum_{i=1}^{n}(x_i-\hat \theta)^2/n \bigg).$$
Based on the posterior distribution, the one-sided posterior probability of the null hypothesis is ${\rm PoP}_1 = \Pr({\theta} \le {0}|{D})$.

As an alternative to Jeffreys' prior distribution, we also consider a normal-inverse-gamma prior distribution for $\theta$ and $\nu$,
$
(\theta,\nu) \sim {\rm N} \mbox{--IG}(\theta_0,\nu_0,\alpha,\beta)
$,
which belongs to the conjugate family of prior distributions for the normal likelihood function.
As a result, the corresponding posterior distribution is also a normal-inverse-gamma prior distribution,
$$
(\theta,\nu)|D \sim {\rm N}\mbox{--IG}\bigg(\frac{\theta_0\nu_0 + n\hat\theta}{\nu_0 + n},\nu_0+n,\alpha+\frac{n}{2},\beta+ \frac{1}{2}\sum_{i=1}^{n}(x_i-\hat \theta)^2 + \frac{n\nu_0}{\nu_0+n}\frac{(\hat \theta - \theta_0)^2}{2} \bigg).
$$

For a two-sided hypothesis test, the $p$-value is
\begin{align*}
p \mbox{-value}_2 &= 2-2F_{t_{n-1}}(|T|)\\
   &=  2[1-{\rm max}\{ F_{t_{n-1}}(T),F_{t_{n-1}}(-T) \}].
\end{align*}

Similarly, we define the two-sided posterior probability as
$${\rm PoP}_2 = 2[1-{\rm max}\{\Pr({\theta} > {0}|{D}), \Pr({\theta} < {0}|{D})\}].$$

\begin{figure}[htb]
\begin{center}
\includegraphics[height=5cm,width=5cm]{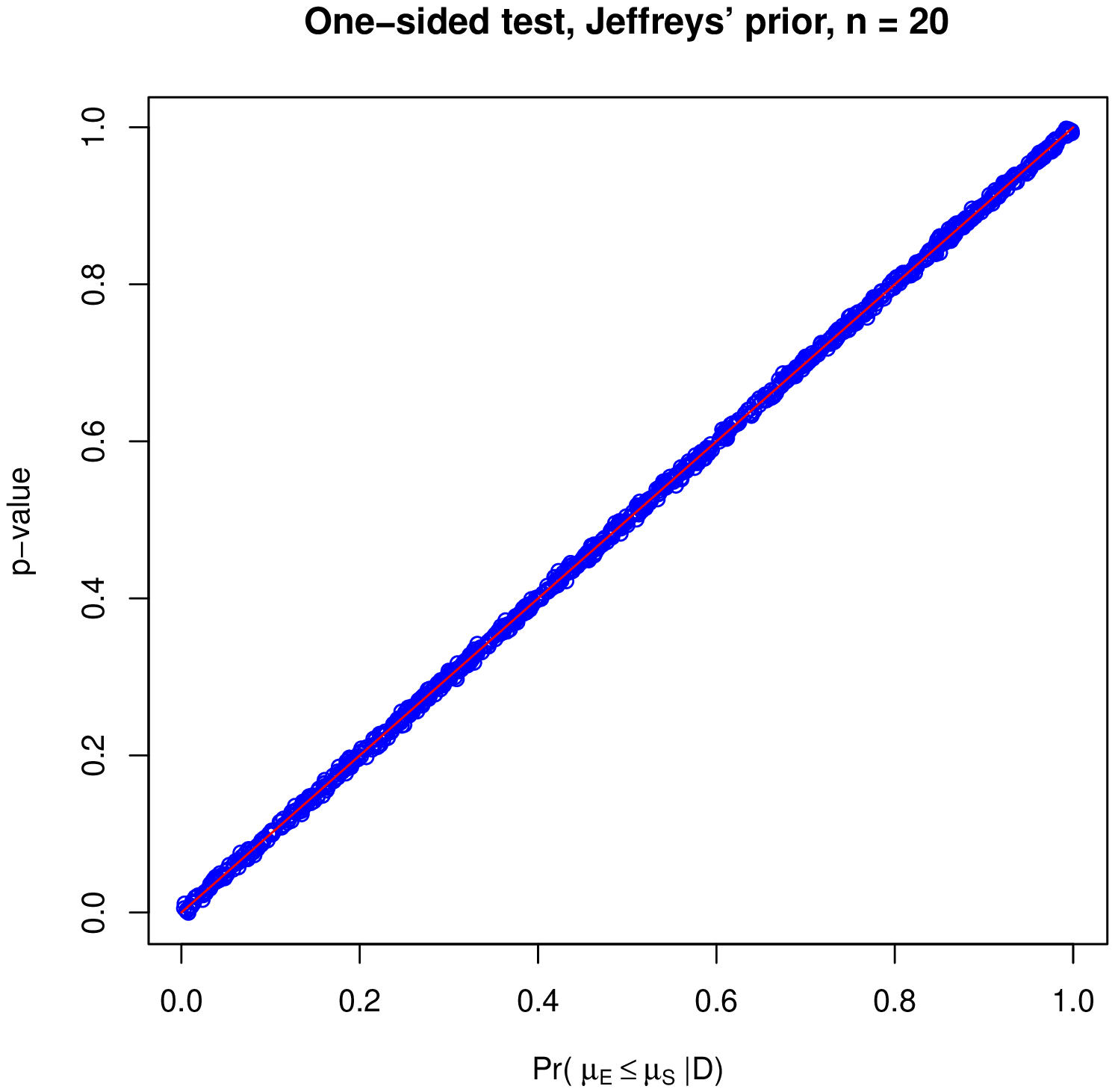}
\includegraphics[height=5cm,width=5cm]{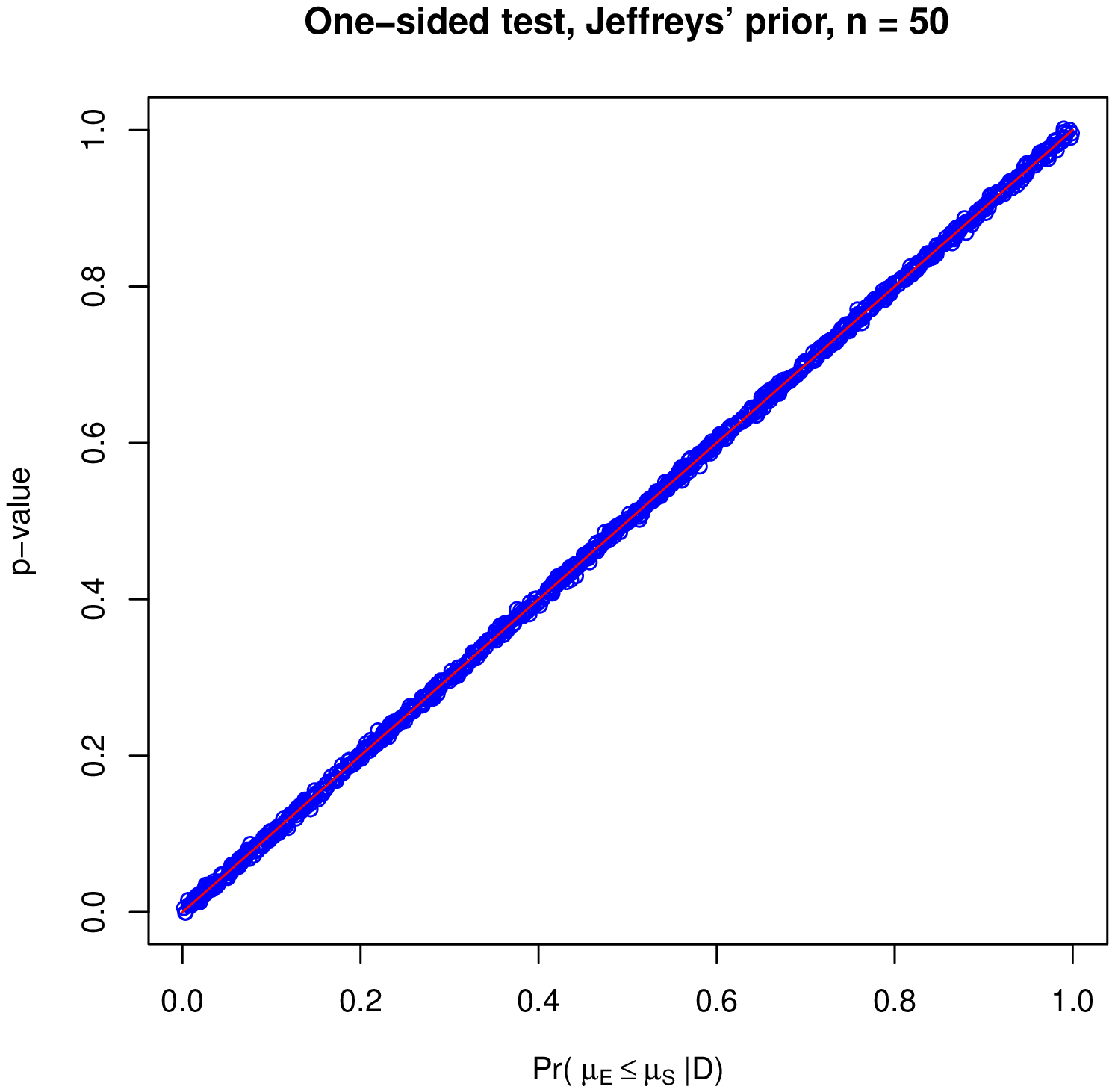}
\includegraphics[height=5cm,width=5cm]{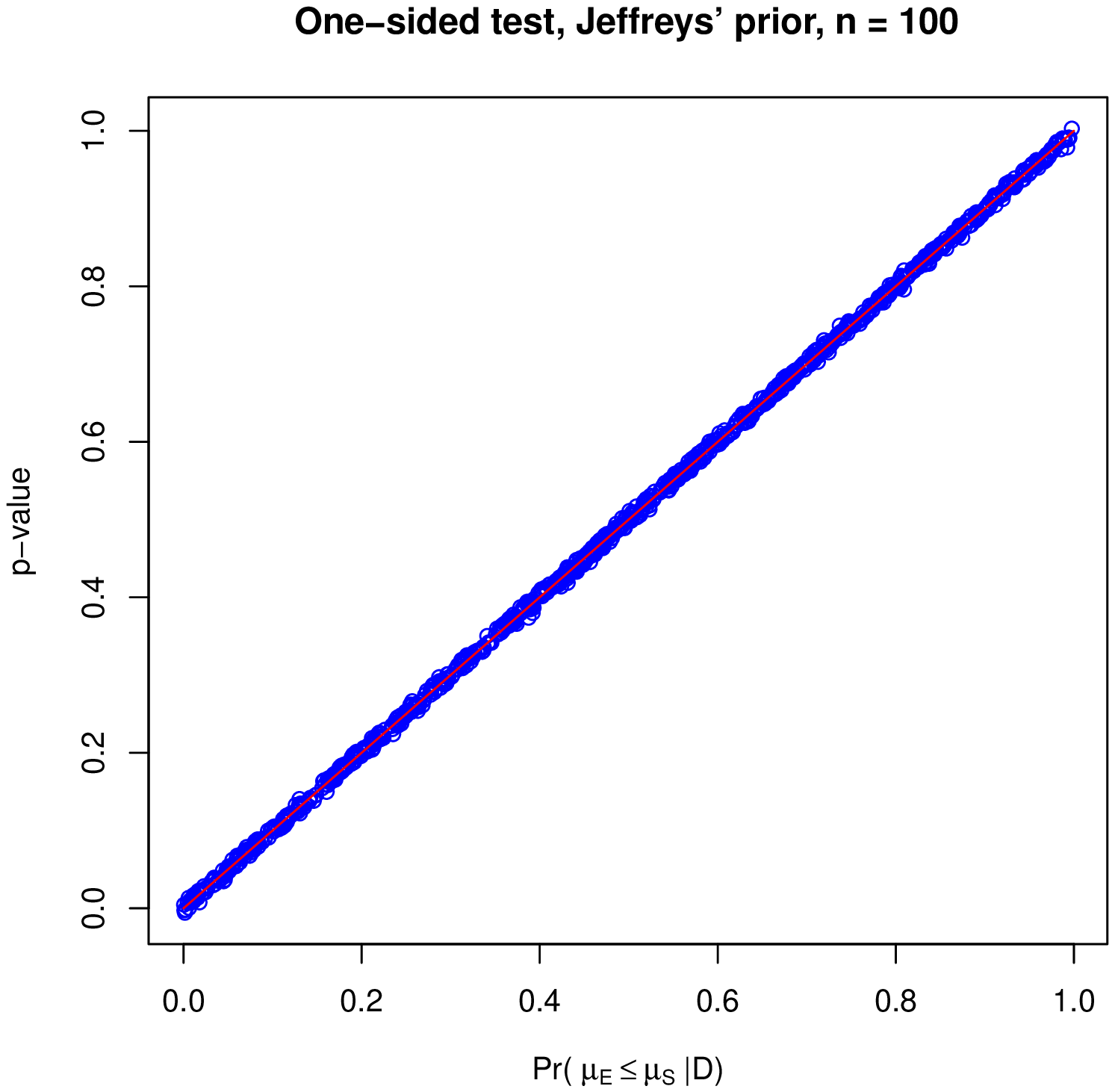}\\
\includegraphics[height=5cm,width=5cm]{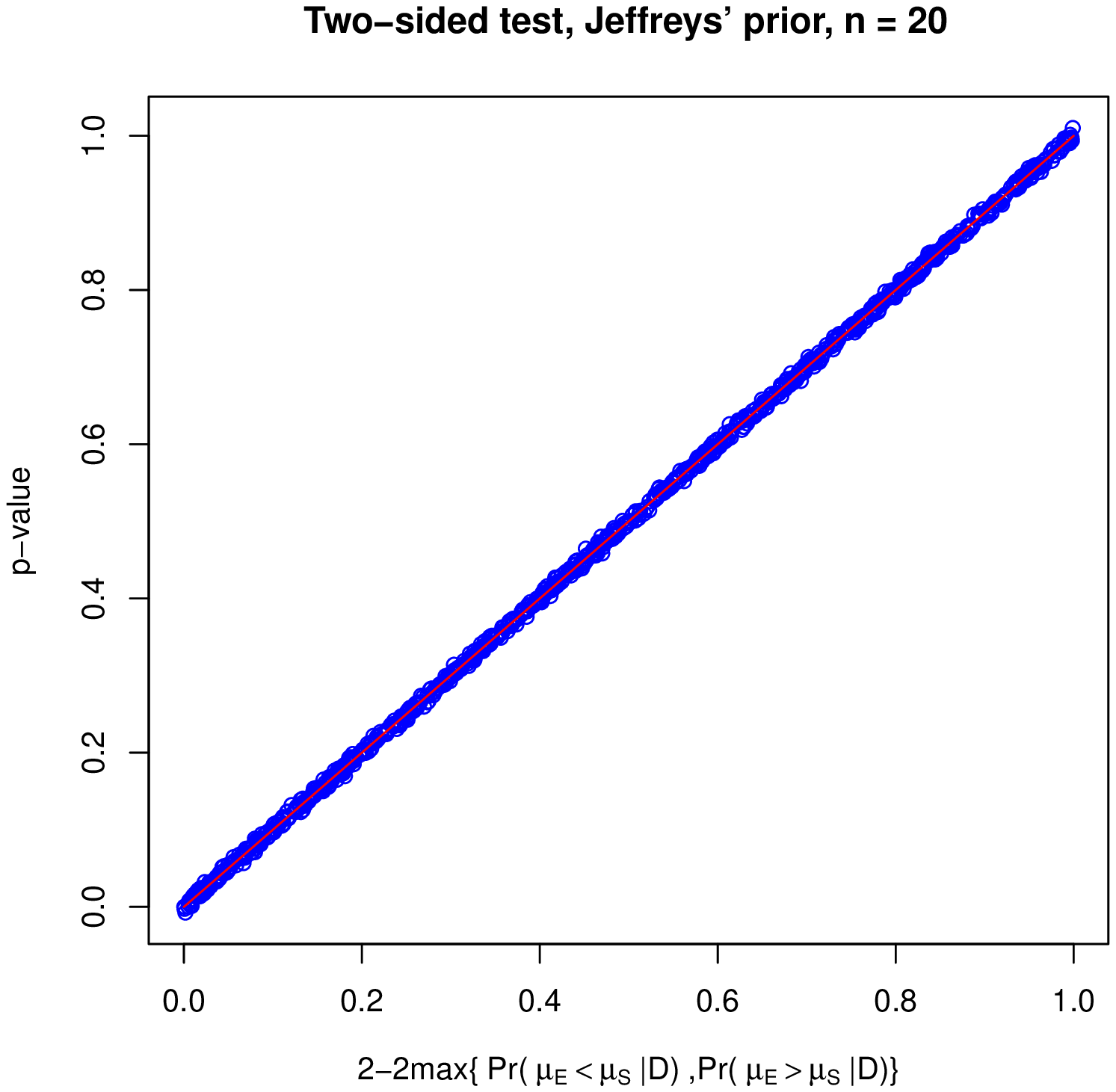}
\includegraphics[height=5cm,width=5cm]{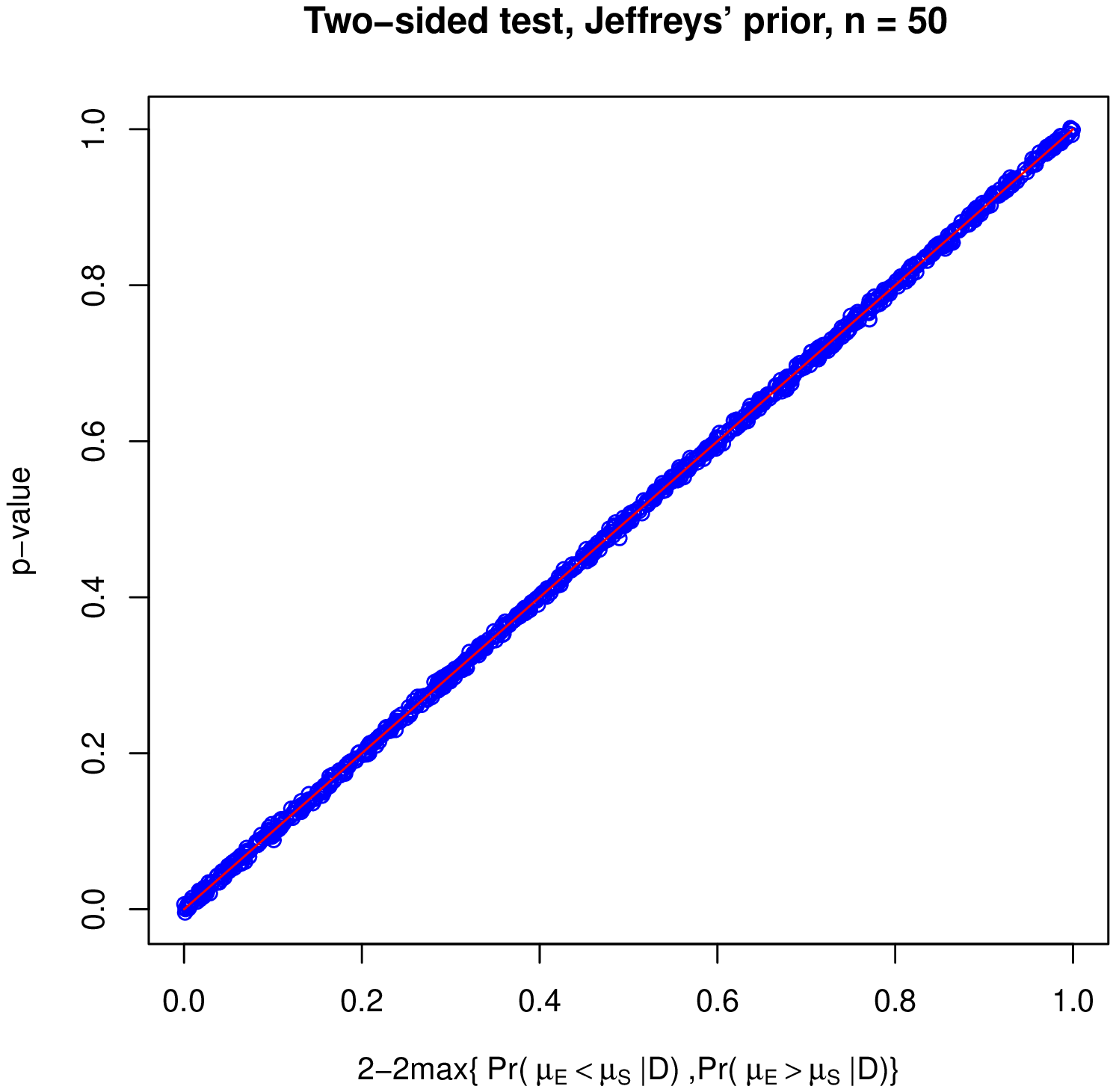}
\includegraphics[height=5cm,width=5cm]{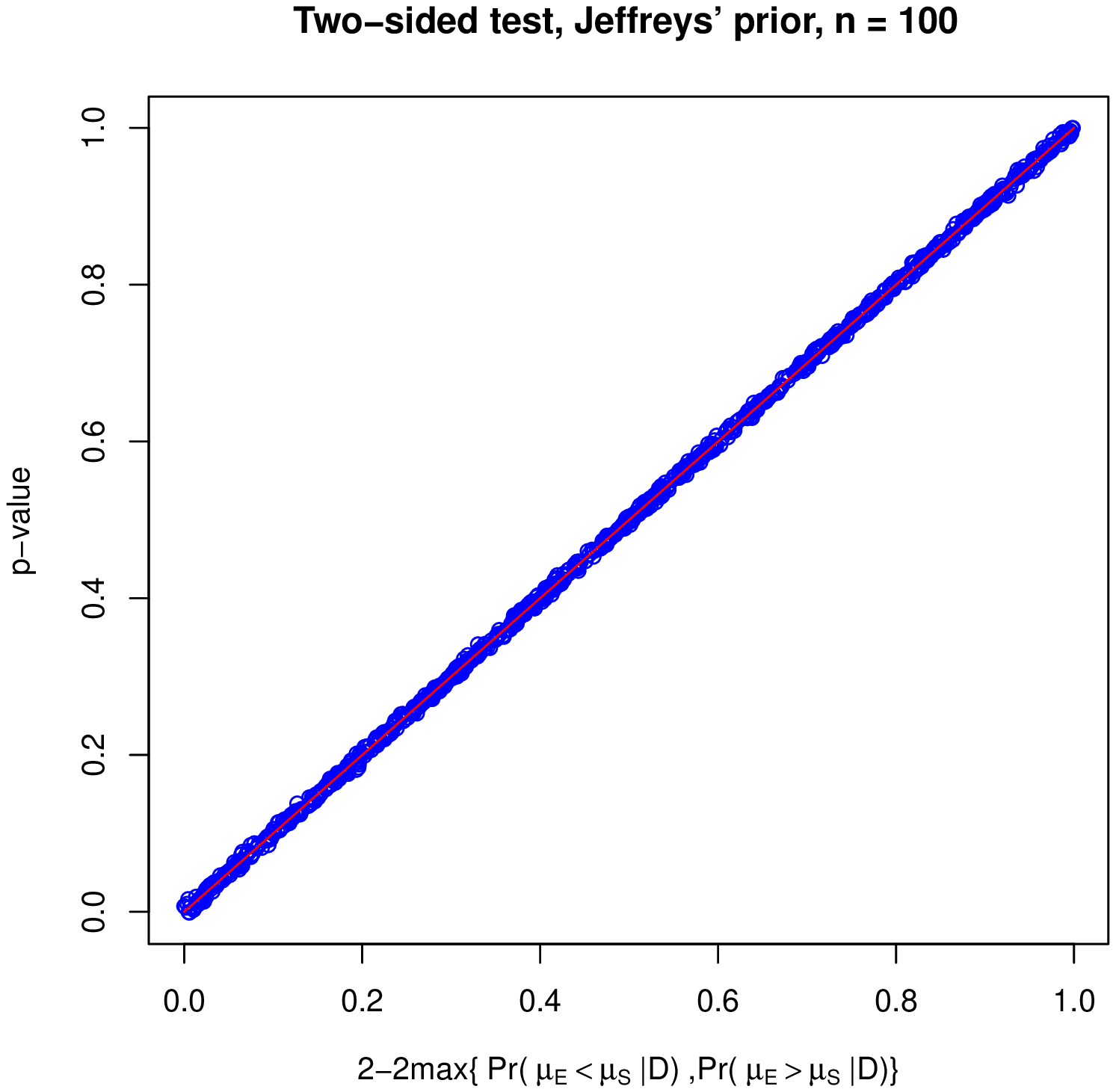}\\
\includegraphics[height=5cm,width=5cm]{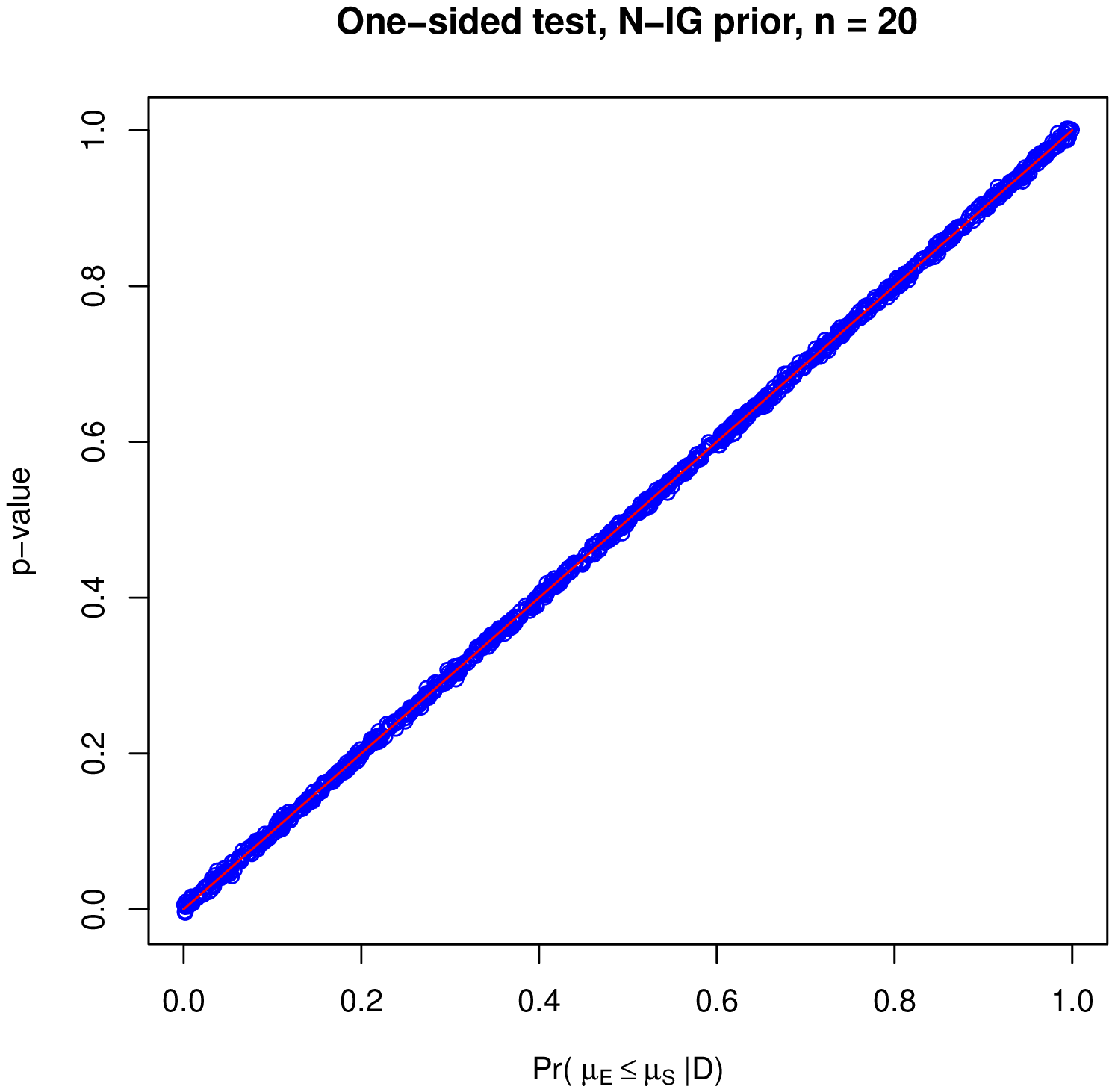}
\includegraphics[height=5cm,width=5cm]{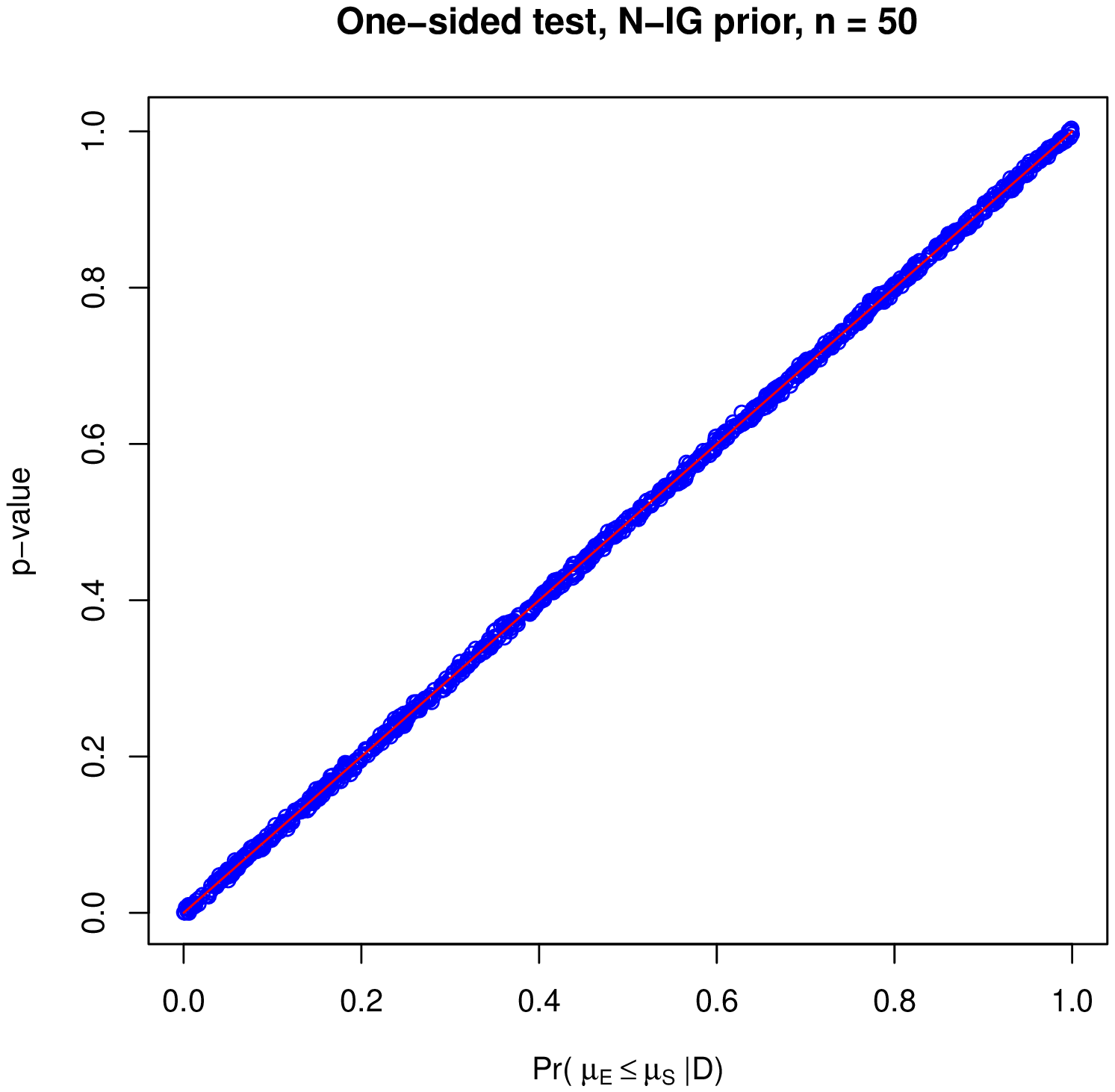}
\includegraphics[height=5cm,width=5cm]{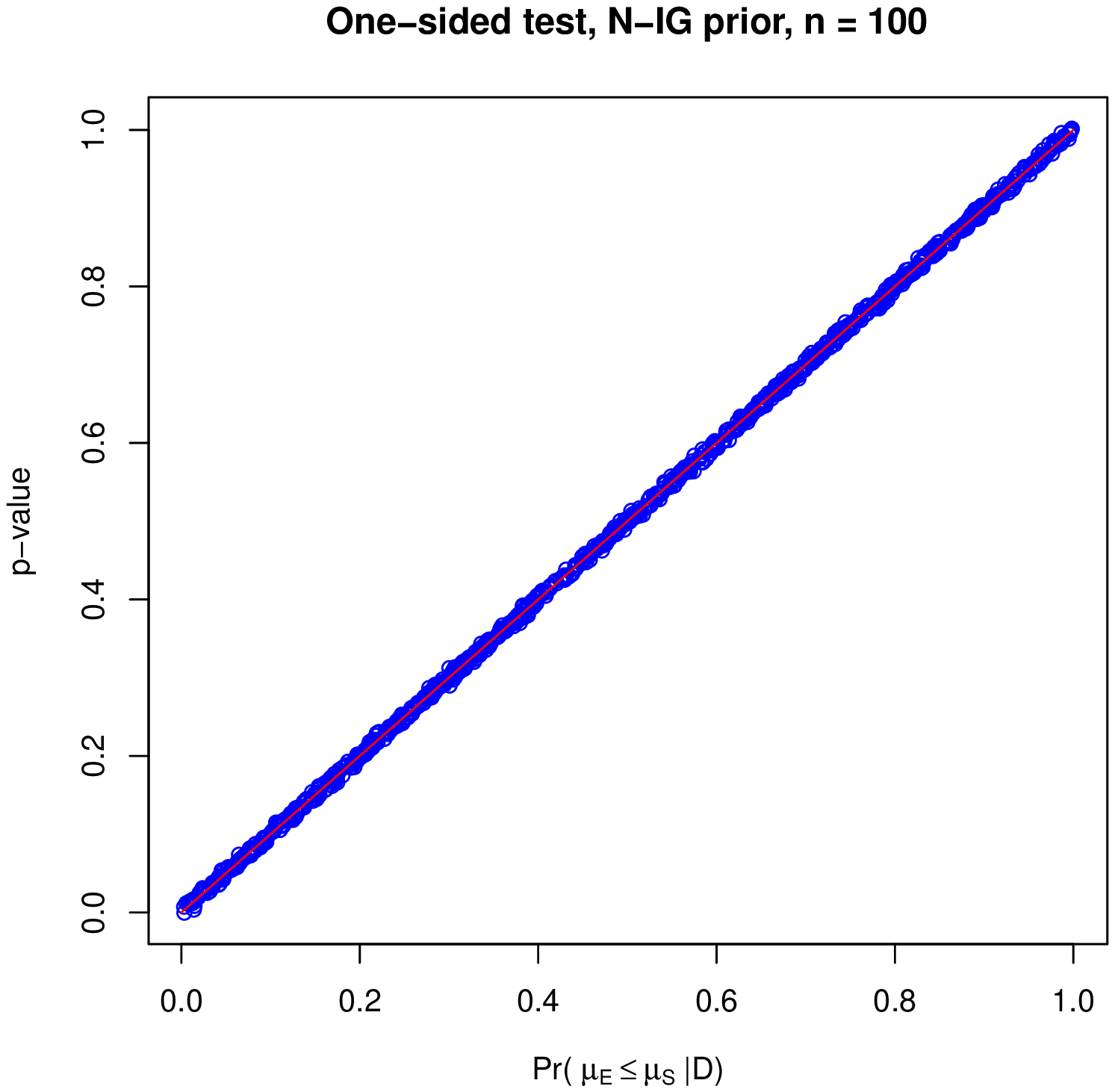}\\
\includegraphics[height=5cm,width=5cm]{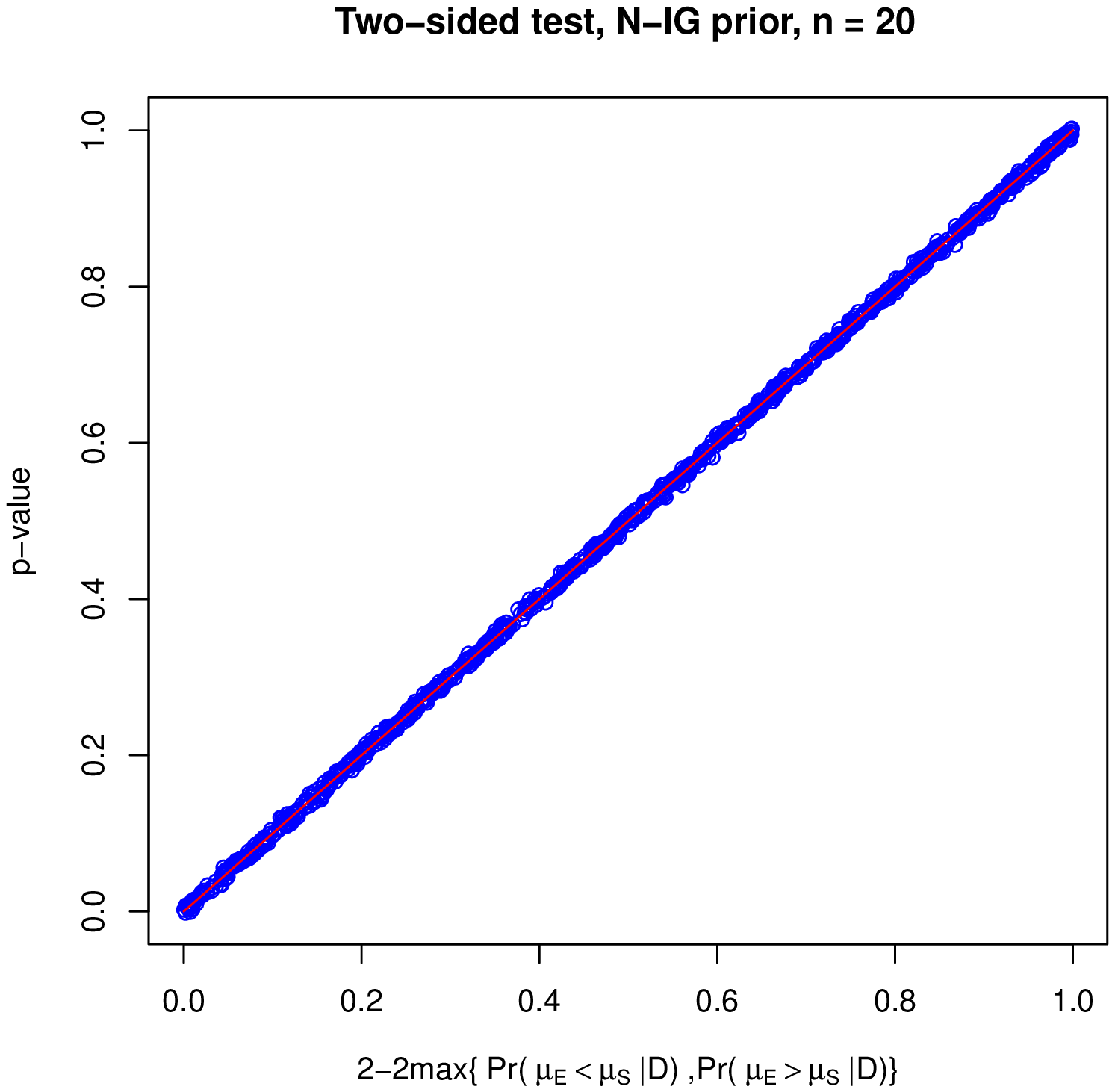}
\includegraphics[height=5cm,width=5cm]{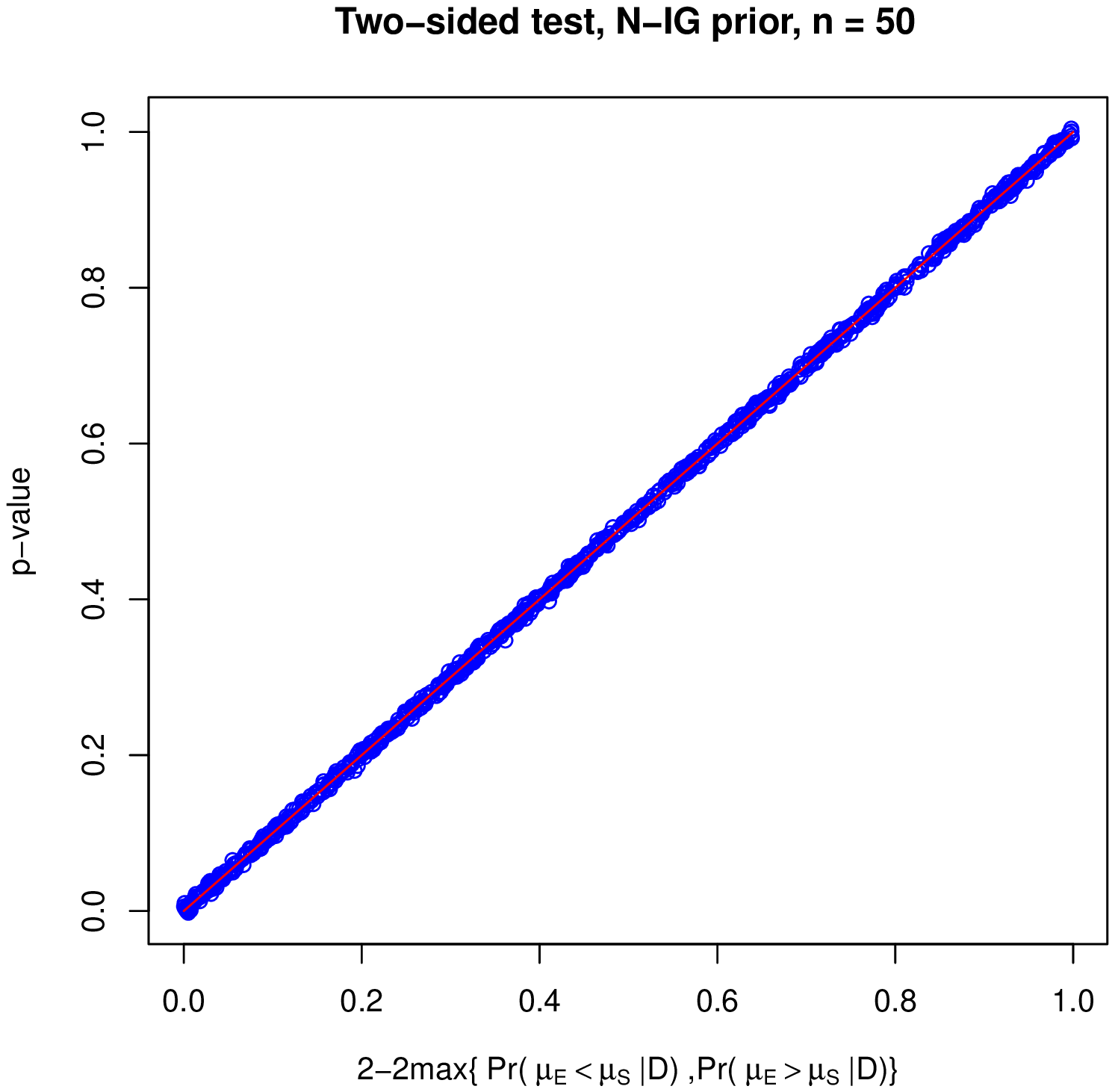}
\includegraphics[height=5cm,width=5cm]{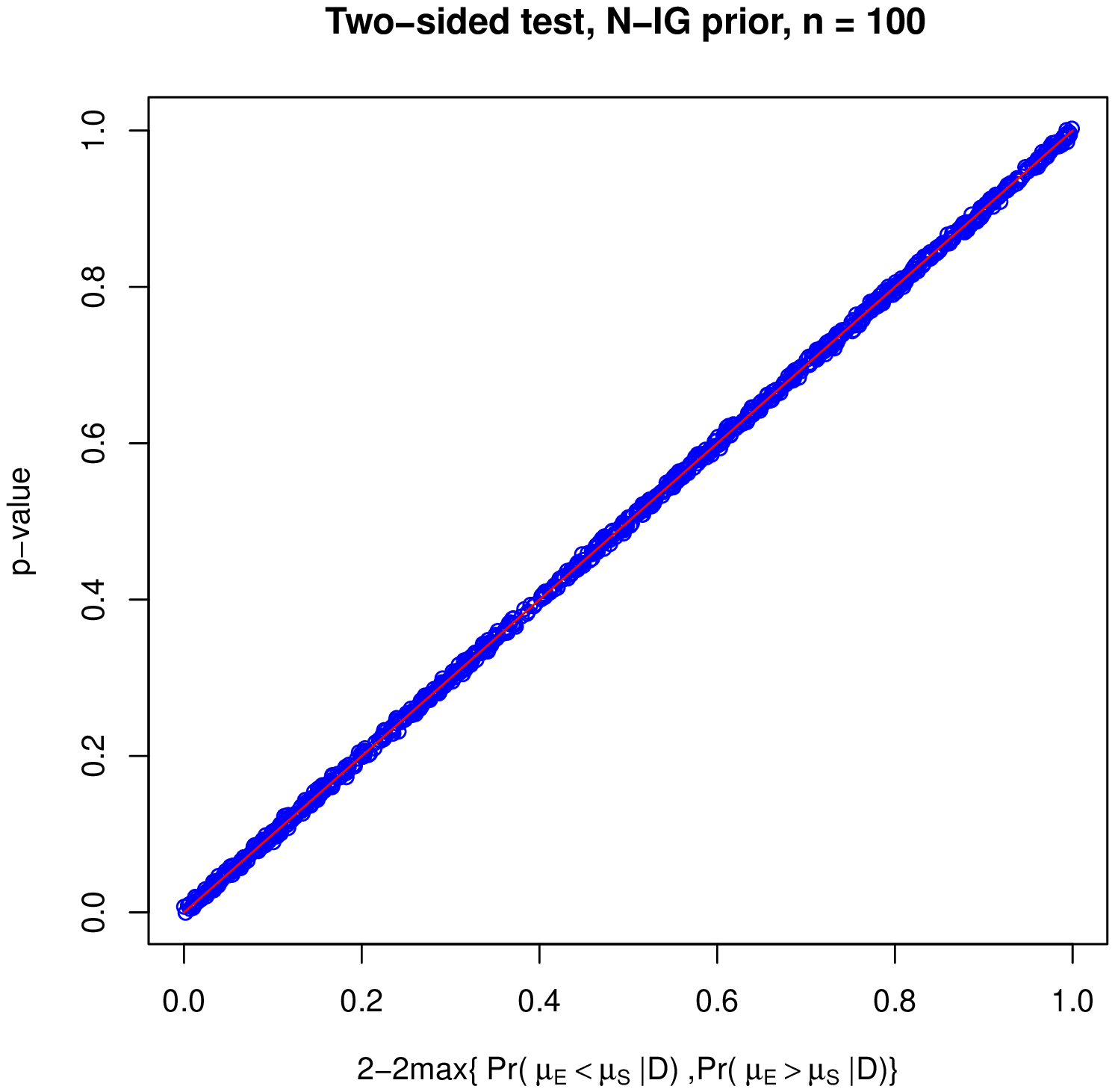}
\end{center}
\caption{The relationship between $p$-value and the posterior probability over 1000 replications under one-sided and two-sided hypothesis tests with normal outcomes assuming Jeffreys' prior and vague normal-inverse-gamma prior under sample size of 20, 50 and 100, respectively.
}
\label{normal_jeff}
\end{figure}

In a numerical study, we simulate a large number of trials, and for each replication we compute the posterior probability $\Pr(\theta \le 0|D)$ and  $p$-value.
To ensure that the simulated $p$-values can cover the entire range of $(0,1)$, we generate values of $\theta$ from ${\rm N}(0,0.05)$ and $\nu$ from ${\rm N}(1,0.05)$ truncated at zero.
To construct a vague normal-inverse-gamma prior distribution, we take $\theta_0 = 0$, $\nu_0 = 100$, and $\alpha=\beta=0.01$. Under Jeffreys' prior and the vague normal-inverse-gamma prior distributions, the equivalence relationships between $p$-values and the posterior probabilities $\Pr(\theta \le 0|D)$ are shown in Figure \ref{normal_jeff}, with sample size of 20, 50 and 100, respectively.

\begin{figure}[htb]
\begin{center}
\includegraphics[height=5cm,width=5cm]{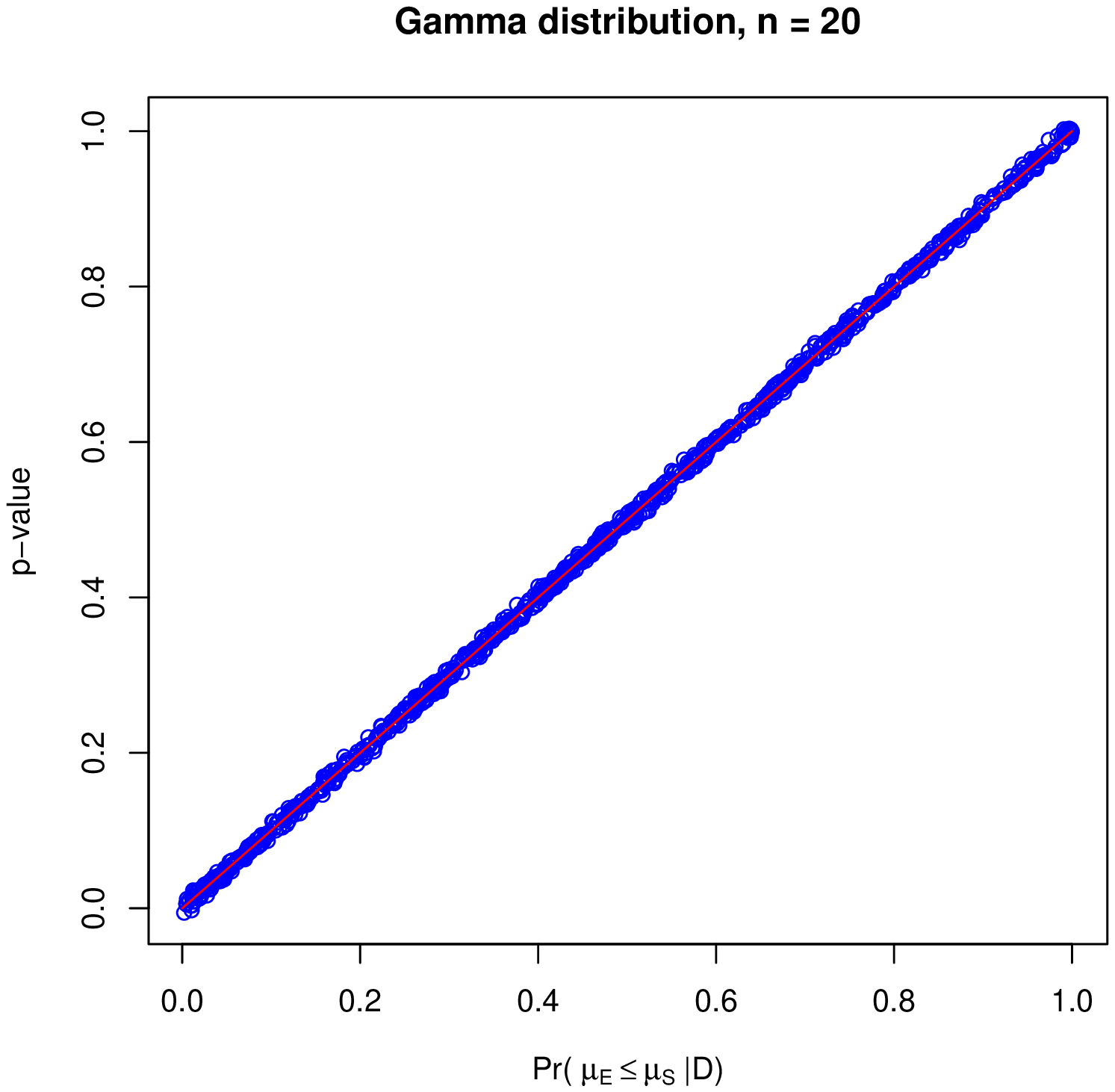}
\includegraphics[height=5cm,width=5cm]{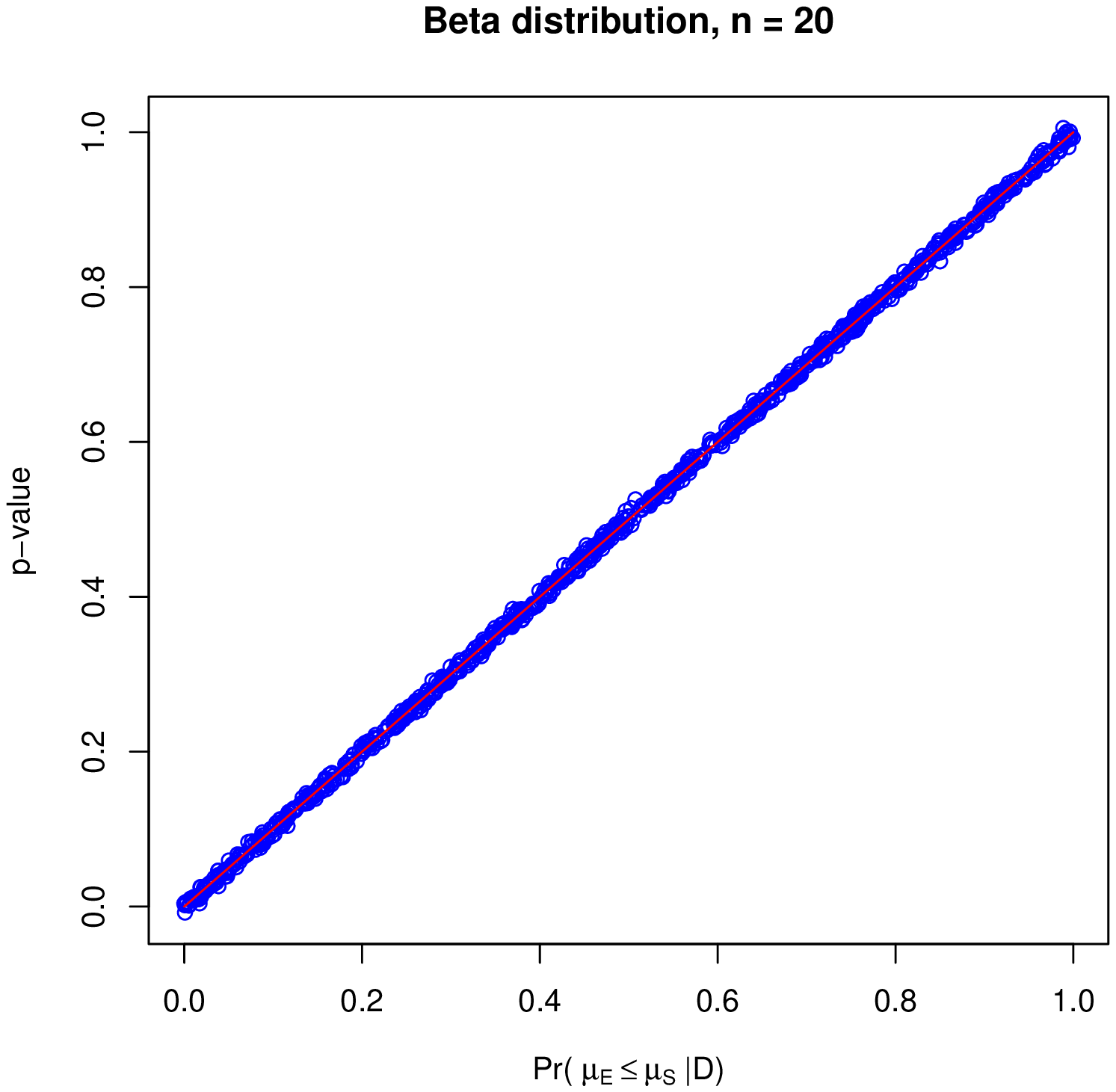}
\includegraphics[height=5cm,width=5cm]{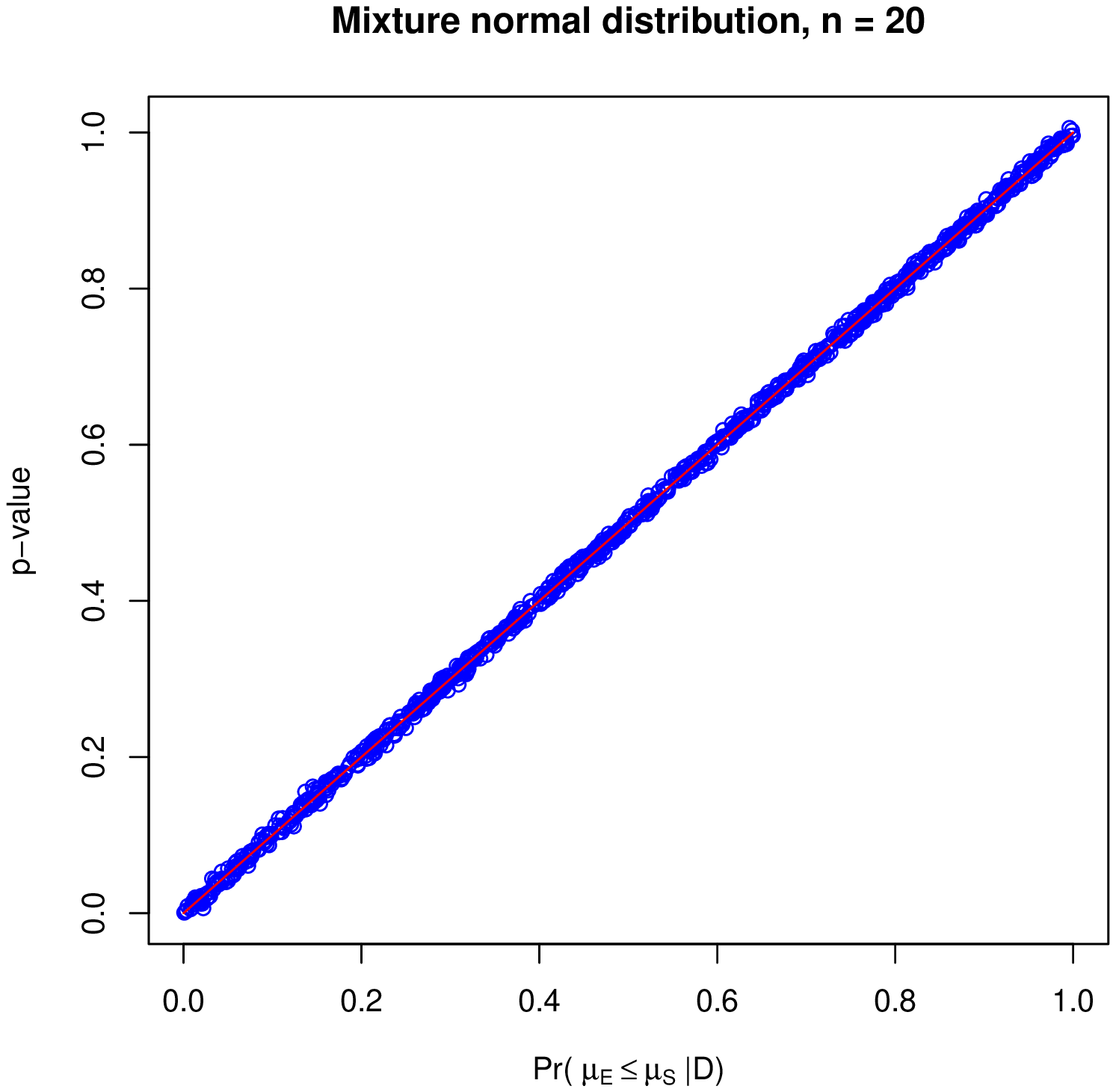}\\
\includegraphics[height=5cm,width=5cm]{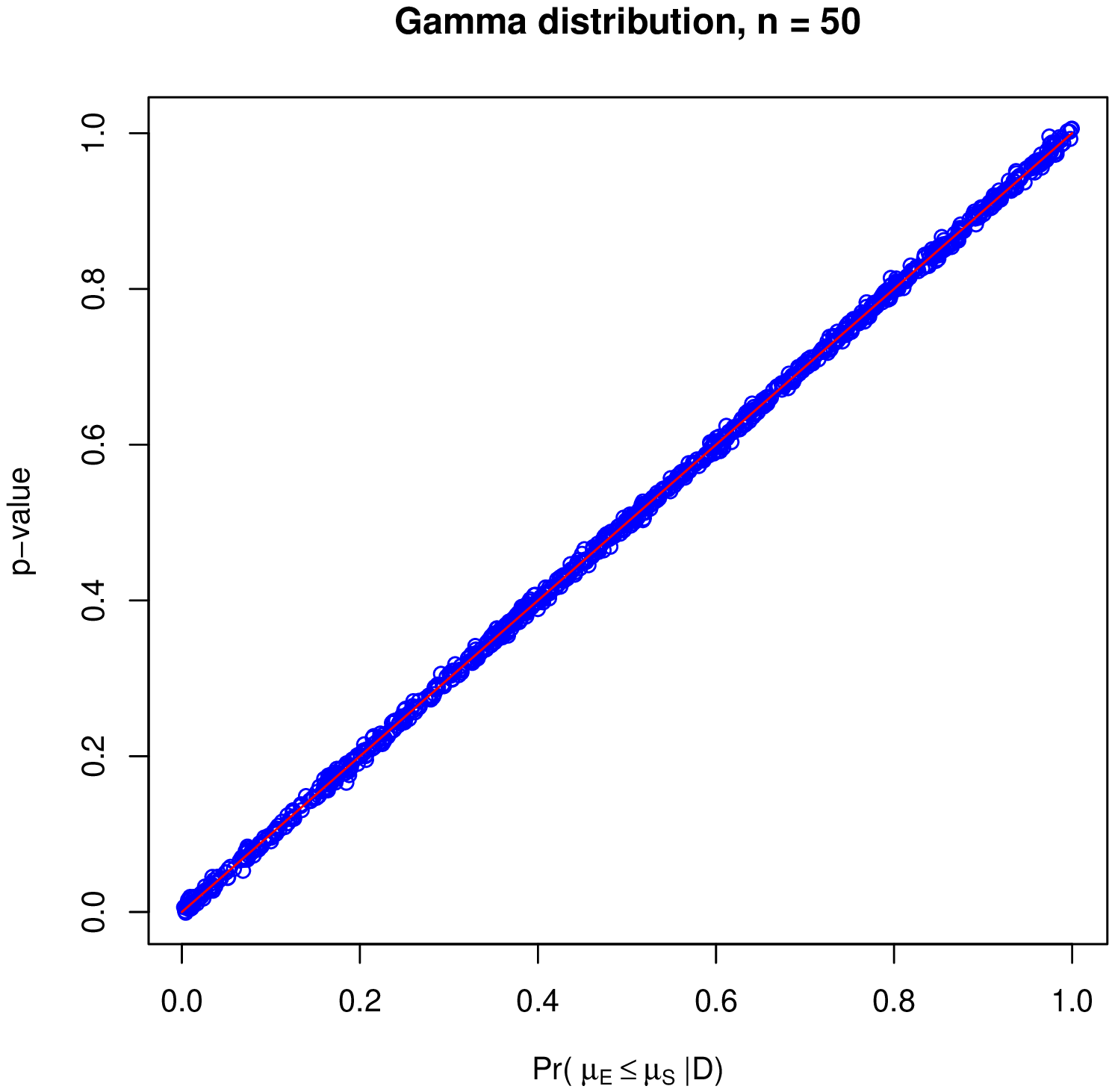}
\includegraphics[height=5cm,width=5cm]{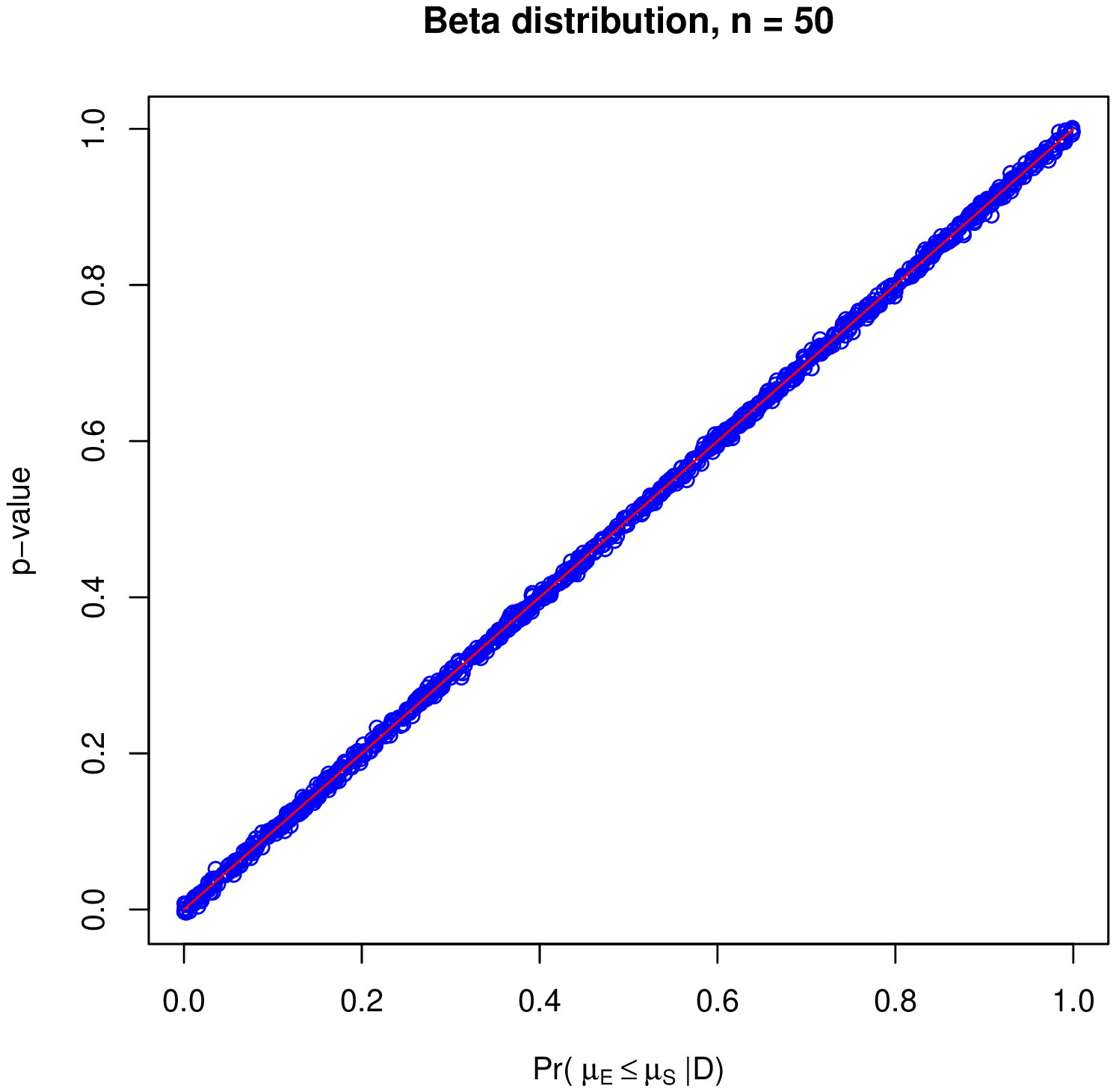}
\includegraphics[height=5cm,width=5cm]{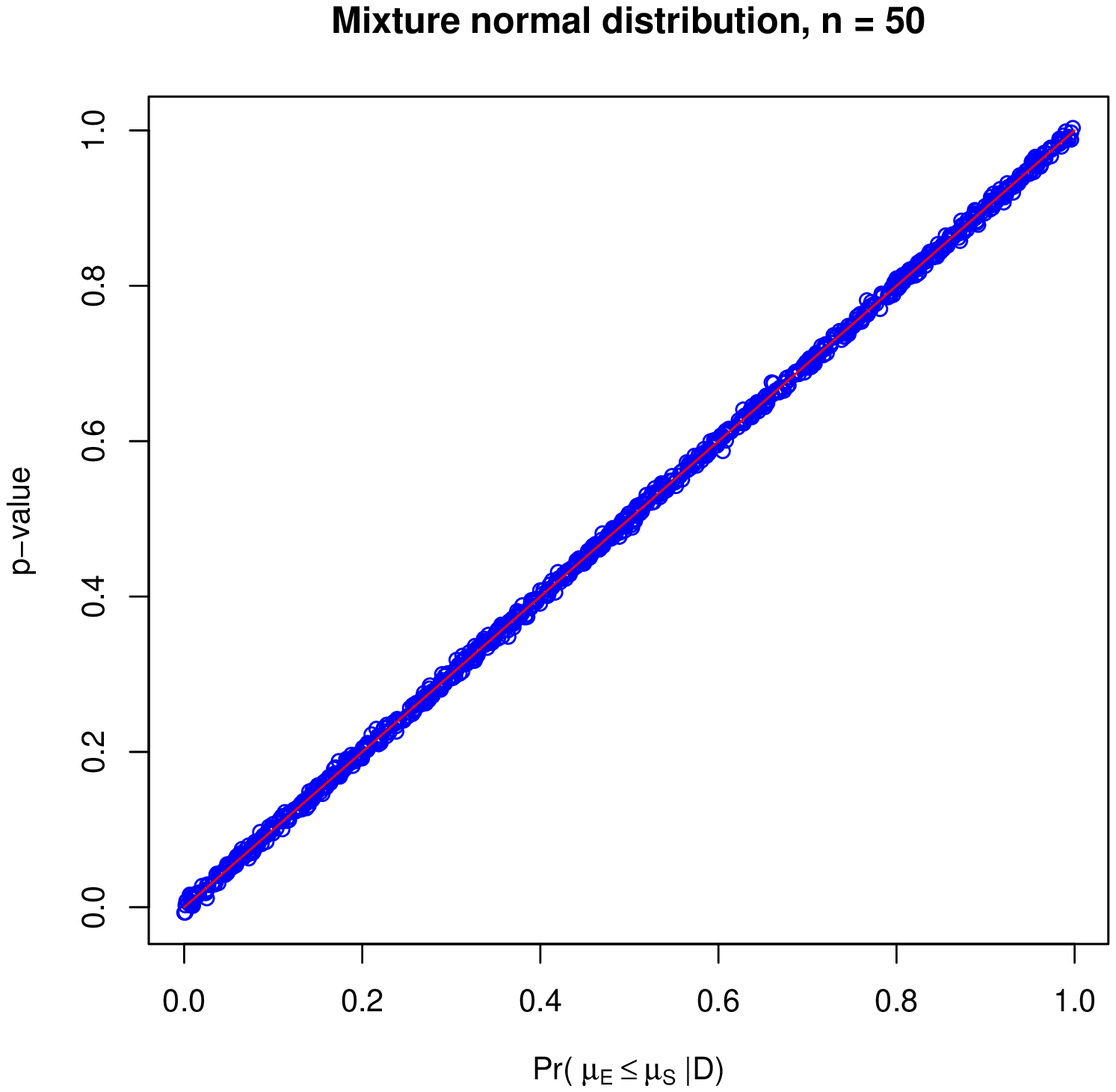}
\end{center}
\caption{The relationship between $p$-value and the posterior probability over 1000 replications under one-sided hypothesis tests with outcomes generated from Gamma, Beta and mixture normal distributions, assuming Jeffreys' prior for the normal distribution under sample size of 20 and 50, respectively.
}
\label{normal_gamma}
\end{figure}

In addition, we generate values of $x_i$ from a Gamma$(2,0.5)$ distribution, a Beta$(0.5,0.5)$ distribution, as well as a mixture of normal distributions of N$(-1,1)$ and N$(1,1)$ with equal weights. To ensure that the simulated $p$-values can cover the entire range of $(0,1)$, the simulated values of $x_i$ are further deducted by the mean value of the corresponding distribution plus a uniform random variable. Under Jeffreys' prior, the equivalence relationships between $p$-values and the posterior probabilities $\Pr(\theta \le 0|D)$ are shown in Figure \ref{normal_gamma}.

\begin{figure}[htb]
\begin{center}
\includegraphics[height=6.5cm,width=6.5cm]{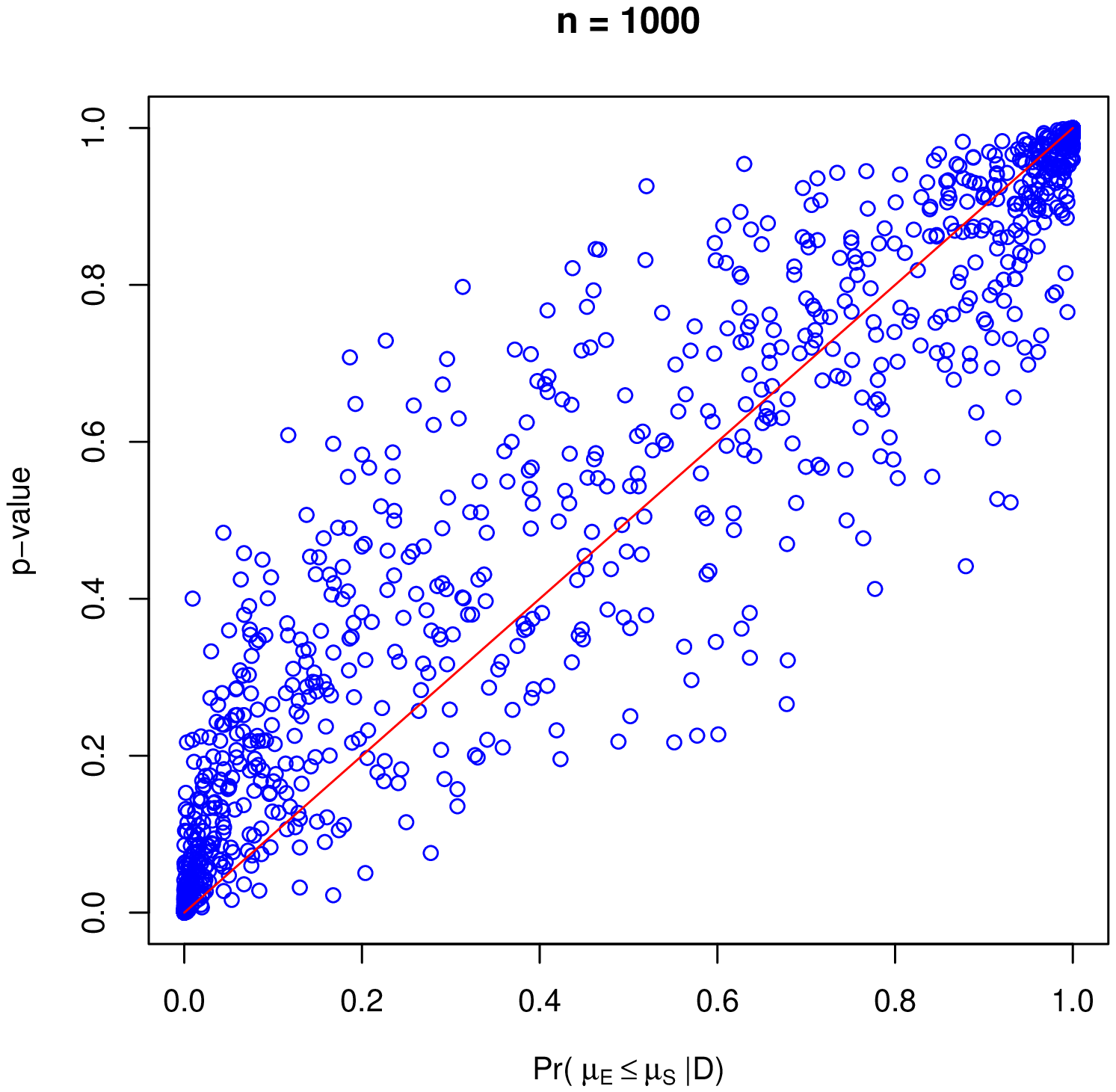}
\includegraphics[height=6.5cm,width=6.5cm]{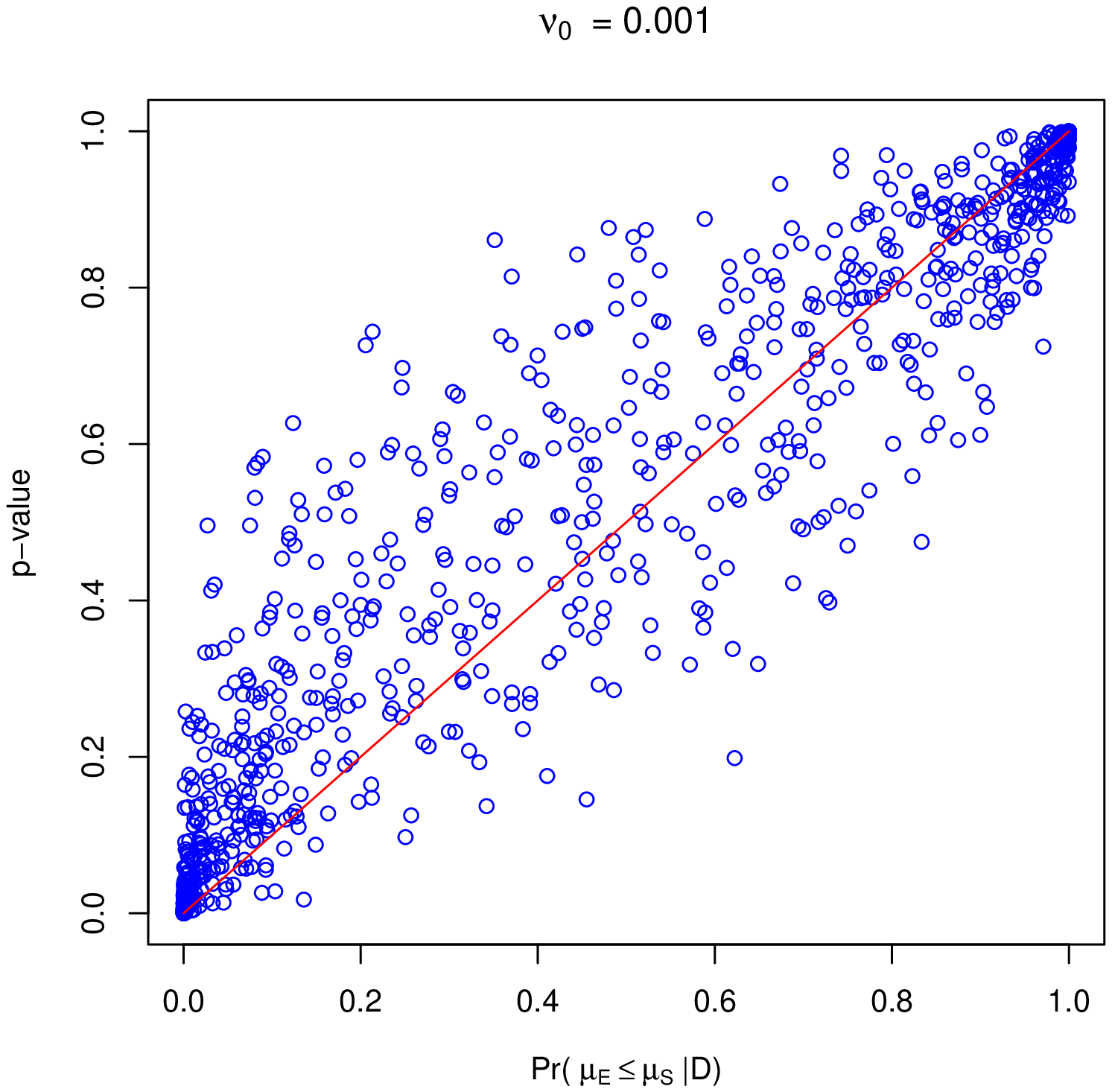}
\\
\includegraphics[height=6.5cm,width=6.5cm]{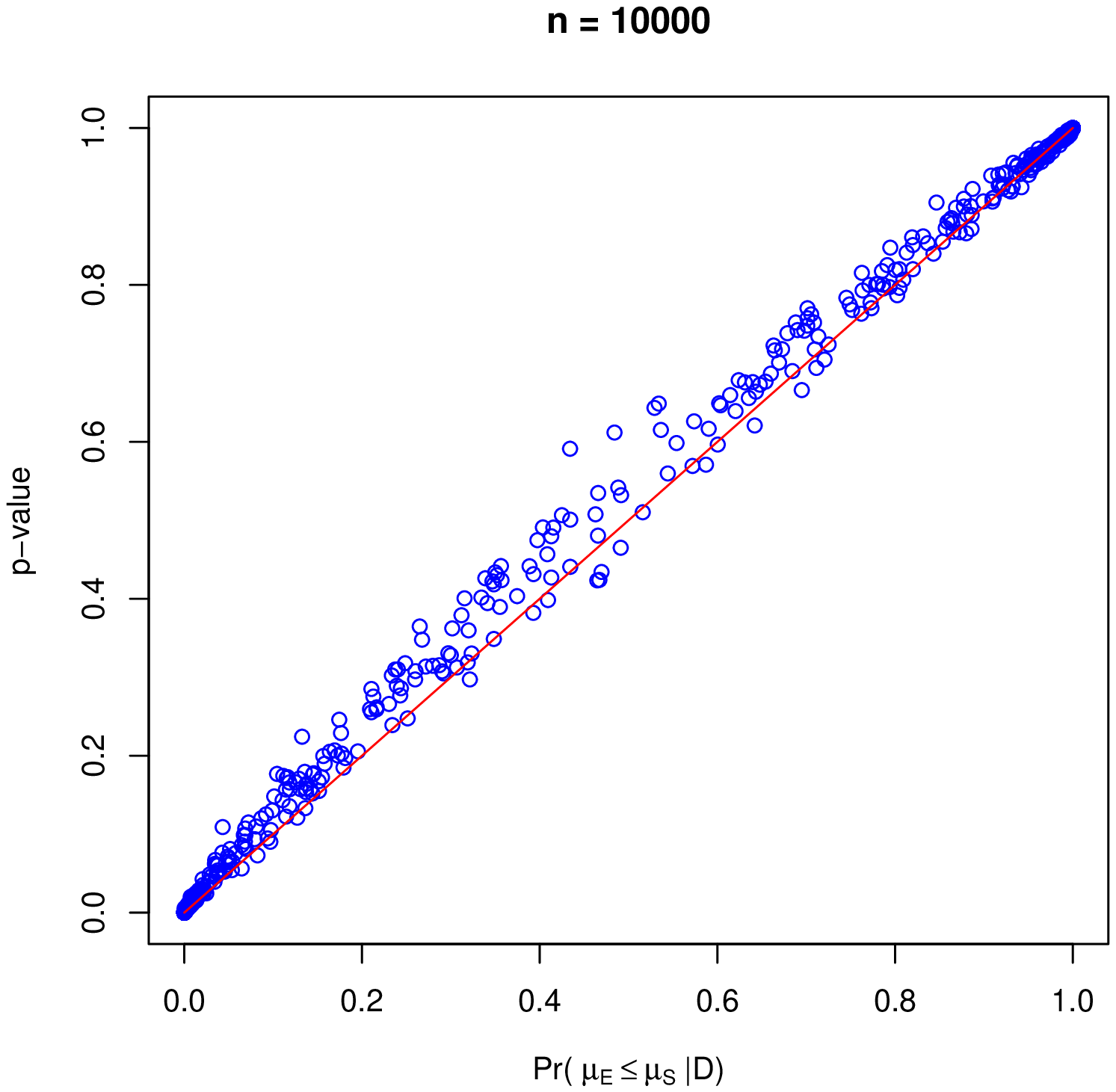}
\includegraphics[height=6.5cm,width=6.5cm]{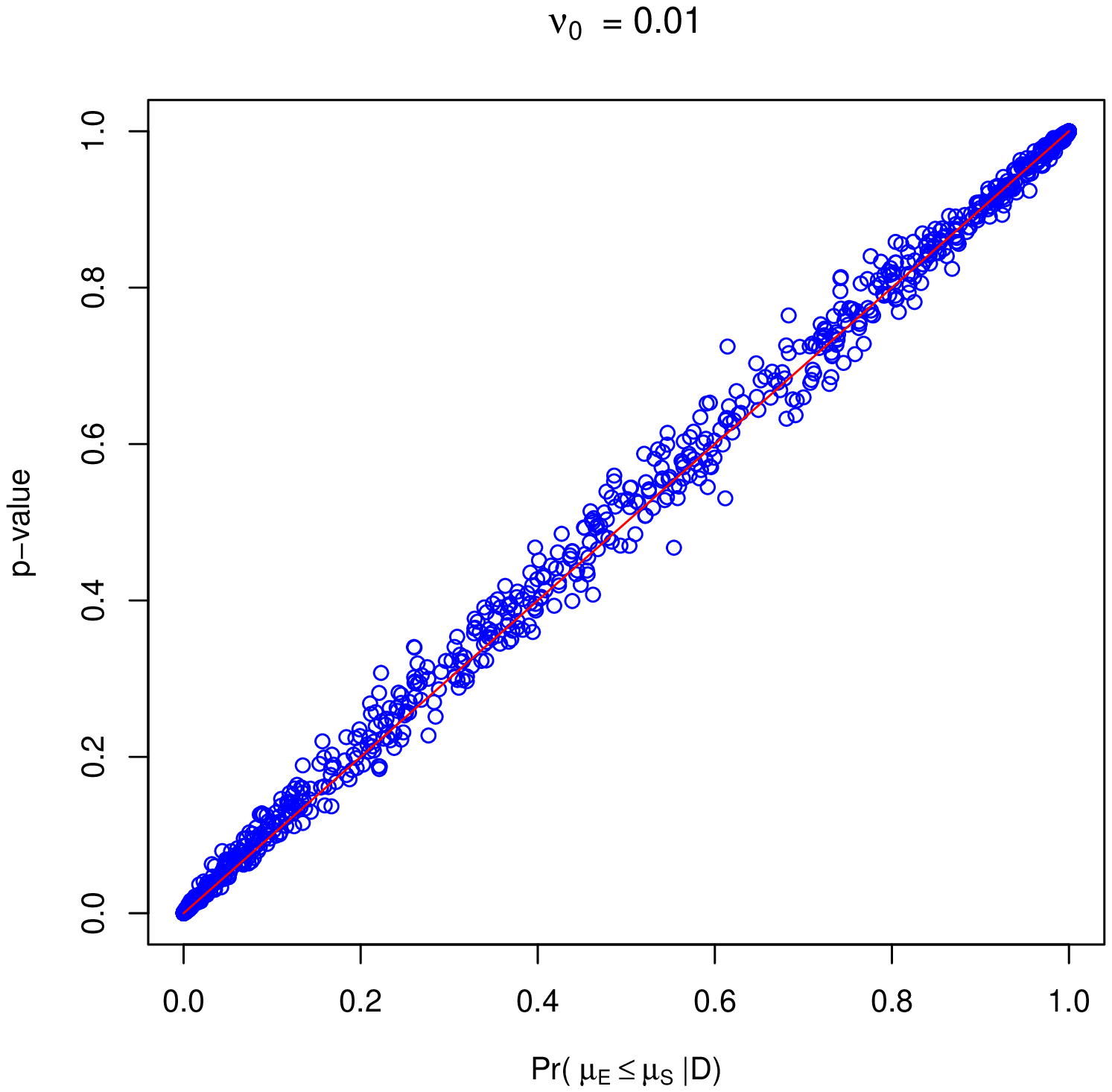}
\\
\includegraphics[height=6.5cm,width=6.5cm]{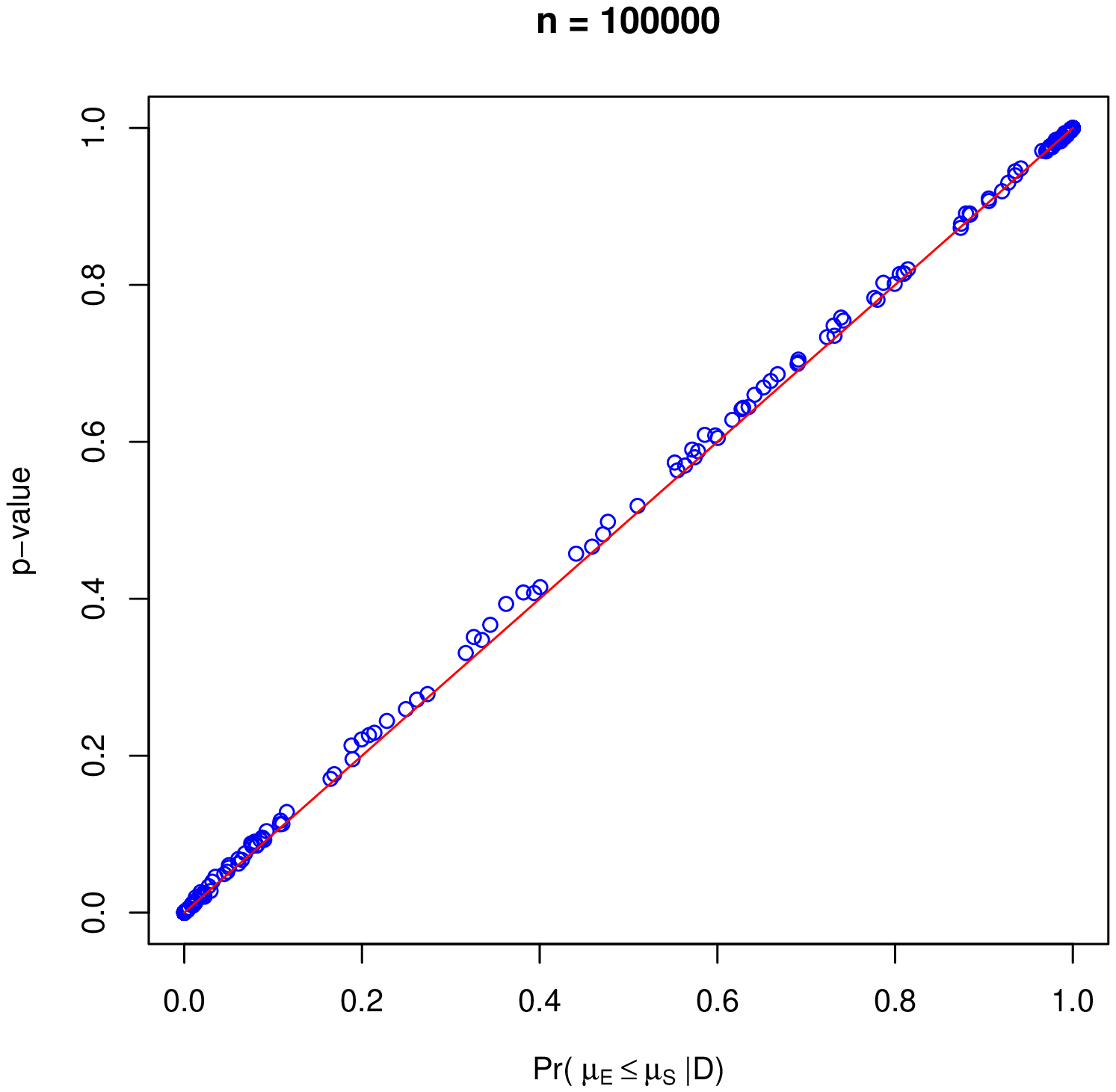}
\includegraphics[height=6.5cm,width=6.5cm]{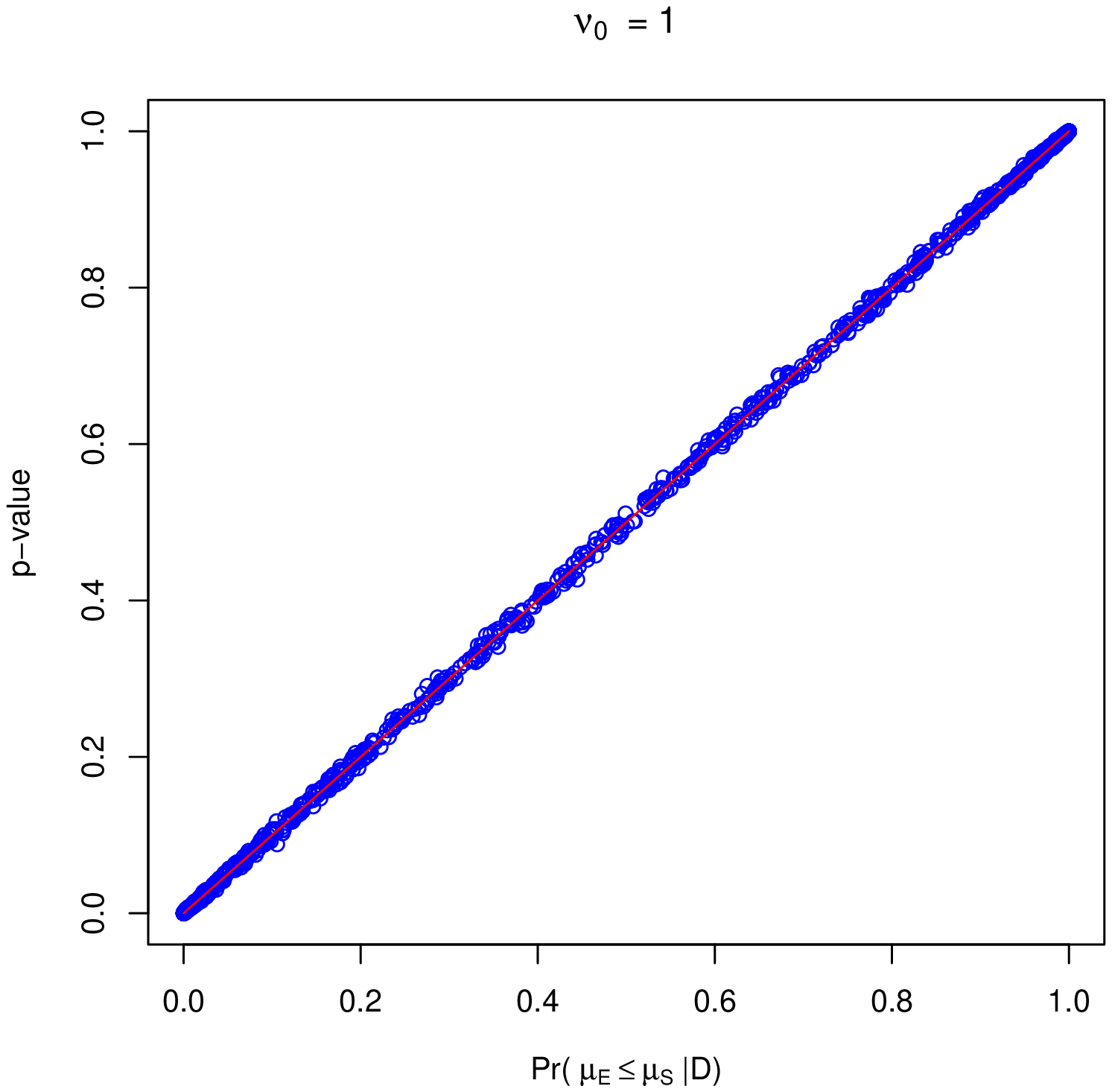}
\end{center}
\caption{The relationship between $p$-value and the posterior probability $\Pr(\mu_E \le \mu_S|D)$ over 1000 replications under one-sided hypothesis tests with normal outcomes; left panel: assuming a fixed informative normal-inverse-gamma prior under increasing sample sizes of 1000, 10000 and 100000 (from top to bottom), right panel: assuming
a fixed sample size of 1000 with an increasing prior variance of 0.001, 0.01 and 1 (from top to bottom).}
\label{informative}
\end{figure}

To study the effect of informative prior and sample size on the relationship between $p$-values and the posterior probabilities, we construct an informative prior distribution on $\theta$ by setting $\theta_0 = \theta + 0.01$, $\nu_0 = 0.01$, and $\alpha=\beta=0.01$. Under such an informative prior distribution, the relationships between $p$-values and the posterior probabilities $\Pr(\theta \le 0|D)$ under increasing sample sizes are shown in Figure \ref{informative}. As sample size increases, the equivalence relationship is gradually established.
Moreover, we consider the case where the sample size is fixed but the prior variance increases, i.e., we take $\theta_0 = \theta + 0.01$ and let $\nu_0$ change from 0.001 to 1. As shown in Figure \ref{informative}, as the prior distribution becomes less informative, the equivalence relationship becomes more evident.

%
%In the case with a single-arm trial, we are interested in
%comparing the mean of the outcome $\mu_E$ and a prespecified threshold value.
%The frequentist $p$-value under the $t$-test and Bayesian posterior probability can be derived along similar lines.

\section{Hypothesis Test for Multivariate Normal Data}

In
hypothesis testing on the mean vector of a multivariate normal random variable, we consider $\bX \sim {\rm N}_p(\bmu,\bSigma)$, where $p$ is the dimension of the multivariate normal distribution. For the ease of exposition, the covariance matrix $\bSigma$ is assumed to be known.
Let $D = \{\bX_1,\ldots,\bX_n\}$ denote the observed multivariate vectors, let $\bar \bX = \sum_{i=1}^{n}\bX_i/n$ denote the sample mean vector, and thus $\bar \bX \sim {\rm N}_p(\bmu,\bSigma/n)$.

Consider the one-sided hypothesis test,
\begin{equation*}\label{Hypothesis}
H_0\mbox{:} \ \bc_k^{\top}\bmu \le 0 \ {\rm for} \ {\rm some} \ k = 1,\ldots,K \quad  {\rm versus} \quad H_1\mbox{:} \ \bc_k^{\top}\bmu > 0 \ {\rm for} \ {\rm all} \ k = 1,\ldots,K,
\end{equation*}
where $\bc_1,\ldots,\bc_K$ are $K$ prespecified $p$-dimensional vectors.
The likelihood ratio test statistics (Sasabuchi, 1980) are
\begin{equation}\label{sasa}
Z_k =  \frac{\bc_k^{\top} {\bar {\bX}}}{\sqrt{\bc_k^{\top} { \bSigma}\bc_k/n}}, \ \  k = 1,\ldots,K,
\end{equation}
and the corresponding $p$-values are
$$
p \mbox{-value}(k)_1  = 1 - \Phi(Z_k).
$$
The null hypothesis is rejected if all of the $K$ $p$-values are smaller than $\alpha$.

In the Bayesian paradigm, we assume a conjugate multivariate normal prior distribution for $\bmu$, $
\bmu \sim {\rm N}_p(\bmu_0,\bSigma_0)$.
The corresponding posterior distribution is $
\bmu|D \sim {\rm N}_p(\bmu_n,\bSigma_n)$,
where
\begin{align*}
  \bmu_n & = \bSigma_0\left(\bSigma_0+ \frac{\bSigma}{n}\right)^{-1}{\bar \bX} + \frac{1}{n} \bSigma \left(\bSigma_0+ \frac{\bSigma}{n}\right)^{-1}\bmu_0, \\
  \bSigma_n & = \frac{1}{n}\bSigma \left(\bSigma_0+ \frac{\bSigma}{n}\right)^{-1}\bSigma.
\end{align*}
The one-sided posterior probability corresponding to $\bc_k$ is
$$
{\rm PoP}(k)_1 = \Pr(\bc_k^{\top}\bmu \le 0 | D).
$$

For two-sided hypothesis testing (Liu and Berger, 1995), we are interested in
\begin{eqnarray*}\label{Hypothesis}
&&H_0\mbox{:} \ \bc_k^{\top}\bmu \le 0 \ {\rm for} \ {\rm some} \ k = 1,\ldots,K, {\rm and} \\
&& \ \ \ \ \ \  \bc_k^{\top}\bmu \ge 0 \ {\rm for} \ {\rm some} \ k = 1,\ldots,K \\
&&{\rm versus}  \\
&&H_1\mbox{:} \ \bc_k^{\top}\bmu > 0 \ {\rm for} \ {\rm all} \ k = 1,\ldots,K, \ {\rm or} \\
&& \ \ \ \ \ \ \bc_k^{\top}\bmu < 0 \ {\rm for} \ {\rm all} \ k = 1,\ldots,K.
\end{eqnarray*}
Based on (\ref{sasa}), the $p$-values are given by
$$
p \mbox{-value}(k)_2  = 2 - 2\Phi(|Z_k|) = 2[1-{\rm max}\{ \Phi(Z_k),\Phi(-Z_k) \}].
$$
The null hypothesis is rejected if all of the $K$ $p$-values are smaller than $\alpha$.
Similar to the univariate case, we define the two-sided posterior probability,
$${\rm PoP}(k)_2 = 2[1-{\rm max}\{\Pr(\bc_k^{\top}\bmu > 0 | D), \Pr(\bc_k^{\top}\bmu < 0 | D)\}].$$

\begin{figure}[htb]
\begin{center}
\includegraphics[height=8cm,width=8cm]{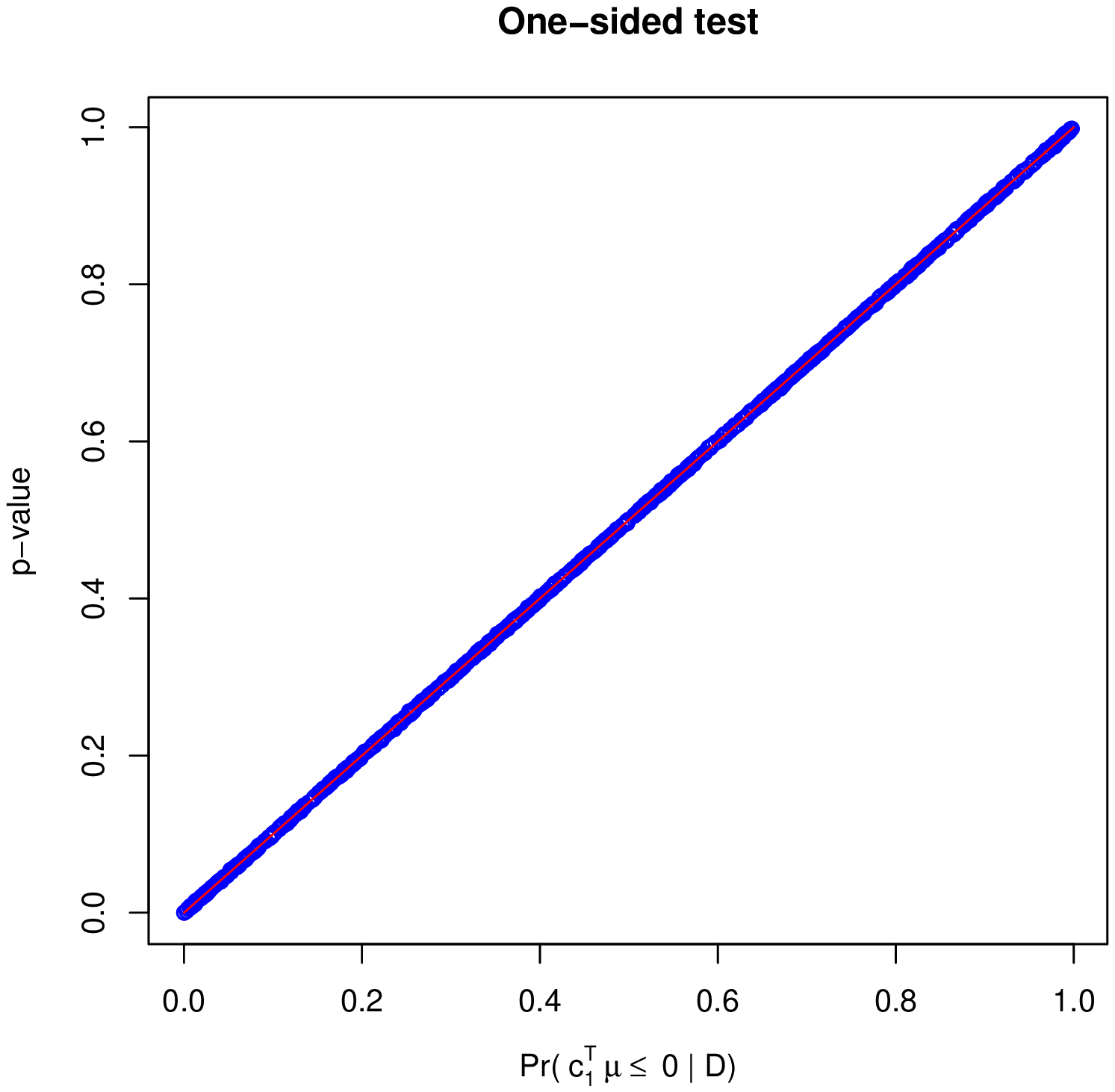}
\includegraphics[height=8cm,width=8cm]{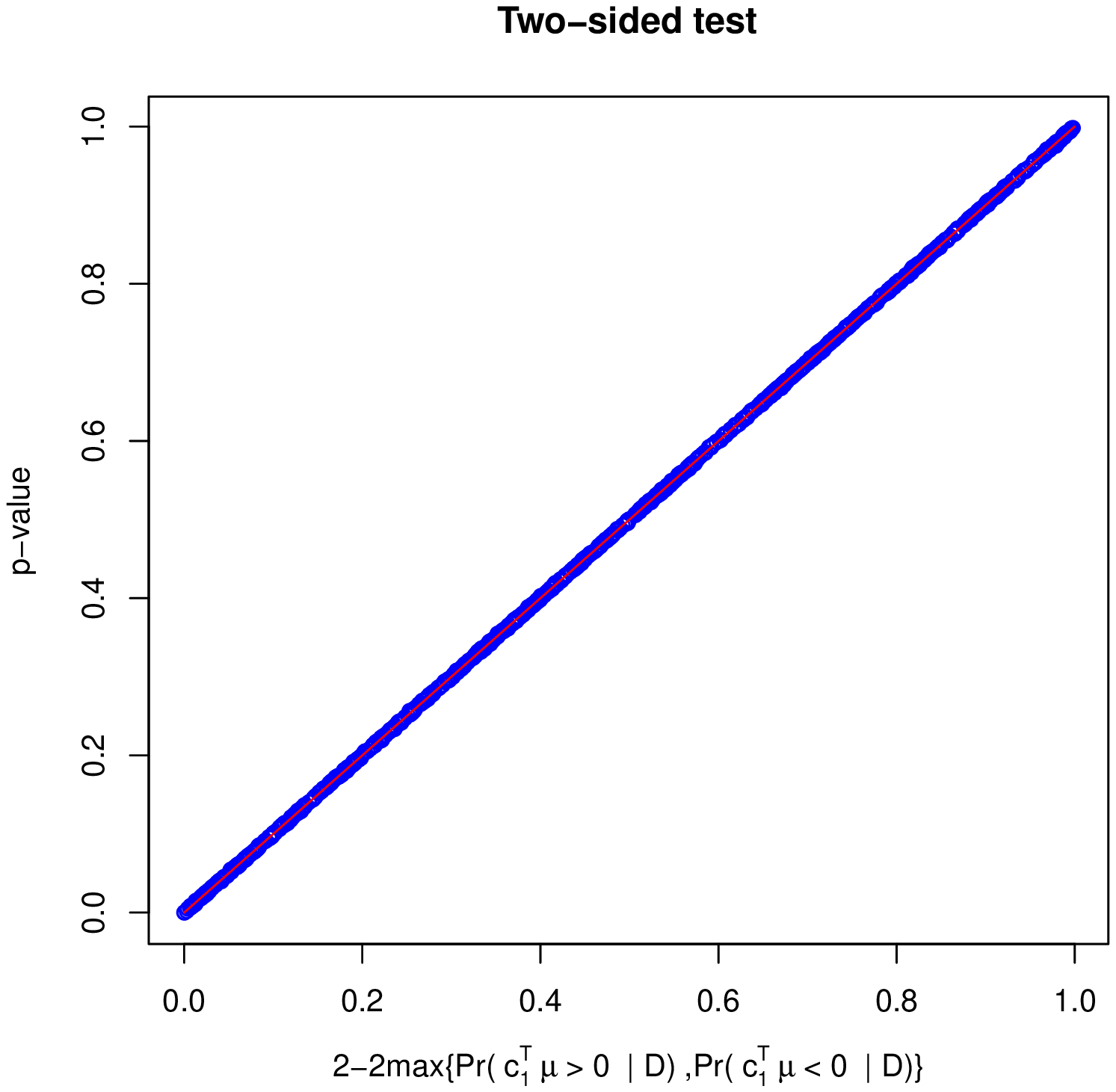}\\
\includegraphics[height=8cm,width=8cm]{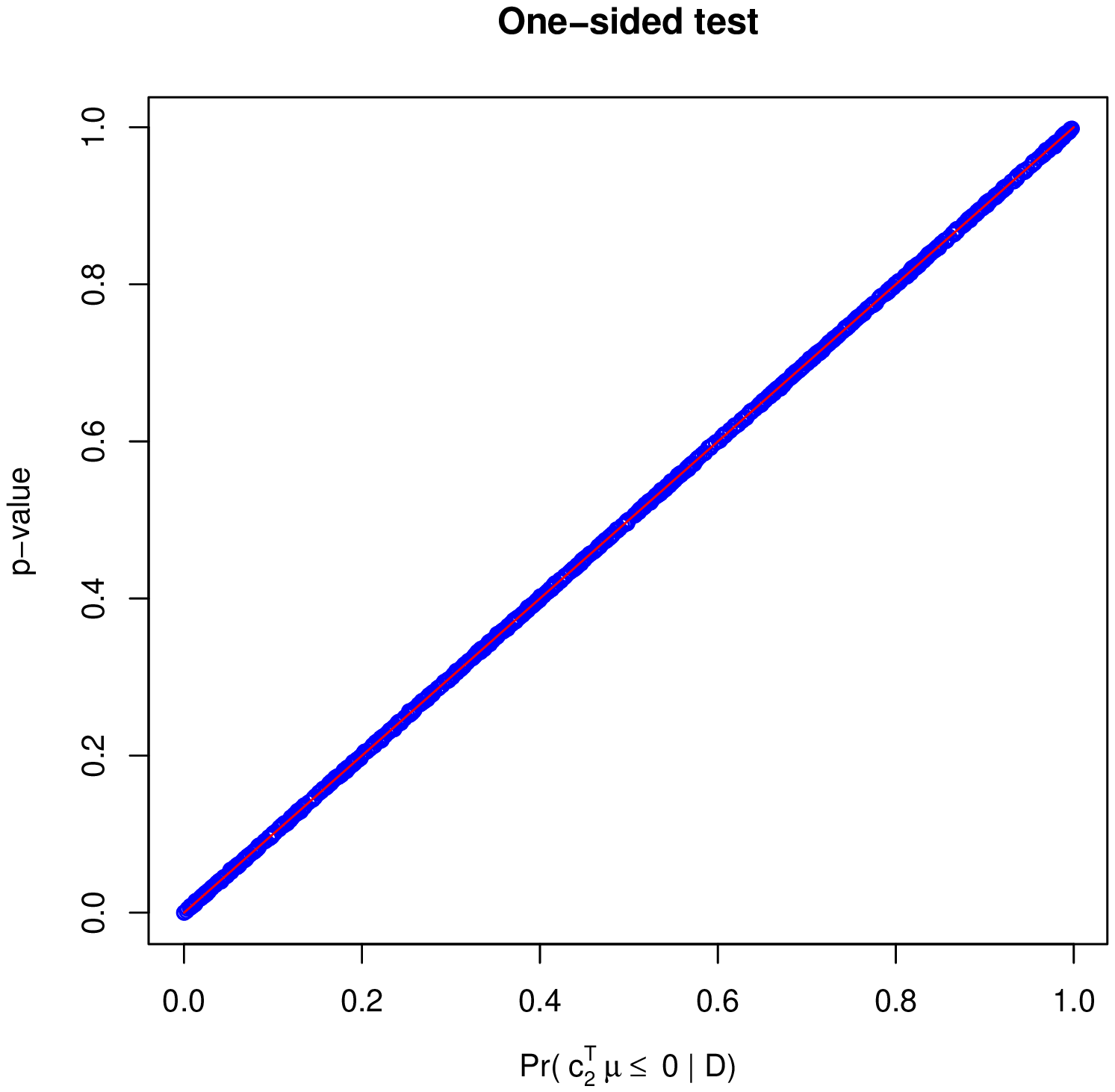}
\includegraphics[height=8cm,width=8cm]{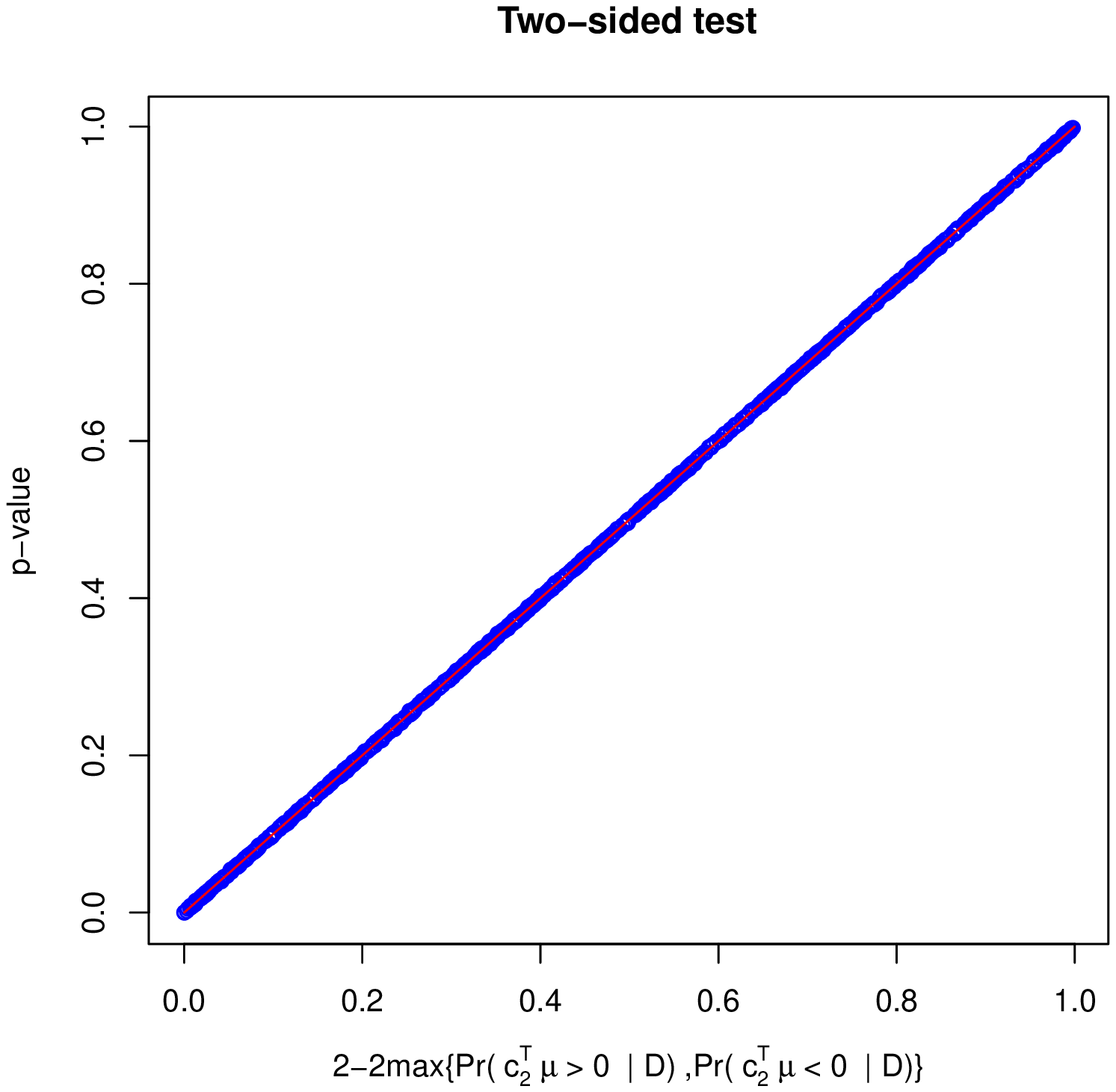}
\end{center}
\caption{The relationship between $p$-value and the posterior probability over 1000 replications under one-sided and two-sided hypothesis tests with multivariate normal outcomes under sample size of 100.
}
\label{multi}
\end{figure}

 In a numerical study, we compute the posterior probabilities of $\bc_k^{\top}\bmu \le 0$ for $k = 1,\ldots,K$, and compare them with the corresponding $p$-values. We take $K=2$ and $\bc_k$ to be a unit vector with 1 on the $k$th element and 0 otherwise, and assume a vague normal prior distribution for $\bmu$, i.e., $\bmu_0 = {\bf 0}$ and $\bSigma_0 = 1000 \bI_p$, where $\bI_p$ is a $p$-dimensional identity matrix. The relationship between the posterior probabilities and $p$-values is shown in Figure \ref{multi}, which is very similar to that in the univariate setting, which again demonstrates their equivalence.

\section{Discussion}
Berger and Sellke (1987) studied the point null for two-sided hypothesis tests, and noted discrepancies between the frequentist test and the Bayesian test based on the posterior probability.
The major difference between their work and the equivalence relationship between the posterior probability and $p$-value established here lies in the assumption of the prior distribution. Berger and Sellke (1987) assumed a point mass prior distribution at the point null hypothesis, which violates the regularity condition of continuity in Dudley and Haughton (2002), leading to the discrepancy between the posterior probability and $p$-value.
The equivalence relationship between the posterior probability and $p$-value for one-sided tests can be established from the theoretical results of Dudley and Haughton (2002), where the posterior probability of a half space is proven to converge to the standard normal CDF transformation of the likelihood ratio test statistic. A future direction of research is on more complex composite hypotheses tests involving multivariate normal outcomes.
Berger (1989) and Liu and Berger (1995) constructed a uniformly more powerful test than the likelihood ratio test for multivariate one-sided tests involving linear inequalities. Follman (1996) proposed a simple alternative to the likelihood ratio test. It would be of interest to study the relationship of these tests with the Bayesian counterparts based on posterior probabilities.

%
%On the other hand, if we assume a non-informative prior distribution $N(\theta_0,\sigma_0^2)$ for $\theta$,
%the likelihood can be expressed as $\prod_{i=1}^n \phi(y_{Ei}-y_{Si};\theta,2\sigma^2)$, where $\phi(\cdot;\theta,2\sigma^2)$ denotes the normal density function with mean $\theta$ and variance $2\sigma^2$. Based on the conjugacy of a normal prior distribution under a normal likelihood, the posterior distribution of $\theta$ follows $N(\theta_{*},\sigma_{*}^2)$, where
%\begin{align*}
%\theta_*&=\bigg(\frac{\theta_0}{\sigma_0^2}+\frac{\sum_{i=1}^n y_{Ei}-y_{Si}}{2\sigma^2}\bigg)\sigma_{*}^2,\\
%\sigma_{*}^2&=\bigg(\frac{1}{\sigma_0^2}+\frac{n}{2\sigma^2}\bigg)^{-1}.
%\end{align*}
%

%\subsection{One-Sample Hypothesis Test for Normal Outcomes}

%
%
%

%
%

%
%

 %

\newpage
\section*{References}
\begin{description}

\item
Bayarri, M. J. and Berger, J. O. (2004). The interplay of Bayesian and frequentist analysis. {\em Statistical Science} {\bf 19}, 58--80.

\item
Berger, J. O. (2003). Could Fisher, Jeffreys and Neyman have agreed on testing? (with discussion) {\em Statistical Science
}  {\bf 18}, 1--32.

\item
Berger, J. O. and Delampady M. (1987). Testing precise hypotheses. {\em Statistical Science} {\bf 2}, 317--335.

\item
Berger, J. O. and Sellke, T. (1987). Testing a point null hypothesis: the irreconcilability of P values and evidence. {\em Journal of the  American Statistical Association} {\bf 82}, 112--122.

\item
Benjamin, D. J., Berger, J. O., Johannesson, M., Nosek, B. A., Wagenmakers, E., et al. (2017). Redefine statistical significance. {\em Nature Human Behaviour} {\bf 2}, 6--10.

\item
Briggs, W. M. (2017). The substitute for p-values. {\em Journal of the American Statistical Association}  {\bf 112}, 897--898.

\item
 Berger, R. L. (1989). Uniformly more powerful tests for hypotheses concerning linear inequalities and normal means. {\em Journal of the American Statistical Association} {\bf 84}, 192--199.

\item
Casella, G. and Berger, R. L. (1987). Reconciling Bayesian and frequentist evidence in the one-sided testing problem. (with discussion) {\em Journal of the  American Statistical Association} {\bf 82}, 106--111.

\item
Concato, J. and Hartigan, J. A. (2016). P values: from suggestion to superstition. {\em Journal of Investigative Medicine} {\bf 64}, 1166--1171.

\item
 Colquhoun, D. (2014). An investigation of the false discovery rate and the misinterpretation of p-values. {\em Royal Society of Open Science} {\bf 1}, 140--216.
\item
Cumming, G. (2014). The new statistics: why and how. {\em Psychological Science} {\bf 25}, 7--29.

\item
Donahue, R. M. J. (1999). A note on information seldom reported via the P value.
{\em The American Statistician}  {\bf 53}, 303--306.

\item
Dudley, R. M. and Haughton, D. (2002). Asymptotic normality with small relative errors of
  posterior probabilities of half-spaces.
 \emph{ The Annals of Statistics} {\bf 30}, 1311--1344.

\item
Fidler, F., Thomason, N., Cumming, G., Finch, S., Leeman, J. (2004). Editors can lead researchers to confidence intervals, but can't make them think: Statistical reform lessons from medicine. {\em Psychological Science} {\bf 15}, 119--126.

\item
Follmann, D. (1996). A simple multivariate test for one-sided alternatives.
 {\em Journal of the American Statistical Association} {\bf 91}, 854--861.

\item
Gill, J. (2018). Comments from the New Editor. {\em Political Analysis} {\bf 26}, 1--2.

\item
Goodman, S. N. (1999). Toward evidence-based medical statistics. 1: the p value fallacy. {\em Annals of Internal Medicine} {\bf Volume 130}, 995--1004.

\item
Hubbard, R. and Lindsay, R. M. (2008). Why P values are not a useful measure of evidence in statistical significance testing. {\em Theory $\&$ Psychology} {\bf 18}, 69--88.

\item
Hung, H. J., O'Neill, R. T., Bauer, P., Kohne, K. (1997). The behavior of the p-value when the alternative hypothesis is true. {\em Biometrics} {\bf 53}, 11--22.

\item
Ioannidis, J. P. (2005). Why most published research findings are false. {\em PLoS Medicine} {\bf 2}, 124.

\item
Jager, L. R. and Leek, J. T. (2014).
An estimate of the science-wise false discovery rate
and application to the top medical literature. {\em Biostatistics} {\bf 15}, 1--12.

%\item
%Johnson, V. E. (2005). Bayes factors based on test statistics.
%{\em Journal of the Royal Statistical Society: Series B (Statistical Methodology)} {\bf 67}, 689--701.

\item
Johnson, V. E. (2013).
Revised standards for statistical evidence.
{\em Proceedings of the National Academy of Sciences} {\bf 110}, 19313--19317.
%
%\item
%Johnson, V. E. (2013b).
%Uniformly most powerful Bayesian tests. {\em Annals of Statistics} {\bf 41},
%1716--1741.

\item
Lee, J. J. (2010). Demystify statistical significance--time to move on from the p-value to Bayesian analysis. {\em Journal of the National Cancer Institute} {\bf 103}, 16--20.

\item
Leek, J., McShane, B. B., Gelman, A., Colquhoun, D., Nuijten, M. B., Goodman, S. N. (2017). Five ways to fix statistics. {\em Nature} {\bf 551}, 557--559.

\item
Lehmann, E. L. and Romano, J. P. (2005). Testing Statistical Hypotheses. New
York: Springer.

\item
Lindley, D. V. (1957). A statistical paradox. {\em Biometrika}  {\bf 44}, 187--192.

\item
Liu, H. and Berger, R. L. (1995). Uniformly more powerful, one-sided tests for hypotheses about linear inequalities.
{\em The Annals of Statistics} {\bf 23}, 55--72.

\item
McShane, B. B., Gal, D., Gelman, A., Robert, C., Tackett, J. L. (2018). Abandon statistical significance. arXiv:1709.07588

\item
Meng, X. L. (1994). Posterior predictive p-values.
{\em The Annals of Statistics} {\bf 22}, 1142--1160.
\item
Murtaugh, P. A. (2014). In defense of P values.
{\em Ecology} {\bf 95}, 611--617.

\item
Nuzzo, R. (2014).
Statistical errors: P values, the `gold standard'
of statistical validity, are not as reliable as many scientists assume.
{\em Nature} {\bf 506}, 150--152.

\item
Ranstam, J. (2012). Why the P-value culture is bad and confidence intervals a better alternative. {\em Osteoarthritis Cartilage} {\bf 20}, 805-808.

\item
Rosenthal, R. and Rubin, D. B. (1983). Ensemble-adjusted p values. {\em Psychological Bulletin} {\bf 94}, 540--541.

\item
Royall, R. M. (1986). The effect of sample size on the meaning of significance tests. {\em The American Statistician} {\bf 40}, 313--315.

\item
Rubin, D. B. (1984). Bayesianly justifiable and relevant frequency calculations for the applies statistician.
{\em The Annals of Statistics} {\bf 12}, 1151--1172.

\item
Rubin, D. B. (1998).
More powerful randomization-based p-values in double-blind trials with non-compliance.
{\em Statistics in Medicine} {\bf 17}, 371--385.

\item
Sackrowitz, H. and Samuel-Cahn, E.  (1999). P values as random variable-expected P values. {\em The American Statistician} {\bf 53}, 326--331.

\item
Sasabuchi, S. (1980). A test of a multivariate normal mean with
composite hypotheses determined by linear inequalities. {\em Biometrika}
{\bf 67}, 429--439.

\item
Savalei, V. and Dunn, E. (2015). Is the call
to abandon p-values the red herring of
the replicability crisis? {\em Frontiers in Psychology} {\bf 6}, 245.

\item
Schervish, M. J. (1996). P values: what they are and what they are not. {\em The American Statistician}  {\bf 50}, 203--206.

\item
Sellke, T., Bayarri, M. J., and Berger, J. O. (2001). Calibration of p-values for testing precise null hypotheses. {\em The American Statistician} {\bf 55}, 62--71.

\item
 Simmons, J. P., Nelson, L. D., Simonsohn, U. (2011). False-positive psychology: undisclosed flexibility in data collection and analysis allows presenting anything as significant. {\em Psychological Science} {\bf 22}, 1359--1366.

\item
Trafimow, D., Amrhein, V., Areshenkoff, C. N.,
Barrera-Causil, C. J., Beh, E. J., et al. (2018). Manipulating the
alpha level cannot cure significance testing. {\em Frontiers in Psychology} {\bf 9}, 699.

\item
Trafimow, D. and Marks, M. (2015). Editorial. {\em Basic and Applied Social Psychology} {\bf 37}, 1--2.

\item
Wagenmakers, E. J. (2007). A practical solution to the pervasive problems of p values. {\em Psychonomic Bulletin $\&$ Review} {\bf 14}, 779--804.

\item
 Wasserstein, R. L. and Lazar, N. A. (2016). The ASA's statement on p-values: context, process, and purpose. {\em The American Statistician} {\bf 70}, 129--133.

\end{description}

\newpage

\newpage

\newpage

\newpage

\end{document}